 \documentclass[final,3p,times,sort&compress]{elsarticle}

\usepackage{graphicx}
\usepackage{amssymb}
\usepackage{amsthm,amsbsy,stmaryrd}
\usepackage{xfrac,mathrsfs}
\usepackage{amsmath,amssymb,color}
\usepackage{empheq}
\usepackage{etoolbox}
\usepackage{commath}
\usepackage{bigints}
\usepackage{relsize}
\usepackage{lineno}
\usepackage{placeins} 
\usepackage{tikz}
\usepackage{booktabs,ctable,multirow,longtable} 
\usepackage[subrefformat=simple]{subfig}
\usepackage[percent]{overpic}
\usepackage[export]{adjustbox}
\usepackage{graphicx, array, blindtext}
\usepackage{mathtools}
\usepackage{float}
\usepackage{lineno}
\usepackage{multirow}
\usepackage{changepage}
\usepackage{array} 
\usepackage{makecell} 
\usepackage{adjustbox}
\usepackage{bbm}
\usetikzlibrary{shapes,positioning}
\usepackage{algorithm,algorithmic}
\usepackage{tikz}
\usepackage{ctable}
\usepackage[utf8]{inputenc}
\usepackage{pgfplots,pgfplotstable}
\pgfplotsset{
        table/search path={Tables},
    }
\pgfplotsset{compat=1.9}
\usetikzlibrary{backgrounds,automata}
\usepackage{bm}
\usetikzlibrary{calc}
\usepackage[export]{adjustbox}

\makeatletter
\newcommand\xcoord[2][center]{{%
    \pgfpointxy{1}{1}%
    \@tempdima=\pgf@x
    \pgfpointanchor{#2}{#1}%
    \@tempdimb=\pgf@x
    \pgfmathparse{\@tempdimb/\@tempdima}%
    \num{\pgfmathresult}%
}}
\newcommand\ycoord[2][center]{{%
    \pgfpointxy{1}{1}%
    \@tempdima=\pgf@y
    \pgfpointanchor{#2}{#1}%
    \@tempdimb=\pgf@y
    \pgfmathparse{\@tempdimb/\@tempdima}%
    \num{\pgfmathresult}%
}}
\makeatother

\makeatother

\usepackage{hyperref}

\graphicspath{{./Images/}} 

\newlength{\padding}  
\newlength{\actualHeight}
\newlength{\initialHeight}           

\newsavebox{\imgboxA}
\newsavebox{\imgboxB}

{\end{figure}}

{\end{minipage}\hspace{\padding}\ignorespaces}

{\end{minipage}}

\newcolumntype{P}[1]{>{\centering\arraybackslash}m{#1}}

\theoremstyle{remark}

\DeclareMathAlphabet{\mathpzc}{OT1}{pzc}{m}{it}
\DeclareMathAlphabet{\mathcalligra}{OT1}{calligra}{m}{it}

\newcommand{\bs}[1]{\boldsymbol{#1}}

\DeclareSubrefFormat{mysub}{{#2}}
\begin{document}

\begin{frontmatter}


\title{Fracture in concrete: X-ray tomography with in-situ testing, digital volume correlation and phase-field modeling}




\author[1]{A. Mishra}
\author[1]{P. Carrara\corref{cor2}}
\ead{pcarrara@ethz.ch}
\author[2]{M. Griffa}
\author[1]{L. De Lorenzis}
\cortext[cor2]{Corresponding author}


\address[1]{Department of Mechanical and Process Engineering, ETH Z\"{u}rich}
\address[2]{Swiss Federal Laboratories for Materials Science and Technology, Empa}

\begin{abstract}

We test and simulate the mesoscopic cracking behavior of specimens made of a standard concrete mixture. To this end, we combine stable wedge-splitting fracture experiments performed during X-ray tomography, their analysis with digital volume correlation providing the full three-dimensional displacement field, and phase-field cohesive fracture modeling. In our computations, we apply the measured boundary conditions and model the actual heterogeneous material structure at the mesoscopic scale. Within the phase-field model, we explicitly distinguish among (thus individually represent) the mesostructural features of distinct material phases with size above a threshold of 1 mm, while we homogenize pores and finer aggregates below this threshold within the cementitious mortar matrix, with material parameters characterized accordingly. We compare experimental and numerical results in terms of both local and global quantities. 

\end{abstract}




\end{frontmatter}


\section{Introduction}
The macroscopic behavior of concrete is significantly influenced by its highly heterogeneous structure, which is composed of a random distribution of air voids/porous patches and aggregates with varying shapes and sizes, ranging from few tens of microns to few centimeters, embedded in a cementitious mortar matrix \cite{Pietro:2016,Pietro:2018,Yang:2017,Thilakarth:2020}. In this work, we focus on the  \textit{mesoscale}, which we define as the scale at which the aggregates and voids with size above an appropriate threshold are explicitly resolved, while those with size below the threshold are homogenized within the cementitious mortar matrix. Clearly, the choice of the size threshold is to some extent arbitrary and depends on the characteristic length(s) involved in the phenomena of interest. 
 Fracture of concrete is heavily influenced by mesoscopic details, such as the shape and location of aggregates and pores, and its comprehensive prediction at the mesoscale requires the accurate and robust description of complex processes such as crack initiation, propagation, branching, and merging in 3D. For calibration and validation of fracture models, accurate experimental data are of crucial importance \cite{Jailin:2017,Carpiuc:2015,MichelG:2012}. 

The most common experimental setups used to study concrete fracture include notched and un-notched three-point bending \cite{Malvar}, unconfined uniaxial compression, Brazilian \cite{Carneiro:1943}, L-shaped panel \cite{Winkler} and double notch tests \cite{Nooru:1993} among others (see \cite{Carpiuc:2015,Hoover:2013} and references therein). Traditionally, in such tests pointwise displacements are measured using linear variable displacement transducers and/or extensometers, while force values are delivered by load cells. More recently, these measuring techniques have been increasingly complemented by full-field, non-contact ones such as \textit{digital image correlation} (DIC), which delivers the full displacement field on a portion of the specimen's free surface, conveniently treated to create a high-contrast speckle pattern \cite{Sutton:1983, Pan2009, Hild:2006}. During a DIC measurement, digital cameras record images of the region of interest during a mechanical test, then a suitably parameterized transformation is applied to the image of the deformed surface with the aim of recovering the reference image, i.e., that of the undeformed surface. The maximization of a cross-correlation score between reference and deformed images enables the identification of the optimal parameters of the transformation, delivering the desired displacement field \cite{Sutton:1983}. This procedure is also often termed \textit{image registration} in the computer vision and image processing fields. To avoid a displacement resolution of integer multiples of the pixel size, the pixel values of the images are interpolated using a so-called \textit{sub-pixel interpolation}, a procedure that enables to reach a precision of 1/10-1/20 of the physical pixel size, i.e. \textit{sub-pixel precision}. DIC has been employed in several studies, such as \cite{Wu:2017,Bu:2020,Carpuic:2017,Carpiuc:2015,Nguyen2016_1}, to study concrete fracture from a macroscopic perspective, i.e., considering a nominally homogeneous material. 

The displacement fields provided by DIC are limited to the imaged portion of the specimen's free surface. This limitation can be overcome by adopting \textit{X-ray computed tomography} (X-CT)  \cite{Buljac2018,Bay1999}. X-CT is a non-destructive computational imaging technique able to retrieve the 3D heterogeneous structure of a sample (namely, the \textit{tomogram}) based on the X-ray attenuation of the constituent materials. For each tomogram, a set of radiographs, i.e., 2D projection images, needs to be acquired at several (typically on the order of 10$^3$) projection angles. The obtained tomogram constitutes a 3D map of the X-ray linear attenuation coefficient at each material point. The acquisition of several tomograms at different stages of a mechanical test, a procedure usually referred to as \textit{in-situ testing tomography}, allows to observe the evolution of the 3D geometry of the sample and, once registered, enables the estimation of the displacement vector field, similarly to what is achieved in 2D by DIC \cite{Bay1999,Buljac2018,Roux:2008}. The 3D DIC counterpart is denoted as \textit{digital volume correlation} (DVC) \cite{Buljac2018,Bay1999} and allows the measurement of the 3D displacement vector field with a sub-pixel precision of up to 1/10 of the physical voxel size. Among other advantages, DVC delivers the complete experimental boundary conditions of the test, which are crucial to ensure a reliable comparison between experimental and numerical results \cite{Buljac:2017,Madi:2013}. Moreover, the displacement field obtained after registration can be applied with opposite sign to the deformed tomogram and the result subtracted from the reference one. This difference (typically its absolute value) provides the \textit{correlation residuals} which, except for artifacts due to bad correlation, stem from two major contributions. The first is the intrinsic image noise and artifacts, due to, e.g., noise of the X-ray detector and interactions between X-rays and matter (e.g., scattering) not accounted for during image reconstruction. The second contribution is due to features that appear during the test, namely, features present in the deformed image but not in the reference one that, hence, cannot be registered. An example of such features are cracks formed during the test, which, thanks to the sub-voxel precision, can be detected from their early onset, i.e., for crack widths well below the image resolution \cite{Bay1999,Buljac2018,Roux:2008}. 

 In the literature, numerous contributions are dedicated to modeling of concrete fracture. Some models \cite{Hai:2024,Wu:2017,Lorentz:2017,Bazant:1983,Bocca:1991,Comi:2001, Schroeder:2022,Bazant:1987} adopt a phenomenological and macroscopic standpoint, which, although capable of describing the overall behavior \cite{Lopez:2007}, cannot account for the effects due to the mentioned material heterogeneities \cite{Pietro:2016,Yang:2017,Pietro:2018}. To capture such effects, contributions such as \cite{Li:2021,Su:2010,Yang:2008,Baxter:2001} adopt stochastic, spatially varying fracture properties, or artificially generated aggregates with idealized shapes which are packed in the computational domain following an a priori defined probability distribution  \cite{Lopez:2007,Wriggers:2006,Zhang:2017,Thilakarth:2020,Lopez:20071}. In both cases, a realistic concrete response cannot be accurately reproduced, either because of an inaccurate  description of the material properties variations or due to the oversimplified geometry and limited packing capability of the approaches involving idealized particles \cite{Pietro:2016}. 
 
 X-CT can solve this problem by providing the actual heterogeneous material phase distribution of the tested specimens \cite{Pietro:2018,Withers:2021,Yang:2017}. For this reason, fracture modeling of cementitious materials based on X-ray tomograms has recently gained interest, see e.g. \cite{Nguyen2015_1, Nguyen2016,ren2015two, Yang:2019, Huang:2015,Ren:2018, Nguyen:2016a,Tsitova:2021, Nguyen2016_1,Yang:2017, Tsitova:2022,Ren:2014}. However, among these contributions, 
 \cite{Yang:2017,Yang:2013,Huang:2015,Ren:2018, Nguyen:2016a,Tsitova:2021, Nguyen2016_1, Tsitova:2022} have used in-situ test setups with unstable crack propagation branches, which precludes the observation of the post-peak behavior, where most of the crack propagation takes place. 
Moreover, the mix designs used in these investigations often do not represent real concrete. For example, the studies in \cite{Tsitova:2022, Tsitova:2021} have involved a low volume fraction of sand with maximum aggregate sieve size of 2 mm while the finest fraction with size $\le$200 $\mu$m was not included. \cite{Nguyen:2016a} studied lightweight concrete, where the explicitly resolved aggregates were expanded polystyrene beads with diameter ranging from 1 mm to 2 mm. 
Also, the specific choice of the raw materials in \cite{Tsitova:2022, Tsitova:2021, Nguyen:2016a} alleviated some difficulties in the identification of the different phases in the X-ray tomograms (a task denoted as \textit{segmentation}), which systematically arise when working with standard concrete mixes. These difficulties are due, on the one hand, to the typically similar X-ray attenuation coefficient of aggregates and cement paste, and, on the other hand, to the dense aggregate packing of real concrete mixes \cite{Pietro:2016,Pietro:2018,Poinrad:2011, Stamati:2018,Stamati_2021}.

The phase-field approach is becoming increasingly popular for modeling concrete fracture, due to its flexibility and strong theoretical background \cite{Ambati:2014, Wu:2017, Wu:2021, Lorentz:2017, Li:2021, Li:2019, Hai:2024}. Originally proposed in \cite{Bourdin:2000} as a regularization of the variational brittle fracture theory of Francfort and Marigo \cite{Francfort:1998}, phase-field fracture models can be also interpreted as a special class within the family of gradient damage models. The approach regularizes a sharp crack through the spatial evolution of a continuous damage parameter, which varies smoothly to describe the transition from sound to fully cracked material. Since damage evolution is dictated by the minimization of the total energy \cite{Bourdin:2000}, the approach elegantly models complicated fracture processes without requiring crack tracking algorithms or the a priori knowledge of the crack path. This makes it attractive not only for brittle fracture but also for ductile fracture \cite{Ambati:2015, Alessi2018}, fatigue \cite{Carrara:2020,Mesgarnejad2019,Heinzmann2024}, anisotropic \cite{Nagaraja:2023} or cohesive fracture \cite{Verhoosel:2013,WU2017,Nguyen2016,Chen:2022,Chen2021, Conti:2016,Lorentz:2011} among many other problems.

In this paper, we aim to study fracture of ordinary concrete, accounting for its heterogeneity, with a combination of X-CT, in-situ testing, DVC and phase-field modeling. We consider a standard concrete mix design, involving ordinary Portland cement and quartzite aggregates. In imaging as well as in modeling, we resolve the heterogeneous material structure at the \textit{mesoscale} (hence, we speak of concrete \textit{mesostructure}) as follows. We explicitly resolve the heterogeneities above a size threshold of 1 mm, with aggregates of size up to 11.25 mm, while we homogenize pores and finer aggregates below this threshold within the cementitious mortar matrix. We design the experimental characterization of the material parameters to ensure consistency with this mesostructural viewpoint. We adopt a wedge splitting test (WST) setup \cite{Bruhweiler:1990,Neuner:2022,Yang:2020} to perform stable mode-I crack propagation tests inside an X-ray tomograph. For each test, several tomograms are obtained which, after DVC analysis, deliver the 3D displacement field and the crack evolution, from the initial stage of linear elastic behavior to complete failure. Real boundary conditions are also extracted and used as input in the numerical computations, which are performed with the phase-field approach to cohesive fracture.

This paper is organized as follows. Section~\ref{S:pf_phase_field} introduces the adopted phase-field model for cohesive fracture. Section~\ref{S:pf_mat_char}  illustrates the experimental characterization of material and model parameters. The in-situ experiments and the related DVC analyses are detailed in Sections~\ref{S:pf_WST_res} and \ref{S:pf_dvc_an}, respectively, while Section~\ref{S:pf_validation} illustrates the numerical computations and the comparison between experimental and numerical results. The main conclusions are finally summarized in Section~\ref{S:pf_conclusions}.

\section{Modeling approach}
\label{S:pf_phase_field}
This section presents the main modeling and computational choices within the phase-field approach to fracture for the numerical prediction of the specimen behavior (to be illustrated later). In the following, we assume linearized kinematics, rate independence and negligible body forces, along with quasi-static and isothermal conditions. Additionally, each constituent of the concrete mesostructure is considered initially linearly elastic, isotropic and homogeneous.

\subsection{General formulation}
\label{S:pf_PF}
We describe the concrete mesostructure as composed of air pores/porous patches, aggregates, mortar matrix and an interface transition zone (ITZ) between aggregates and matrix. As mentioned earlier, we explicitly resolve all heterogeneities (pores and aggregates) with size above 1~mm, while the fine aggregates and the pores below such threshold are homogenized in the mortar matrix. The ITZ is modeled as a thin continuum phase around the explicitly resolved aggregates  \cite{Wang:2021}.
 While cracking is expected to take place mainly within the ITZ and the mortar matrix, it is allowed within the aggregates as well.

\begin{figure} 
   \centering
   \subfloat[]{
    \begin{tikzpicture}
   \node [anchor=north west,inner sep=0] (sharp) at (0,0){\includegraphics[width=0.3\textwidth]{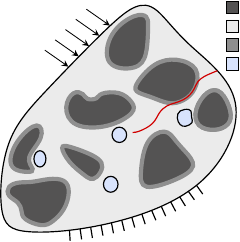}};
\draw (1.2,-.5) node {\footnotesize $\boldsymbol{\bar t}$};
\draw (.8,-1.5) node {\footnotesize $\partial\Omega_N$};
\draw (3.75,-4.5) node {\footnotesize $\boldsymbol{\bar u}$};
\draw (4.6,-4) node {\footnotesize $\partial\Omega_D$};
\draw (5.28,-0.18) node {$\footnotesize\Omega_a$};
\draw (5.3,-0.56) node {$\footnotesize\Omega_m$};
\draw (5.26,-0.94) node {$\footnotesize\Omega_i$};
\draw (5.4,-1.32) node {\footnotesize Pores};
\draw (3,-2.39) node {$\footnotesize\Gamma_c$};

   \end{tikzpicture}\label{fig:sharp}
   }\hspace{2cm}
   \subfloat[]{    \begin{tikzpicture}
   \node [anchor=north west,inner sep=0] (regul) at (0,0){\includegraphics[width=0.3\textwidth]{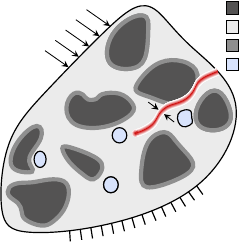}};
\draw (1.2,-.5) node {\footnotesize $\boldsymbol{\bar t}$};
\draw (.8,-1.5) node {\footnotesize $\partial\Omega_N$};
\draw (3.75,-4.5) node {\footnotesize $\boldsymbol{\bar u}$};
\draw (4.6,-4) node {\footnotesize $\partial\Omega_D$};
\draw (5.28,-0.18) node {$\footnotesize\Omega_a$};
\draw (5.3,-0.56) node {$\footnotesize\Omega_m$};
\draw (5.26,-0.94) node {$\footnotesize\Omega_i$};
\draw (5.4,-1.32) node {\footnotesize Pores};
\draw (2.85,-2.9) node {$\footnotesize\alpha$};
\draw (2.98,-2.31) node {$\footnotesize4\ell$};
   \end{tikzpicture}\label{fig:regul}
   }
        \caption{Schematic representation of the specimen domain: (a) sharp representation of the crack $\Gamma_c$ and (b) its phase-field regularization.}
        \label{fig:regularisation}
\end{figure}
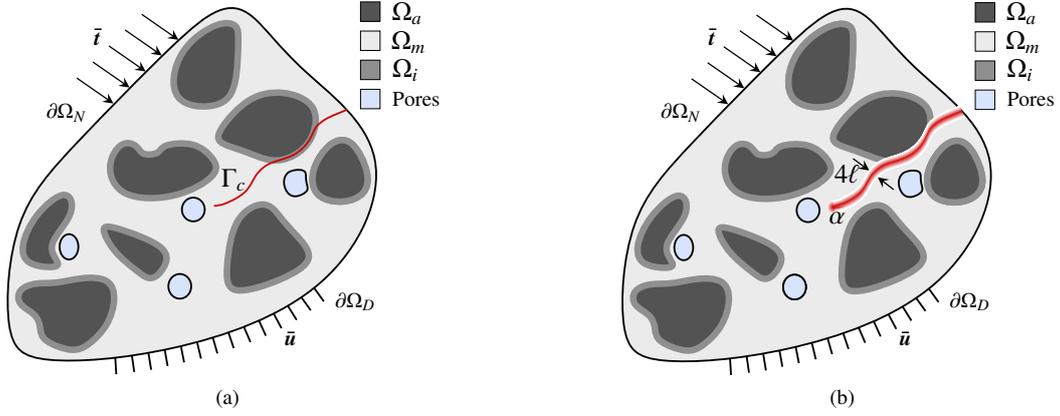

 We denote the specimen domain as $\Omega=\bigcup_{I}\Omega_I$ with $I=m,\,a,\,i$, where $\Omega_m$, $\Omega_a$ and $\Omega_i$ are the (disjoint) portions of $\Omega$ occupied by mortar matrix, aggregates and ITZ, respectively  (Fig.~\ref{fig:regularisation}). Note that the matrix and the ITZ subdomains are not simply connected but may contain pores. 
 The displacement in $\Omega$ is denoted as $\bs u(\bs x)$,  where $\bs{x}$ is the spatial coordinate vector. The strain tensor is $\bs \varepsilon(\bs x)={\nabla}^s\bs u(\bs x)$ where ${\nabla}^s(\pmb{\bullet})$ is the symmetric gradient of $(\bullet)$, while $\bs \sigma(\bs x)$ denotes the Cauchy stress tensor. The boundary $\partial \Omega$ has outward normal unit vector $\bs n$ and is decomposed into a Neumann part $\partial \Omega_N$, with prescribed tractions $\bar{\bs t}(\bs x)$, and a Dirichlet part $\partial \Omega_D$, with imposed displacements $\bar{\bs u}(\bs x)$, such that  $\partial \Omega = \partial \Omega_D \cup \partial \Omega_N$ and $\partial \Omega_D \cap \partial \Omega_N = \emptyset$. The pore space boundaries are considered as internal boundaries subjected to homogeneous Neumann boundary conditions (i.e., they are treated as free boundaries). The damage variable $\alpha(\bs x)$  approximates the sharp cracks  (i.e., $\Gamma_c$ in Fig.~\ref{fig:sharp}) in a smeared fashion (Fig.~\ref{fig:regul}) and varies smoothly between $0$ at the intact state and $1$ at the completely damaged state, within a support whose width is controlled by a length scale parameter $\ell$.

The governing equations for the phase-field model at the discrete load step $n$ are derived starting from the regularized total energy functional
  
  \begin{equation}
 \label{eq:pf_penalty_reg_total_energy}
 \begin{aligned}
 	\mathcal{E}_n(\bs{u},\alpha) = \int_{\Omega}\underbrace{a\left(\alpha(\bs x),\bs{x}\right)\psi^+\left(\bs{\varepsilon}(\bs{u}(\bs x)),\bs{x}\right)+\psi^-\left(\bs{\varepsilon}(\bs{u}(\bs x)),\bs{x} \right)\vphantom{\int_{\Omega}\dfrac{G_{f}
 	\left(\bs x\right)}{c_w}\left(\dfrac{w\left(\alpha(\bs x)\right)}{\ell}+\ell|\nabla\alpha(\bs x)|^2\right)}}_{\psi(\bs{\varepsilon}, \alpha,\bs{x})}d\bs{x}
 	&+\ \int_{\Omega}\underbrace{\dfrac{G_{f}
 	\left(\bs x\right)}{c_w}\left(\dfrac{w\left(\alpha(\bs x)\right)}{\ell}+\ell|\nabla\alpha(\bs x)|^2\right)}_{\varphi(\alpha, \nabla\alpha,\bs{x})}d\bs x \\&
 \quad \quad \quad	-\underbrace{\int_{\partial\Omega_{N}}\bar{\bs{t}}_n\cdot\bs{u}(\bs x)ds}_{\mathcal{L}_n(\bs u)} 
 +\underbrace{\dfrac{\rho}{2}\int_{\Omega}\langle \alpha(\bs x) \rangle^2_-d\bs x\vphantom{\int_{\partial\Omega_{N}}\bar{\bs{t}}\cdot\bs{u}ds}}_{\text{penalty}}.
 \end{aligned}
  \end{equation}


\noindent Here $\psi(\bs{\varepsilon}, \alpha, \bs{x})$ is the elastic strain energy density, $\varphi(\alpha, \nabla\alpha,\bs{x})$ is the fracture energy density, $\mathcal{L}_n(\bs u)$ is the work of the external traction at the current load step, $\bar{\bs{t}}_n$, and the last penalty term enforces non-negativity of $\alpha$ \cite{Gerasimov:2019} and depends on a penalty parameter $\rho$ discussed in  Sect.~\ref{S:pf_num_aspects}. 

The elastic strain energy density $\psi(\bs{\varepsilon}, \alpha, \bs{x})$ is defined phase-wise and depends on the damage variable $\alpha$ through a monotonically decreasing degradation function $a\left(\alpha,\bs x\right)$ that is introduced in Sect.~\ref{S:pf_degradation_func}.
 To allow for an asymmetric cracking behavior under tensile and compressive stress states and prevent unphysical crack interpenetration upon closure, the undegraded total strain energy density is decomposed into active and inactive parts \cite{Amor:2009,Miehe:2010,Vicentini2024}.
The active part, $\psi^+\left(\bs{\varepsilon}(\bs x),\bs x\right)$, is assumed to drive the damage evolution, whereas the inactive part, $\psi^-\left(\bs{\varepsilon}(\bs x),\bs x\right)$, represents the residual energy in fully damaged conditions \cite{DeLorenzis:2021,Vicentini2024}.  
  While many energy decompositions (or splits) have been proposed in the literature \cite{DeLorenzis:2021}, we focus here on two common variants. The first is the volumetric-deviatoric decomposition \cite{Amor:2009}, referred to as \texttt{\texttt{vol/dev}} split
 \begin{equation}
 	\label{eq:pf_vol_Dev}
  \begin{aligned}
      \psi^{+}\left(\bs{\varepsilon}(\bs x),\bs x\right)=\dfrac{1}{2}\kappa(\bs x)\left\langle tr\left(\bs{\varepsilon}(\bs x)\right)\right\rangle_{+}^2+\mu(\bs x) \left(\bs{\varepsilon}_{d}(\bs x):\bs{\varepsilon}_{d}(\bs x)\right),\quad \psi^{-}\left(\bs{\varepsilon}(\bs x),\bs x\right)=\dfrac{1}{2}\kappa(\bs x)\left\langle tr\left(\bs{\varepsilon}(\bs x)\right)\right\rangle_{-}^2,
  \end{aligned}
 \end{equation}
the second is based on a \texttt{spectral} decomposition of the strain tensor \cite{Miehe:2010} and reads 
 \begin{equation}
 	\label{eq:pf_spectral_Split}
  \begin{aligned}
 \psi^{\pm}\left(\bs{\varepsilon}(\bs x),\bs x\right)=\dfrac{1}{2}\lambda(\bs x)\left\langle tr\left(\bs{\varepsilon}(\bs x)\right)\right\rangle_{\pm}^2+\mu (\bs x)tr\left(\bs{\varepsilon}^2_{\pm}(\bs x)\right), \quad  \text{with}\quad \bs{\varepsilon}_{\pm}(\bs x)=\sum_{a=1}^3\langle\varepsilon^a(\bs x)\rangle_{\pm}\bs{v}^a(\bs x)\otimes\bs{v}^a(\bs x).
 \end{aligned}
 \end{equation}
In the above equations,
 $tr(\bullet)$ denotes the trace of tensor $(\bullet)$, $\bs \varepsilon_{d}(\bs x)$ the deviatoric component of the strain tensor, $\langle(\bullet)\rangle_{+}$ and  $\langle(\bullet)\rangle_{-}$ give the positive and negative part of $(\bullet)$, $\varepsilon^a$ are the eigenvalues and $\bs v^a$ the eigenvectors of the strain tensor, and $\lambda(\bs x)$, $\mu(\bs x)$ and $\kappa(\bs x)$ are phase-wise Lam{\'e} parameters and bulk moduli given by 
 \begin{equation}
\label{eq:pf_lame_par}
        \lambda(\bs x) =
        \begin{cases}
         \lambda_{m}\,, & \quad\text{if} \quad \bs x \in \Omega_m\,, \\
        \lambda_{a}\,, & \quad \text{if} \quad \bs x \in \Omega_a\,, \\
        \lambda_{i}\,, & \quad \text{if} \quad \bs x \in \Omega_i \,,
        \end{cases}
      \quad \quad \mu(\bs x) =
        \begin{cases}
           \mu_{m}\,, & \quad\text{if} \quad \bs x \in \Omega_m\,, \\
        \mu_{a}\,, & \quad \text{if} \quad \bs x \in \Omega_a\,, \\
        \mu_{i}\,, & \quad \text{if} \quad \bs x \in \Omega_i \,.
        \end{cases}
        \quad \text{ and} \quad \kappa(\bs x) =
        \begin{cases}
           \kappa_{m}\,, & \quad\text{if} \quad \bs x \in \Omega_m\,, \\
        \kappa_{a}\,, & \quad \text{if} \quad \bs x \in \Omega_a\,, \\
        \kappa_{i}\,, & \quad \text{if} \quad \bs x \in \Omega_i \,.
        \end{cases}
        \end{equation}
 \noindent where $(\lambda_m,\,\mu_m)$, $(\lambda_a,\,\mu_a)$ and $(\lambda_i,\,\mu_i)$ are the Lam{\'e} parameters and $\kappa_m$, $\kappa_a$, $\kappa_i$ the bulk moduli corresponding to matrix, aggregates and ITZ, respectively. Clearly, only two elasticity parameters can be fixed independently for each phase.

The fracture energy density $\varphi(\alpha, \nabla\alpha,\bs{x})$ consists of a local contribution, depending on the monotonically increasing \textit{dissipation function} $w\left(\alpha \right)$ (Sect.~\ref{S:pf_degradation_func}), and a non-local contribution depending on the gradient of the damage variable. The normalization constant $c_w$ ensures that the dissipated energy per unit surface of the regularized crack corresponds to the fracture toughness $G_{f}(\bs x)$ \cite{Gerasimov:2019,Braides:1998}. The spatial dependence of $G_{f}$ is introduced to distinguish between different phases, namely 
  \begin{equation}
        \label{eq:pf_Gf}
        G_f(\bs x) =
        \begin{cases}
        G_{f,m}\,, & \text{if}\quad \bs x \in \Omega_m\,,\\
        G_{f,a}\,, & \text{if} \quad\bs x \in \Omega_a\,,\\
        G_{f,i}\,, & \text{if} \quad \bs x \in \Omega_i\,,
        \end{cases}
             \end{equation}
where $G_{f,m}$, $G_{f,a}$ and $G_{f,i}$ are the fracture toughnesses of matrix, aggregates and ITZ, respectively. 
 
 \subsection{Choice of degradation and dissipation functions}
 \label{S:pf_degradation_func}
\begin{figure}[!htb]
\centering
\includegraphics[width=0.35\textwidth]{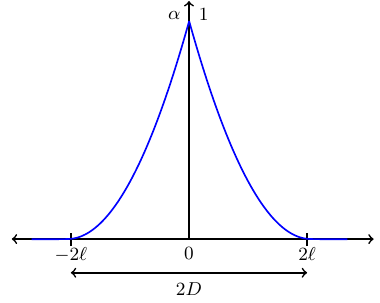}  
     \caption{Optimal damage profile at failure obtained using the dissipation function \eqref{eq:pf_dissipation_func}.}\label{fig:pf_dam_profile}
 \end{figure}

The degradation and dissipation functions are important ingredients of the phase-field fracture formulation and their choice influences the damaging behavior \cite{Pham:2010,Pham:2011,Kuhn:2015,WU2017}.  In this work, we adopt the linear dissipation function
  \begin{equation}
  \label{eq:pf_dissipation_func}
  	w(\alpha)= \alpha,
  \end{equation}
\noindent which ensures the existence of a linear elastic stage before damage initiation \cite{Pham:2011}. Then $c_w = 8/3$, while the half-width of the support of the damage variable at full damage in a one-dimensional setting is $D=2\ell$  (Figs.~\ref{fig:regul} and \ref{fig:pf_dam_profile}).

\begin{figure*}[!htb]
       \centering
                           \subfloat[]{
                \includegraphics[width=0.42\textwidth]{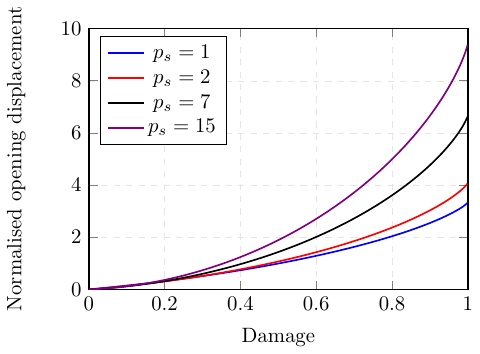}
               \label{fig:pf_displ_alpha}}
               \subfloat[]{
                \includegraphics[width=0.42\textwidth]{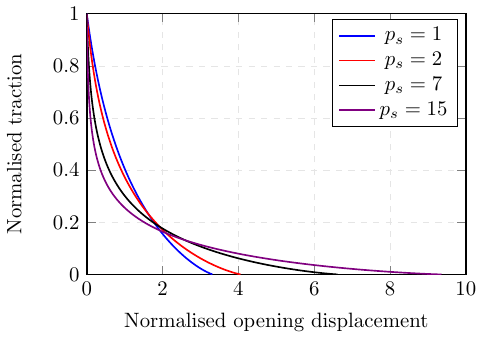}
               \label{fig:pf_peffect}}
               \hspace{0.05cm}
                     \caption{Cohesive-like law obtained from \eqref{eq:pf_cohesive_lorentz} for different values of the shape parameter $p_s$: (a) relation between the damage variable $\alpha$ and the displacement jump across the crack, and (b) traction-separation law.}
\label{fig:pf_1d_bar}
\end{figure*}

behavior of concrete
         \begin{equation}
          \label{eq:pf_degradation_func}\begin{aligned}
        	a(\alpha,\bs x)=\dfrac{(1-\alpha)^2}{(1-\alpha)^2+P(\alpha, \bs x)} +\eta_l \quad\text{ with } \quad P(\alpha,\bs x)=\dfrac{3E(\bs x)G_{f}(\bs x)}{4f_{t}^2(\bs x)\ell}\alpha(1+p_s\alpha),
         \end{aligned}
        \end{equation}
        
 \noindent where the parameter $\eta_l$ avoids numerical instabilities by providing a small residual stiffness at complete degradation \cite{Bourdin:2000}. Also,  $E(\bs x)$  and $f_t(\bs x)$ read 
 
 \begin{equation}
        \label{eq:pf_ym}
      E(\bs x) =
        \begin{cases}
         E_{m}\,, & \quad\text{if} \quad x \in \Omega_m\,, \\
        E_{a}\,, & \quad \text{if} \quad x \in \Omega_a\,, \\
        E_{i}\,, & \quad \text{if} \quad x \in \Omega_i \,,
        \end{cases}
      \quad \text{ and} \quad f_t(\bs x) =
        \begin{cases}
           f_{t,m}\,, & \quad\text{if} \quad x \in \Omega_m\,, \\
        f_{t,a}\,, & \quad \text{if} \quad x \in \Omega_a\,, \\
        f_{t,i}\,, & \quad \text{if} \quad x \in \Omega_i \,.
        \end{cases}
            \end{equation}

  \noindent  with $E_m$, $E_a$ and $E_i$ indicating the Young's moduli of mortar, aggregates and ITZ, respectively. Similarly, $f_{t,m}$, $f_{t,a}$ and $f_{t,i}$ are the tensile strengths of mortar, aggregates and ITZ, respectively, while $p_s\geq1$ is a model parameter that governs the post-peak response \cite{Lorentz:2012}. 
  Studying the response of a long one-dimensional bar under tension using \eqref{eq:pf_dissipation_func}-\eqref{eq:pf_degradation_func} in \eqref{eq:pf_penalty_reg_total_energy}, it can be analytically shown that as $\ell\to0$ the post-peak behavior converges toward the following cohesive-like traction-separation law \cite{Lorentz:2011,Lorentz:2012}
  \begin{equation}\label{eq:pf_cohesive_lorentz}
    \begin{split}
    	\bar{\tau}(\alpha) = &\frac{1-\alpha}{\sqrt{1+p_s\alpha}} , \quad 
   		 \bar\delta(\alpha)= 3 \bar\tau(\alpha)\int_{0}^{\alpha}\dfrac{\bar{\alpha}(1+p_s\bar{\alpha})}{(1-\bar{\alpha})^2}\bar{G}(\alpha,\bar{\alpha})^{-1/2}d\bar{\alpha} \\
   		\\	&\text{ with }\quad \bar{G}(\alpha,\bar{\alpha})=4\bar{\alpha}\bigg[1-\bigg(\dfrac{1-\alpha}{1-\bar{\alpha}}\bigg)^2\dfrac{1+p_s\bar{\alpha}}{1+p_s\alpha}\bigg],
     \end{split}
    \end{equation}
  where  $ \bar\delta(\alpha) $  and $ \bar\tau(\alpha)$ respectively denote the non-dimensional displacement jump $\delta(\alpha) $ and traction $\tau(\alpha)$ across the crack, given by
   	   	\begin{equation}\label{eq:pf_cohesive_lorentz2}
   		\displaystyle  \bar\tau(\alpha) = \frac{\tau(\alpha)}{f_t\,,} \qquad \text{and}\qquad \displaystyle \bar\delta(\alpha)=\frac{\delta(\alpha)f_t}{G_f}\,.
   	\end{equation}	
 \noindent Note that, for a given $p_s$, the displacement jump $\bar\delta(\alpha)$ depends only on the damage variable $\alpha$, which thus represents a measure of crack opening \cite{Lorentz:2011}. From \eqref{eq:pf_cohesive_lorentz}-\eqref{eq:pf_cohesive_lorentz2}, since for a fully formed crack ($\alpha=$1) it is $\tau= 0$, we obtain the critical or ultimate opening value $\delta_u$ as \cite{Lorentz:2012}
	   	   	\begin{equation}\label{eq:pf_critical_delta}
   		\displaystyle  \delta_u=\frac{3\pi G_f}{4 f_t}\sqrt{p_s+1}\,.
   	\end{equation}	
Fig.~\ref{fig:pf_1d_bar} shows the law \eqref{eq:pf_cohesive_lorentz} for different values of $p_s$, highlighting the role of this parameter in controlling the shape of the post-peak softening branch \cite{Lorentz:2012}. Reducing $p_s$ yields  a more brittle  response and a smaller ultimate displacement jump \cite{Lorentz:2011}.
 
\begin{figure*}[!htb]
       \centering
                \includegraphics[width=0.42\textwidth]{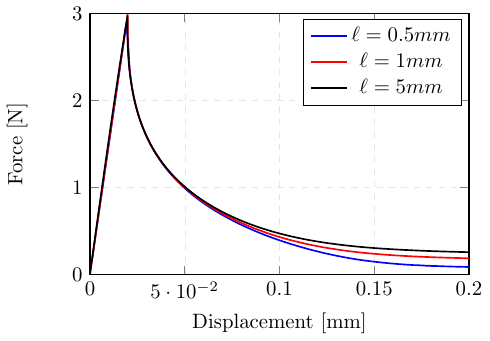}
                     \caption{Force-displacement curve  for a bar of length 200 mm with different values of $\ell$ using the cohesive-like degradation function \eqref{eq:pf_degradation_func}.}
\label{fig:pf_leffect}
\end{figure*}

An important property of the model \eqref{eq:pf_degradation_func} is that the response obtained for a long one-dimensional bar under tensile loading is independent of $\ell$, provided that the latter is sufficiently small compared to the characteristic size of the domain \cite{Lorentz:2012}. In particular, this holds true for the peak stress (Fig.~\ref{fig:pf_leffect}), so that $\ell$ is a pure numerical parameter. These results are derived analytically for the one-dimensional bar under tension \cite{Lorentz:2011}, however numerical analyses confirm their validity also in higher dimensions  \cite{Lorentz:2012,Lorentz:2017}. For heterogeneous domains such as the ones analyzed here, this implies that a single $\ell$ can be adopted for all materials, leading to similar damage profiles across different phases. 

\subsection{Governing equations}
\label{S:pf_gov_Eq}
The governing equations for the phase-field model are derived as the necessary conditions for the minimality of (\ref{eq:pf_penalty_reg_total_energy}) under the irreversibility constraint for the damage variable
 \cite{Pham:2011,Tanne:thesis}. With the standard procedure of calculus of variations, we obtain the balance of linear momentum, complemented by the boundary conditions
\begin{equation}
\label{eq:pf_balance_eq}
\nabla \cdot \bs{\sigma}\left(\bs x\right) = 0\ \ \text{in}\ \ \Omega , \quad\text{with} \quad \begin{cases} \bs{\sigma}(\bs x) \cdot \bs{n}  = \bar{\bs{t}}_n & \text{on} \quad \partial \Omega_{\text{N}} \,
              \\ \bs{u} = \bar{\bs{u}}_n\,  & \text{on} \quad \partial \Omega_{\text{D}}\,\end{cases},
              \end{equation}
              
\noindent where $\bar{\bs{u}}_n$ is the imposed displacement at the current load step $n$ and 
\begin{equation}
\label{eq:pf_Cauchy_stress}
\bs{\sigma}(\bs x) =a(\alpha,\bs x) \dfrac{\partial \psi^+ }{ \partial\bs \varepsilon}\left(\bs x\right)  + \dfrac{\partial \psi^-}{\partial\bs \varepsilon}\left(\bs x\right),
\end{equation}
and the damage evolution equations

\begin{equation}
\label{eq:pf_damage_evolution}
\begin{split}
&\alpha(\bs x)\geq  \alpha_{n-1}(\bs x),\\
 	\quad a’(\alpha(\bs x),\bs x)\psi^+\left(\bs{\varepsilon},\bs x\right)+\dfrac{G_{f}(\bs x)}{c_w}\bigg( \dfrac{w'(\alpha(\bs x))}{\ell}-&2\ell \Delta\alpha(\bs x)\bigg) \geq 0, \quad \\
 	\Bigg[a’(\alpha(\bs x),\bs x)\psi^+\left(\bs{\varepsilon},\bs x\right)+\dfrac{G_{f}(\bs x)}{c_w}\bigg( \dfrac{w'(\alpha(\bs x))}{\ell}-&2\ell \Delta\alpha(\bs x)\bigg) \Bigg](\alpha(\bs x) - \alpha_{n-1}(\bs x)) = 0, 
  	\end{split}
 \end{equation}
 valid in $\Omega$, supplemented by the natural boundary conditions 
 \begin{equation}
 \label{eq:pf_damage_bc}
 	\nabla\alpha\cdot\bs n \geq0 \quad \text{ and }  \quad (\nabla\alpha\cdot\bs n)(\alpha - \alpha_{n-1})=0 \quad \quad\text{ on }\partial\Omega\,,
 \end{equation}
\noindent where $a' = \partial a/\partial\alpha$ and $w'=\partial w/\partial\alpha$. In this work, irreversibility of damage is enforced following \cite{Miehe:20102}, by replacing the active part of the elastic strain energy density with a history variable $\mathcal{H}(\bs{x})$ defined as 
\begin{equation}
\label{eq:pf_history_var}
\mathcal{H}(\bs{x})=\underset{history}{\max}\left\{ \psi^+\left(\bs \varepsilon(\bs x),\bs x\right)\right\}\,.
\end{equation}
Thus, (\ref{eq:pf_damage_evolution})$_b$ in the form of equality (when damage evolves) is replaced by its approximate version
\begin{equation}
\label{eq:pf_damage_evolution_approx}
 \quad a’(\alpha)\mathcal{H}(\bs{x})+\dfrac{G_{f}(\bs x)}{c_w}\left( \dfrac{w'(\alpha)}{\ell}-2\ell \Delta\alpha\right) = 0\,. \quad 
 \end{equation}

 \subsection{Numerical aspects}
 \label{S:pf_num_aspects}

The coupled system of non-linear equations (\ref{eq:pf_balance_eq}) and (\ref{eq:pf_damage_evolution_approx}) can be solved either in a monolithic \cite{Gerasimov:2015,Farrell:2015} or in a staggered fashion  \cite{Bourdin:2000}. Due to the non-convexity of the total energy functional with respect to $\bs u$ and $\alpha$, in the absence of specific remedies \cite{Gerasimov:2015,Farrell:2015}, convergence difficulties are encountered using a monolithic strategy. Hence, in our computations we adopt a staggered solution procedure, which involves the alternate solution of the equilibrium and damage equations (each in turn solved via a Newton-Raphson iterative scheme) until the residuals fall below a  tolerance $tol_{\text{stag}}$ \cite{Bourdin:2000}. In the following, we use $tol_{\text{stag}}=10^{-4}$, whereas the tolerance for the nested Newton-Raphson  loops  is set as $tol_{\text{NR}}=10^{-5}$. Within each Newton-Raphson loop, the linear system of equations is solved using a conjugate gradient method with no preconditioning.

To avoid numerical instabilities at  complete degradation, we set $\eta_l = 10^{-5}$ when using the \texttt{spectral} split. However, as explained in \cite{Amor:2009}, a larger value of residual stiffness is required with the \texttt{vol/dev} split, which through heuristic evaluation we determine as $\eta_l = 10^{-3}$. We remark that the higher residual stiffness in the latter case may lead to a damage profile with a band width $2D$ larger than the optimal one (Fig.~\ref{fig:pf_dam_profile}), leading to a more diffuse representation of the cracks. However, as shown later, in our results these effects remain within reasonable limits. 

The fulfillment of the non-negativity of the damage is not a priori guaranteed by the selected linear dissipation function (\ref{eq:pf_dissipation_func}) \cite{Gerasimov:2019} in conjunction with the history variable approach for irreversibility, therefore the functional (\ref{eq:pf_penalty_reg_total_energy}) is complemented with a penalty term characterized by the parameter $\rho$, which here is empirically chosen as $\rho$= 10$^{3}$. This values is sufficiently high to lead to negligible violation of the non-negativity condition, but still low enough to avoid bad conditioning of the system.

The boundary value problem described in Sect.~\ref{S:pf_gov_Eq} is discretized and solved using the finite element (FE) method. To account for the real concrete mesostructure, each domain is discretized using the \textit{voxel-based} structured mesh of the X-ray tomograms \cite{Lengsfeld:1998} using tri-linear cubic elements whose dimensions must be small enough to sufficiently resolve the phase-field profile at the cracks (Fig.~\ref{fig:pf_dam_profile}). Since the location of the cracks is not known in advance, we adopt a uniform mesh with element size $h\le \ell/2$.  The detailed procedure to generate the voxel-based mesh starting from the tomograms is reported in Sect.~\ref{sct:pf_mesh}. To each node correspond four degrees of freedom, i.e. the three components of the displacement and the scalar damage variable. We perform the computations on the Euler High Performance Cluster of ETH Zürich in a parallelized fashion, using 512 cores and the C++ finite element library deal.ii \cite{Arndt:2021}. The mesh and the system of equations are  distributed among the processors, and the system is solved using the parallel solvers available in the PETSC library \cite{PETSC}.

\section{Calibration of material and model parameters}\label{S:pf_mat_char}

The model described in Section~\ref{S:pf_phase_field} requires the calibration of fourteen parameters, namely two Lamé constants, fracture toughness and tensile strength for the mortar matrix ($\lambda_m,\ \mu_m, G_{f,m},\ f_{t,m}$), the aggregates ($\lambda_a,\ \mu_a, G_{f,a},\ f_{t,a}$) and the ITZ ($\lambda_i,\ \mu_i, G_{f,i},\ f_{t,i}$), and the parameters $\ell$ and $p_s$. In principle, a different value of $p_s$ can be adopted for each mesoscale constituent. However, here we use a single value for all.
The calibration of the aforementioned parameters by means of a set of experimental tests and numerical experiments is the focus of this section. 

\subsection{Concrete mix design}
\label{sct:pf_composition}
Contrary to previously published work, e.g., \cite{Nguyen:2016a,Tsitova:2021,Tsitova:2022},
this study concerns a real-world concrete type, namely a concrete with realistic aggregate types and containing all the aggregate fractions up to a sieve size of $\varnothing=$11.25 mm. The maximum aggregate size is chosen to guarantee a homogeneous particle packing within the WST specimen (Sect.~\ref{sct:test_setup}). Also, we adopt a water-to-cement ratio (by weight) $w/c$ = 0.5.
The concrete mix design for the WST specimens is summarized in Tab.~\ref{tab:pf_conc_vol_frac}.
In the present work, the cut-off size for the heterogeneity features to be explicitly described in the
numerical tests is taken equal to 1 mm, meaning that aggregate particles and pores below 1~mm are homogenized within a single material phase termed here \textit{mortar matrix}. Thus, such matrix includes the cement paste, based on 
an ordinary Portland cement of type CEM I 42.5 N, aggregate particles and pores with sieve size $\varnothing\le$ 1~mm and, specifically
for this study, baryte ($\text{BaSO}_4$) powder. The latter, with sieve size $\varnothing\le$ 10 $\mu$m, is added at mixing time and substitutes
a corresponding volume of aggregate particles.
Following \cite{Pietro:2018}, baryte can replace part of the aggregate content with the purpose to enhance the X-ray attenuation contrast between the matrix and the aggregates. Using a set of preliminary tests, we determined the optimal amount of contrast enhancer as 3.5\% of the total aggregate volume, corresponding to 2.1\% of the total volume of cast material. As reported in \cite{Pietro:2018}, such low amount of baryte does not affect the casting of the material nor the cement hydration (since baryte is thermodynamically extremely stable, with
an extremely low aqueous solubility \cite{Correa:2022}). Therefore, baryte does not influence the mechanical properties of the concrete at any hardening stage, while being crucial for the segmentation of the concrete mesostructure using X-CT  (Sect.~\ref{sct:pf_segmentation}). The same conclusions are drawn in \cite{Shakoorioskooie:2022} for a mix design similar to the one adopted here but for different aggregate types. The baryte powder is dry-mixed with the other solid components and ends up uniformly distributed throughout the matrix, as systematically shown in \cite{Pietro:2018,Shakoorioskooie:2022}. 

\begin{table}[!htb]
    \begin{adjustwidth}{-3cm}{-3cm}
        \footnotesize
        \centering
        \begin{tabular}{c c c c c}
            \toprule
            \multirow{2}{*}{ \textbf{Mesoscale phase}} & \textbf{Total volume fraction} & \multirow{2}{*}{\textbf{Component}} &\multirow{2}{*}{\textbf{Volume fraction}}&\textbf{Weight composition}  \\
            & \textbf{(mix design)} & & & \textbf{(kg/$m^3$)}\\
            \toprule
            \multirow{4}{*}{\text{Mortar matrix}} & \multirow{4}{*}{ $57.4\%$} & \quad\text{Portland cement} & $14.3\%$ & 450 \\
            & &\quad\text{sand }$\leq 1$ mm & $18.5\%$ & 496\\
            & &\quad\text{water} & $22.5\%$ & 225\\
            & &\quad\text{baryte} & $2.1\%$ & 94\\
            \toprule
            \multirow{2}{*}{\text{Aggregates}} & \multirow{2}{*}{ $39.4\%$} & \quad $1-8$ mm & $19.1\%$ & 509\\
            & & \quad$8-11.25$ mm & $20.3\%$ & 540\\
            \toprule
            \quad\text{Entrapped/entrained air}$^{(\dagger)}$ & 3.2\% & - & - & - \\
            \toprule
            \multicolumn{5}{l}{$^{(\dagger)}$ {\footnotesize Value estimated for the mix design.}}  \\
            \end{tabular}
    \end{adjustwidth}
    \caption{Mix design of the studied concrete, in terms of volume fraction and gravimetric amount of each component \textit{per} cubic meter of cast material. The volume fraction refers to the whole volume of cast material.
    The values for the distinct components are aggregated together for each mesoscale material phase segmented in the X-ray tomograms and
    considered as a distinct homogeneous phase in the corresponding computational description.
  }
    \label{tab:pf_conc_vol_frac}
\end{table}

\begin{figure}
\centering
\subfloat[]{
{\footnotesize
\begin{tikzpicture}
  \begin{axis}[
      axis line style = thick,
      width=6cm,
      height=6.5cm,
      enlargelimits=false,
      xmin=-.5, xmax=5.5,
      xtick={0,1,2,3,4,5},
     xticklabels={$<0.063$ \\ $ (BaSO_4)$, $0.063$, $0.125$, $0.25$, $0.5$, $1.0$},
     xticklabel style={text width=1.1cm, align=right, rotate=90},
      ymin=0, ymax=100,
     ytick={0,20,40,60,80,100},
     yticklabels={0\%, 20\%, 40\%,60\%,80\%,100\%},
      xlabel={{\normalsize	 Sieve size [mm]}},
      ylabel={{\normalsize	Vol. fraction [\%]}},
      grid style={line width=.1pt, draw=gray!20, dashed},
      xmajorgrids,
      ymajorgrids,
    ]{
      \addplot[gray!80, mark=*,mark options={thick, scale=0.75, solid}, dash dot] table [col sep=comma, x index=0, y index =1] {sieve_sand.txt};}
    \end{axis}
  \end{tikzpicture}}\label{fig:pf_sieve_sand}}\hspace{15mm}
  \subfloat[]{
{\footnotesize
\begin{tikzpicture}
  \begin{axis}[
      axis line style = thick,
      width=6cm,
      height=6.5cm,
      enlargelimits=false,
      xmin=-.5, xmax=3.5,
      xtick={0,1,2,3},
     xticklabels={$2.0$, $4.0$, $8.0$, $11.25$},
     xticklabel style={text width=1.1cm, align=right, rotate=90},
      ymin=0, ymax=100,
     ytick={0,20,40,60,80,100},
     yticklabels={0\%, 20\%, 40\%,60\%,80\%,100\%},
      xlabel={{\normalsize	 Sieve size [mm]}},
      ylabel={{\normalsize	Vol. fraction [\%]}},
      grid style={line width=.1pt, draw=gray!20, dashed},
      xmajorgrids,
      ymajorgrids,
    ]{
      \addplot[gray!80, mark=*,mark options={thick, scale=0.75, solid}, dash dot] table [col sep=comma, x index=0, y index =1] {sieve_coarse.txt};}
    \end{axis}
  \end{tikzpicture}} \label{fig:pf_sieve_coarse}}
  \caption{Particle size distribution curves in terms of valume fraction adopted for (a) fine aggregates (i.e., with particle sieve size $\varnothing\le$ 1~mm), which are homogenized in the mortar matrix, and (b) coarse aggregates ($\varnothing >$ 1~mm), which are explicitly modeled. 
  The first data point in (a) indicates the content of baryte ($\text{BaSO}_4$) powder, which also contributes to the content of
  particles with size $\varnothing <$ 63$\mu$m.}
\label{fig:pf_sieve}\end{figure}
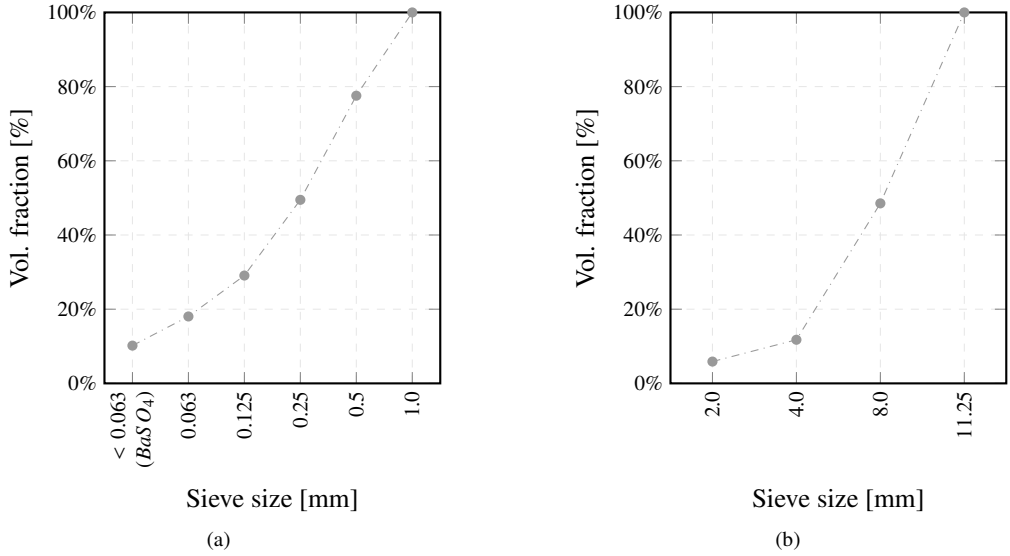

All used aggregates stemmed from monolithic quartzite-like rock pieces extracted from the same quarry (Eberhard Bau AG, Rümlang, Zürich, Switzerland), crushed and sieved to obtain the final aggregate size distribution. The measured size distribution after crushing is illustrated in Fig.~\ref{fig:pf_sieve_sand} for the fraction with sieve size $\varnothing\leq$~1~mm, i.e., the fraction homogenized in the mortar matrix, and in Fig.~\ref{fig:pf_sieve_coarse} for the portion explicitly represented, namely, for 1~mm $\le\varnothing\le$ 11.25~mm.  The exact particle size distribution for the baryte is not available, therefore, its total amount in Fig.~\ref{fig:pf_sieve_sand} is reported as $\varnothing\le$ 63$\mu$m since the particles composing the adoped powder have a diameter smaller than about 10 $\mu$m and of the minimum sieve size. 
Before crushing and sieving, from the original monolithic rock pieces we obtain a set of specimens and test them to measure the aggregate elastic properties, as detailed in Sect.~\ref{sct:pf_el_prop}.

The ITZ between aggregates and cement paste is typically a high porosity zone which forms due to an altered $w/c$ ratio and packing of cement particles close to the aggregate surfaces, locally modifying the chemical composition and porosity of the mortar \cite{Monteiro1985,Maso1992,Carrara2017b,Husem:2003}. The ITZ typical thickness ranges between 20-50 $\mu$m \cite{Yang:1998}, a size requiring a spatial resolution of about 10-20 $\mu$m for its explicit description. Such level of detail is incompatible with the analysis of representative mesostructural volumes with typical size on the order of 20-100 mm and is also orders of magnitude smaller than the selected cutoff length of 1 mm above which we explicitly model heterogeneities.  Nevertheless, since the ITZ is the weakest phase in concrete, due to its lower mechanical characteristics compared to those of cement paste and aggregates, its contribution toward the mesoscale fracture behavior cannot be neglected \cite{Jebli:2018, Scrivener2004, Carrara2017}. This scenario ideally requires the introduction of interfaces, which would however increase the already high computational effort. To circumvent this issue, we rely on a different strategy detailed in the following sections.
 
\subsection{Elastic properties of the mesoscopic phases}
\label{sct:pf_el_prop}
To quantify the linear elastic moduli of the aggregates, a set of nine unconfined, uniaxial compression tests on prisms with size 160$\times$90
$\times$45 mm$^3$ (height $\times$ width $\times$ thickness) are performed following the procedure of the
SIA 262.263 standard \cite{SIA_262_2022}, i.e. the Swiss version of the EN 12390-13:2021 standard \cite{EN12390}.
The obtained average values of Young's modulus $E_a$ and Poisson's ratio $\nu_a$ are summarized in Tab.~\ref{tab:pf_final_pat}.
Before the compressive tests, the specimens are also used to measure the density of the rocks, leading to an average value of $\rho_a=$ 2675 kg/m$^3$. This is the value used for estimating the volumetric contents of aggregate particles reported in Table \ref{tab:pf_conc_vol_frac}.

The Young's modulus of the mortar matrix with the composition summarized in Tab.~\ref{tab:pf_conc_vol_frac} is measured following \cite{SIA_262_2022} by means of unconfined,
uniaxial compression tests on three prismatic beams with size $40\times40\times160$ mm$^3$  (width
$\times$ thickness $\times$ length). Further, three prismatic beams with size $90\times45\times160$ mm$^3$  (width 
$\times$ thickness $\times$ length) are used to measure the Poisson's ratio. All the specimens are cast after mixing under vacuum to reduce the likelihood of pores with size $\varnothing\ge$ 1 mm, and cured for $28$ days.
The average values of Young's modulus $E_m$ and Poisson's ratio $\nu_m$ for the mortar matrix are listed in Tab.~\ref{tab:pf_final_pat}.  Also in this case the mortar density is measured resulting in $\rho_m=$ 2360 kg/m$^3$.

For the ITZ, both elastic and fracture properties depend on several parameters (i.e. thickness, water-to-cement ratio, curing age, aggregate shape and roughness) and cannot be easily measured \cite{Monteiro1985,WangXing:2021,Chen:2024}. Therefore, for the estimate of $E_i$ we rely on a literature review \cite{WangXing:2021,Xiao:2013,Yang:1998, Chen:2024}, where a range of values $E_{i}= 20\%E_{m}-90\%E_{m}$ is reported. We select the upper bound $E_{i}= 90\%E_{m}$, since in the model we  adopt an ITZ thickness larger than the real one (see Sect.~\ref{sct:pf_composition} and Sect.~\ref{sct:pf_mesh}).
 
\subsection{Selection of the length scale parameter $\ell$}\label{S:eff_ell}
 As mentioned in Sect.~\ref{S:pf_degradation_func}, the model obtained using the degradation function \eqref{eq:pf_degradation_func} is  independent of the length scale, provided that this is sufficiently small compared to the size of the smallest feature in the domain.
 However, the adoption of a smaller length scale reduces the support  of the phase-field profile, therefore, it requires the adoption of finer FE meshes with consequent increase of the computational cost. For this reason, we adaptively select the length scale parameter striking a balance between numerical efficiency and accuracy of the results. This task is straightforward in case of homogeneous domains where the domain geometry governs the smallest length of the problem. For heterogeneous materials, the smallest length is often represented by the minimum distance between two inclusions. When the particle packing is very high such as in concrete, this length can vanish in presence of aggregates in contact. In this case, we adopt a pragmatic approach and fix  the smallest desired size of FE $h_{min}$, which in turn  dictates the smallest value of $\ell$ attainable through the relationship $h_{min} \le \ell/2$. The details are reported in Sect.~\ref{sct:pf_bc}.

\subsection{Fracture properties and post-peak response parameter}
\label{sct:pf_tpb_calibration}
\begin{figure*}[!htb]
 \centering
 \begin{tikzpicture}
\node [anchor=north west,inner sep=0] (notched) at (0,0){\includegraphics[width=0.45\textwidth]{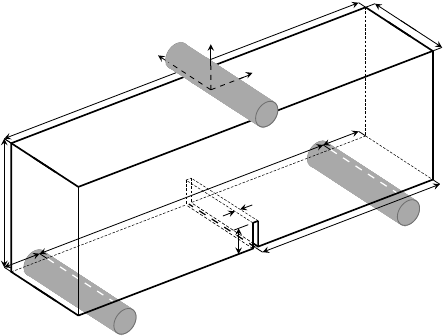}};
\draw (2.5,-.9) node {\scriptsize $x_2$};
\draw (3.7,-.7) node {\scriptsize $x_3$};
\draw (4.45,-1.3) node {\scriptsize $x_1$};
\draw (1.8,-1.5) node[rotate=20] {\scriptsize 160 mm};
\draw (-.1,-3.45) node[rotate=90] {\scriptsize 40 mm};
\draw (2.5,-3.36) node[rotate=20] {\scriptsize 120 mm};
\draw (.56,-4.1) node[rotate=20] {\scriptsize 20 mm};
\draw (5.6,-2.18) node[rotate=20] {\scriptsize 20 mm};
\draw (6.0,-3.75) node[rotate=20] {\scriptsize 80 mm};
\draw (3.55,-4.03) node[rotate=0] {\scriptsize 13 mm};
\draw (6.9,-.3) node[rotate=326] {\scriptsize 40 mm};
\draw (4.2,-3.25) node[rotate=20] {\scriptsize 3 mm};
\end{tikzpicture}
 \caption{Test-setup for the notched TPB test.}\label{fig:pf_tpb_setup}
\end{figure*}
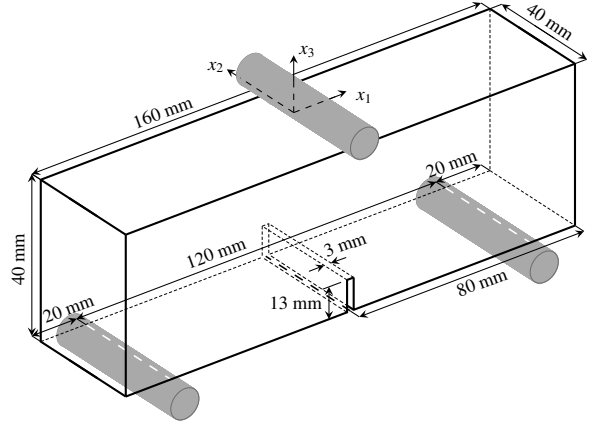

\begin{figure*}[!htb]
       \centering
       \subfloat[]{
                \includegraphics[width=0.42\textwidth]{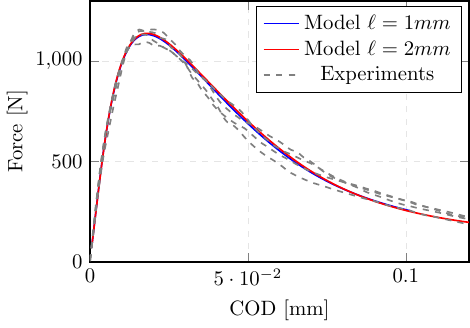}
               \label{fig:pf_cod}}
               \hspace{0.2cm}
                    \subfloat[]{
                     \includegraphics[width=0.42\textwidth]{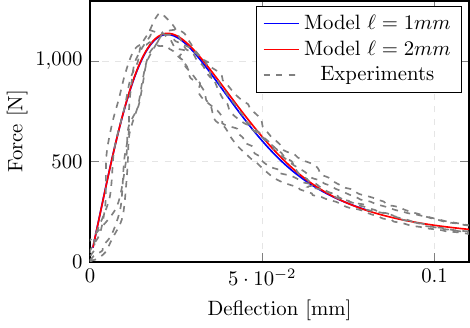}
               \label{fig:pf_deflection}}
                     \caption{Experimental and numerical comparison in terms of (a) load-COD curves and (b) load-deflection curves for the TPB tests using the cohesive phase-field model described in Section~\ref{S:pf_PF}.}
\label{fig:pf_TPB_tests}
\end{figure*}

We characterize the fracture properties of the mortar matrix  by casting five beams with size $160\times40\times40mm^3$ (length $\times$ width $\times$ thickness) with a midspan notch of width $3$ mm and height $13$ mm (Fig.~\ref{fig:pf_tpb_setup}). The specimens are then subjected to three-point bending (TPB).
The tests are performed on a closed loop universal testing machine and are driven by controlling the crack opening displacement (COD), measured with a COD gauge (gauge length 10 mm) placed across the notch. This allows to control the whole test, including the post-peak phase.
Also, stereo DIC analyses are used to measure the beams deflection. Fig.~\ref{fig:pf_TPB_tests} shows the obtained load vs. COD curves (Fig.~\ref{fig:pf_cod}) and load vs. deflection curves (Fig.~\ref{fig:pf_deflection}), demonstrating the quasi-brittle nature of the mortar. Assuming that the work done on the system is dissipated only by crack propagation, the fracture toughness of the mortar can be estimated as
\begin{equation}
\label{eq:pf_initial_Gf}
	G_{f,m}\approx \frac{\mathcal{W} }{\Gamma_c}\,,
\end{equation}
\noindent where $\mathcal{W}$ is the area enclosed by the load vs. deflection curve (i.e., the work done on the system) and $\Gamma_c$ is the nominal crack surface, i.e. the midspan cross-sectional area of the beam excluding the notch area. The average value obtained from the performed five tests is reported in Tab.~\ref{tab:pf_final_pat}.

\begin{figure}[!htb]
\centering
  \begin{tikzpicture}
             \node[anchor=south west,inner sep=0] (main) at (0,0) {\includegraphics[width=0.65\textwidth]{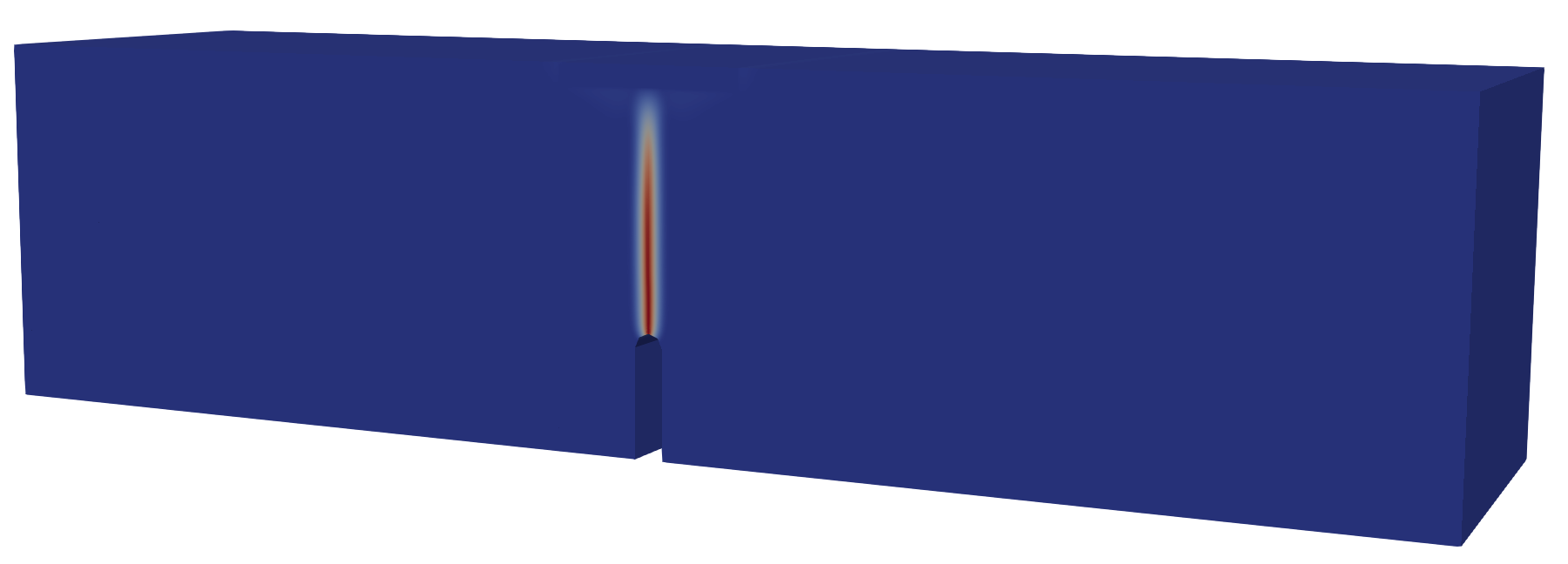}};
            \begin{scope}[x={(main.south east)},y={(main.north west)}]
             \node[anchor=south west] at (1,0.01) {\includegraphics[width=.12\textwidth]{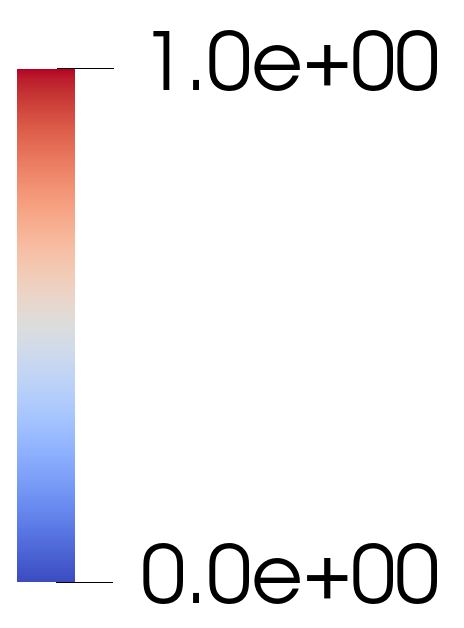}};
  \end{scope}
  \end{tikzpicture}
 \caption{Predicted phase field for the notched TPB test, where red and blue denote fully damaged and intact material states, respectively.}
\label{fig:pf_tpb_pattern}
  \end{figure}

 The tensile strength of the mortar $f_{t,m}$ and the shape model parameter $p_s$ are calibrated by numerically reproducing the notched TPB tests in 3D using the phase-field model described in Sect.~\ref{S:pf_PF} along with the \texttt{spectral} energy split.  As explained in Sect.~\ref{S:pf_num_aspects}, the \texttt{spectral} energy split requires a lower residual stiffness $\eta_l$, thus it presumably leads to more accurate values of $f_{t,m}$ and $p_s$. We remark that, in tests such as the notched TPB, where the crack is mainly subjected to mode-I or opening loading conditions, the predictions obtained with the \texttt{vol/dev} or \texttt{spectral} splits are very similar. In the present case, the domain is homogeneous and composed of mortar matrix, i.e. $\Omega = \Omega_m$. Considering that the heterogeneities with diameter $\varnothing\le1$ mm are homogenized in the mortar matrix and that in this case the smallest length of the geometry is the notch width (i.e., 3 mm), we set the length scale to $\ell=$ 1mm. Since the crack is expected to propagate from the midspan notch, the mesh is locally refined at the center in order to fulfill the requirement $h\le\ell/2$ leading to a total of $\sim 6\cdot10^6$ degrees of freedom. The estimates for both $f_{t,m}$ and $p_s$ are obtained by comparing the experimental and numerical load-displacement curves.
  The optimal values of $f_{t,m}$ and $p_s$ are obtained thruogh a trial-and-error procedure trying to reproduce numerically the esperimental load-deflection and load-COD curves. The obtained values are listed  in Tab.~\ref{tab:pf_final_pat}, while the comparison between experimental and optimized numerical curves is presented in Fig.~\ref{fig:pf_TPB_tests}. An excellent agreement can be observed there, highlighting that the calibrated model is able to accurately represent the mortar behavior. Also, the obtained failure crack pattern in terms of distribution of the damage field is reported in Fig.~\ref{fig:pf_tpb_pattern} and confirms the expected development of a single central crack.

To confirm the $\ell$-independence of the numerical results, the TPB tests are also numerically reproduced using the parameters of Tab.~\ref{tab:pf_final_pat} and $\ell=$ 2 mm. The obtained results are included in  Fig.~\ref{fig:pf_TPB_tests} and show only marginal differences compared to the case with $\ell=$ 1 mm, confirming the purely numerical role of the length scale parameter.

	\begin{table}
\begin{adjustwidth}{-3cm}{-3cm}
\footnotesize
\centering
 \begin{tabular}{c c c c c c c c}
\toprule
\multirow{2}{*}{\textbf{Phase}}& $E$ & $\nu$ & $\lambda$ & $\mu$ & $G_f$ & $f_t$ & $p_s$ \\
 &  ${(GPa)}$ & ${(-)}$ & ${(GPa)}$ & ${(GPa)}$ & ${(N/mm)}$ & ${(MPa)}$ & ${(-)}$ \\

\toprule
\textbf{Mortar matrix}& $35 $&  $0.16 $ & $7.1$ & $15.1$ & $0.06$ & $5.0$ & \multirow{3}{*}{$2$} \\
\textbf{Aggregates}  &$162$& $0.17$ & $35.66$  & $69.23$ & $0.3$ & $25.0$ &  \\
\textbf{ITZ}  &$31.5$& $0.16$ & $6.38$& $13.58$ & $0.03$ & $3.5$ &  \\

 \toprule
  \end{tabular}
    \end{adjustwidth}
    \caption{Adopted material and model properties.}
     \label{tab:pf_final_pat}
    \end{table}
    
Next, we estimate the fracture parameters of the aggregates starting from the observation that their resistance is typically higher than that of the mortar matrix. This is also confirmed by the experimental evidence, as better illustrated later in Sect.~\ref{sct:pf_exp_test}. Here, we estimate the fracture properties of the aggregates starting from those of the mortar matrix assuming that the ratio between them is equal to that observed for the elastic properties, namely we set $f_{t,a}=5f_{t,m}$ and $G_{f,a}=5G_{f,m}$ (Tab.~\ref{tab:pf_final_pat}). The obtained estimate of  $f_{t,a}$ is in good agreement with the values reported in \cite{Gupta:1998, Gupta:2000} for quartzite-like rocks. 

The last parameters that need to be calibrated are those of the ITZ. Similarly to the elastic properties, they are estimated considering the available literature where values of $f_{t,i}=30\%f_{t,m}-70\%f_{t,m}$ for the tensile strength and $G_{f,i}=10\%G_{f,m}-50\%G_{f,m} $ for the fracture toughness are reported \cite{Thomas:1963,Tschegg:1995,Rao:2002,Xiao:2013,
Chen:2024,Jebli:2018,Yang:2019,Wang:2021}. 
As for the elastic properties, we adopt here the upper values of the ranges, namely $f_{t,i}=70\%f_{t,m}$ and $G_{f,i}=50\%G_{f,m} $ (Tab.~\ref{tab:pf_final_pat}).
	
\section{In-situ X-CT wedge splitting tests}
\label{S:pf_WST_res}
This section illustrates the  WSTs. We describe the setup, specifically designed for in-situ testing  and DVC analyses, and explain how the heterogeneous meshes are obtained from the X-CT tomograms. 

\subsection{Test setup}\label{sct:test_setup}
\begin{figure}[!htb]
	     \centering
\subfloat[]{\centering
	           \includegraphics[scale=1.1,valign=c]{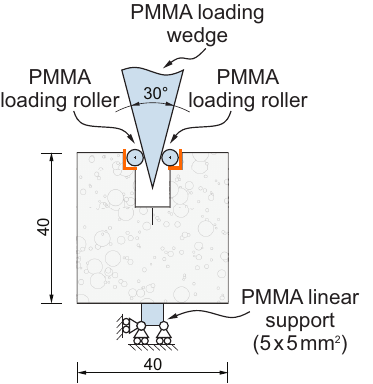}\label{fig:pf_setup_spcm}}\hspace{6mm}
\subfloat[]{ \centering\includegraphics[scale=1.1,valign=c]{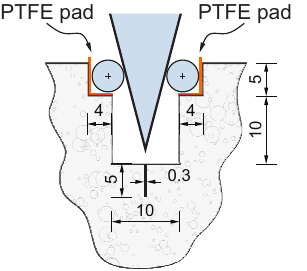} \vphantom{ \includegraphics[scale=1.2,valign=c]{wedge}}\label{fig:pf_Tslot}}
                \caption{Scheme of the in-situ WST setup during X-ray tomography: (a) specimen and loading wedge and (b) detail of the T-slot and of the notch.}
                \label{fig:pf_test_setup}
\end{figure}

The WSTs are performed on three specimens (specimen $\#$1, $\#$2 and $\#$3) casted from the same batch of concrete and with size  of $40\times40\times40$ mm$^3$ (Fig.~\ref{fig:pf_setup_spcm}). The specimens are loaded with a tension-compression-torsion testing stage (Deben OFTT  with a load capacity of 20 kN in tension/compression and 100 Nm in torsion) installed inside the X-ray tomograph Easytom XL
from RX Solutions.
The tomograph has a 230 keV X-ray source with a tungsten filament, 230 W maximum target power, $\pm$20$^\circ$ cone beam angle, a 0.5 mm thick beryllium emission window and a maximum optical resolution of 4 $\mu$m. Its X-ray detector is an amorphous silicon flat panel with 2536 $\times$ 2024 pixels (physical pixel size 124 $\mu$m) covered by a cesium iodide scintillator screen.
The WST setup is schematized in Fig.~\ref{fig:pf_setup_spcm} and relies on a wedge with opening angle $2\gamma=30^{\circ}$ pushed inside a T-slot to apply a splitting force on a notch with a width of $\sim$ 300 $\mu$m and a length of 5 mm. The force is transmitted through two PMMA loading rollers laying against L-shaped PTFE pads (0.5 mm thick) to reduce  friction and limit stress concentrations at the contact surface (Fig.~\ref{fig:pf_Tslot}).  The specimen is supported by a prismatic PMMA linear support bar with cross sectional size 5 $\times$ 5 mm$^2$ and length 40 mm (Fig.~\ref{fig:pf_setup_spcm}). To limit artifacts due to X-ray photon starvation, the components in contact with the specimens are made of PMMA or PTFE, whose X-ray attenuation coefficient is negligible compared to that of concrete \cite{Withers:2021}. 

\begin{figure}[!htb]
	     \centering
	           \includegraphics[width=.7\textwidth]{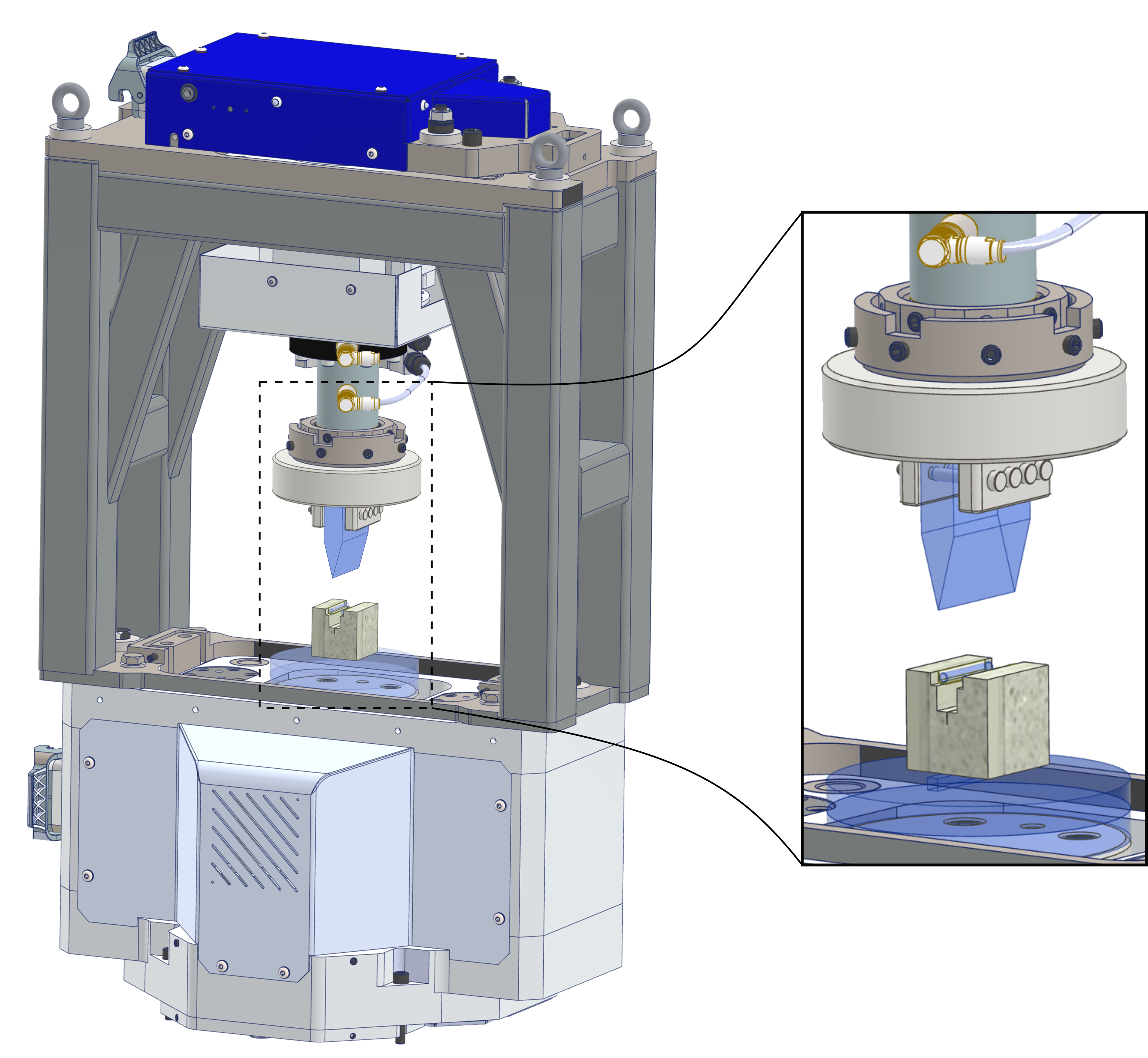}
                \caption{Scheme of the in-situ WST setup during tomography as seen at larger scales, including the loading instrument (Deben open frame stage OFTT 20kN), the whole WST setup and one specimen.}
                \label{fig:pf_test_setup}
\end{figure}



\subsection{Loading and tomographic acquisition procedure}

 The WST induces a dominant mode-I fracture by imposing a monotonically increasing COD through the progressive vertical displacement of the wedge \cite{Neuner:2022,Bruhweiler:1990}, see Fig. \ref{fig:pf_test_setup}. This procedure is adopted to trigger a stable and controllable crack propagation, an aspect particularly important for the X-CT and DVC applicability since it ensures stability and absence of motion for the sample during a measurement time of a few tens of minutes. If this requirement is not satisfied, the volume reconstruction and the associated DVC analyses may be affected by motion artifacts preventing any further investigation. This can be particularly limiting while investigating the failure behavior of brittle or quasi-brittle materials, where fracture propagation, close to the peak load, is often unstable and can lead to an instantaneous transition of the specimen from intact to completely broken. 
 
 To ensure stable tests and reproducible results, the loading and image acquisition procedure is defined based on a set of ex-situ experiments leading to the protocol in Tab.~\ref{tab:pf_load_protocol}. In particular, the X-CT acquisition protocol is tailored to apply DVC analyses.
 It involves two X-CT measurements in the reference configuration, corresponding to an applied load $\le20$ N, and the
 acquisition of about 8-10 tomograms at different load levels during the test (Tab.~\ref{tab:pf_load_protocol}).
 The limited load applied in the reference configuration avoids unwanted motion of the specimens during image acquisition, while repeatedly acquired tomograms allow for the quantification of the measurement uncertainty \cite{Buljac2018} as better detailed in Sect.~\ref{S:pf_dvc_an}.

 During the pre-peak phase the risk of unstable crack propagation is very low, hence we can stop the test more frequently for tomographic acquisition. In this phase, we acquire one tomogram for each load variation of about 50-70 N. Also, a higher tomographic acquisition frequency improves the computational efficiency of the DVC analyses.
 Close to the expected load peak $F_{max}\simeq$ 350 N and in the post-peak phase, due to the increased risk of unstable crack propagation, we significantly reduce the vertical wedge speed and stop the test less frequently, hence perform a measurement every 100-150 N of load variation.

\begin{table}[htb]
\begin{adjustwidth}{-3cm}{-3cm}
\renewcommand{\arraystretch}{1.5}
\footnotesize
\centering
 \begin{tabular}{c c c c}
\hline
\multicolumn{2}{c}{\multirow{2}{*}{ \textbf{Vertical load}} }&\textbf{Loading protocol:} &{\textbf{X-CT acquisitions: }} \\
&&\textbf{vertical wedge speed} & \textbf{image frequency} \\
\hline
\multirow{3}{*}{Pre-peak:} &At $\le20$ N & Reference configuration & 2 tomograms\\
&$<290$ N & $10$ $\mu$m/s & 1 tomogram every 50-70 N $^\dagger$\\
&$>290$ N to the peak & $1$ $\mu$m/s & only 1 tomogram\\
\hline
\multirow{2}{*}{After peak:} &$<50\%$ load drop & $1$ $\mu$m/s &\multirow{2}{*}{1 tomogram every 100-150 N $^\dagger$}\\
 &$>50\%$ load drop & $10$ $\mu$m/s &\\
 \hline
\multicolumn{3}{l}{{$^\dagger$ to be intended as absolute values of load variation.}}
   \end{tabular}
    \end{adjustwidth}
     \caption{Loading and simultaneous imaging protocols for the in-situ WSTs during X-CT.}
     \label{tab:pf_load_protocol}
 \end{table}

The tomographic acquisition time is another aspect to account for to avoid sample motion or failure, due to relaxation at a constant imposed displacement. In general, the X-CT setting can be tuned to reduce the acquisition time at the cost of an increased image noise, which, in turn, leads to a higher uncertainty in the DVC measurements. To find the best compromise between acquisition time and image quality we analyze the signal-to-noise ratio (SNR) of several tomograms of the studied concrete mix obtained with different X-CT settings as detailed in \ref{app:pf_tuning_CT}.
The analysis of the results leads to the settings summarized in Tab.~\ref{tab:pf_adopted_CT_parameters}. A 2$\times$2 hardware pixel binning during the radiographic acquisition and a relatively high number of projection images (called frames) averaged \textit{per} finally stored radiograph contribute to a high SNR \cite{Milena:2022}, while a high value of the X-ray source current, leading to a higher X-ray beam intensity, thus allowing for an increment of the frame rate, and a reduction of the number of radiographs per tomographic turn over 360° specimen rotation yield a reduction of the whole tomographic acquisition time. A thin, i.e., 500 $\mu$m-thick, copper foil is used to high-pass filter the energy spectrum of the X-ray beam by absorbing low-energy photons, limiting the artifacts due to beam hardening \cite{Withers:2021}. This is due to the progressive depletion in the transmitted X-ray beam within the outer part of the specimen of the low-energy photons, which are more likely absorbed compared to the high-energy ones. The physical voxel size achieved with this configuration is about 50 $\mu$m, which constitutes the \textit{tomographic optical resolution}. Note, however, that this does not correspond to the size of the smallest features that can be identified in a tomogram (namely the \textit{physical resolution}), which is usually taken as 2-3 times larger, i.e. 100-150 $\mu$m.

\begin{table}[!htb]
\footnotesize
\renewcommand{\arraystretch}{1.2}
\begin{adjustwidth}{-4cm}{-4cm}
\centering
 \begin{tabular}{c c}
 \hline
\multicolumn{2}{c}{ \textbf{X-CT parameters}}\\
\hline
& setting\\
\hline
Mode & step by step rotation and imaging\\
{Detector mode}&2$\times$2 binning  \\
Detector resolution (pixel)&  1268$\times$1012\\
Source-object distance (mm) & $\sim$235 \\
Source-detector distance (mm) & 1123 \\
Geometrical magnification &  $\sim$4.8$\times$\\
Voxel size  ($\mu$m) & $\sim$50\\
Radiographic frame rate ($s^{-1}$) &10 \\
Frame average & 10 \\
Pre-hardening X-ray source filter &  Copper (500 $\mu$m)\\
Source voltage (kV)  & 140   \\
Source current ($\mu$A) & 500  \\
Source power (W) & 70 \\
Radiographs \textit{per} tomogram & 1206\\
Tomographic signal-to-noise Ratio, SNR ($dB$)& 28 \\
Tomographic acquisition time (minutes) & 25 \\
 \hline
  \end{tabular}
   \end{adjustwidth}
   \caption{Settings adopted for the in-situ WST during X-CT.}
    \label{tab:pf_adopted_CT_parameters}
  \end{table}
  
 To reconstruct each tomogram, we use the software \textit{X-Act UniCT} provided by RX Solutions \cite{X-Act}. It features a Feldkamp-Davis-Kress filtered back-projection algorithm with a Tukey kernel. During the reconstruction, artifacts resulting from the motion of the focal point of the X-ray source or from limited changes in the X-CT geometry during the acquisition are compensated by means of a focal spot/geometry correction procedure \cite{Schulze}. Despite the copper source filter, residual beam hardening artifacts may still appear due to low-energy photons not absorbed by the copper filter.
 Such residual beam hardening is corrected via a digital voxel value compensation based on a best-fitted polynomial beam hardening profile \cite{Barrett:2004}.
To facilitate the detection of the phases in the final images, we apply to the radiographs a weak phase contrast filter \cite{Paganin_2002}.

\subsection{Concrete mesostructure segmentation}
\label{sct:pf_segmentation}

To obtain the  heterogeneous mesostructure geometry of the specimens, we start from the reference tomograms and assign to each voxel the proper material phase (a procedure known as \textit{segmentation}), based on its greyscale value. Finally, for each material phase, we separate and digitally sieve each particle, i.e., each cluster of connected voxels. This overall process is performed within a specific image processing workflow in the software Avizo \cite{Avizo}. First, we denoise the tomograms by using the non-local means filter proposed in \cite{Buades:2011}, which allows preserving the edges between the distinct phases. This filter reduces the image noise by reassigning to each voxel a weighted mean of the original values of neighbouring voxels in a window centered in the computation voxel. In this case, we adopt a convolution search window of 10 voxels and a Gaussian kernel for the weights. Each voxel of the denoised tomogram is then classified (i.e., segmented) into air, aggregate or mortar matrix. We remark that, since the real ITZ thickness is comparable to the voxel size, thus smaller than the physical resolution, the ITZ phase cannot be explicitly segmented from the obtained tomograms. Rather, it is handled following the methodology in Sect.~\ref{sct:pf_mesh}.

The very low linear attenuation coefficient of air creates a sharp image contrast between air and  solid materials, hence, air can be segmented based only on voxel greyscale values, i.e., with a \textit{thresholding} procedure. Conversely, the segmentation of aggregates and mortar matrix is more complex. Although the addition of baryte greatly improves the contrast, the level of noise due to the relatively short acquisition time makes the adoption of a simple thresholding unfeasible. Therefore, we proceed with a multistep thresholding procedure, where the voxels are first separated into a given number of classes based on their greyscale value ranges, leading to a simplified image.
For the tomograms analyzed here we adopt a total of fifteen classes. Next, each class is assigned to the proper mesoscopic phase by means of a thresholding of the simplified image. Although requiring a limited amount of manual intervention, this procedure allows for the different phases to be identified with better accuracy and efficiency. If small voids or mortar inclusions are present in large aggregates, they are automatically identified and removed using a morphological holes filling algorithm. Once the phases are assigned, we shift our attention to identifying each single aggregate particle or pore, namely we perform a \textit{labeling} operation. First, the (simply connected) clusters of voxels belonging to the same materials (i.e., void or aggregate) are identified and tagged as separate entities. However, in the case of the aggregates, some of them include two or more particles in contact or so close to appear as a unique large cluster. To disconnect them, we apply a standard watershed algorithm based on the Euclidean distance transform \cite{Beucher:1993,Meyer1994}. Due to the noise still present in the images, the surfaces of the aggregates and pores result artificially jagged and need to be regularized through morphological closing and dilation transformations using a spherical structural element.  

The final step of the workflow involves a \textit{digital sieving} of all the labeled pore and aggregate particles and the aggregation into the homogenized matrix phase of those with size $\varnothing <$ 1 mm. Here we adopt the approach proposed in \cite{Michele:2016}, which starts by computing the 3D moment of inertia tensor of each particle. Then, the particles are sorted following the size of the corresponding inertia ellipsoid along the second principal axis of the particle. For concrete aggregates, it has been shown that this is the quantity governing the particle size distribution as physically measured with standard gravitational sieving \cite{Erdogan:2007}. The particles are then sorted in ascending size order and the voxels of those smaller than 1 mm are labeled as mortar matrix. 

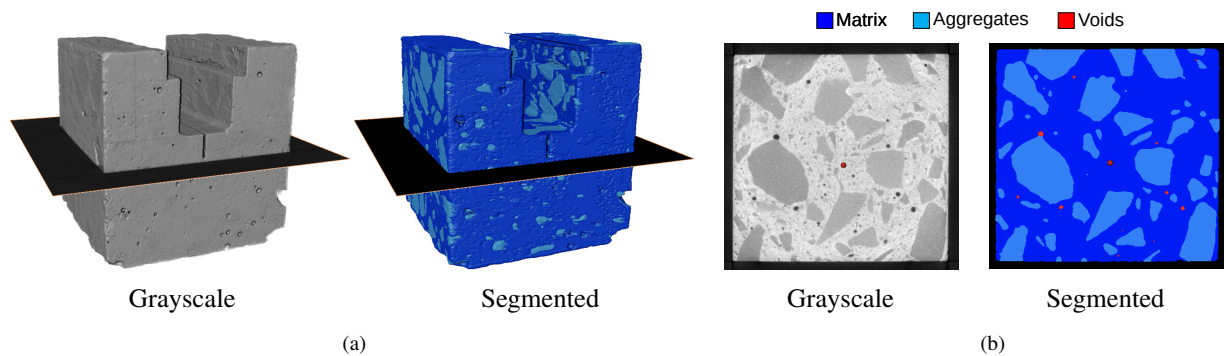
\begin{figure*}[htb]
  \begin{minipage}{\textwidth}%
       \centering
       \subfloat[]{\label{fig:pf_segmentation3d}
            \begin{tikzpicture}      
                \node[anchor=south west,inner sep=0] (main) at (0,0){\includegraphics[width=0.55\textwidth]{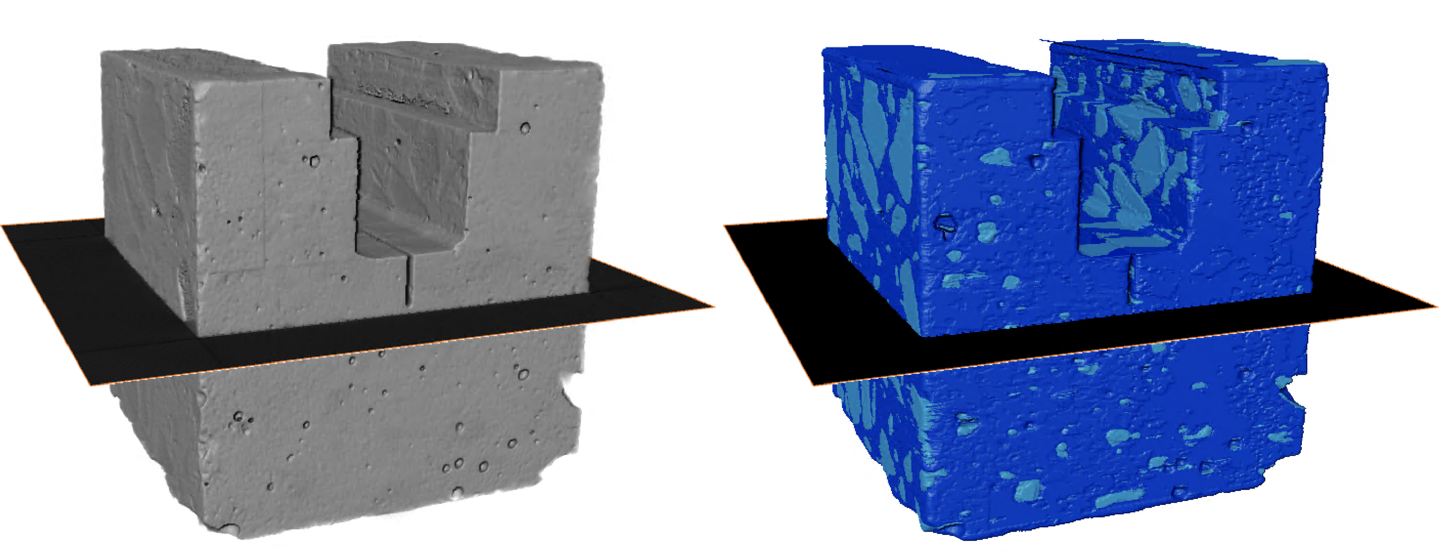}};
                 \begin{scope}[x={(main.south east)},y={(main.north west)}]
                \node[anchor=north] at (0.25,-0.01){Grayscale}; 
\node[anchor=north] at (0.77,-0.01){Segmented}; 
  \end{scope}
 \end{tikzpicture}
}
                    \subfloat[]{\label{fig:pf_segmentation2d}
                \begin{tikzpicture}      
                \node[anchor=south west,inner sep=0] (main) at (0,0){\includegraphics[width=0.45\textwidth,trim={9.5cm 5.5cm 10cm 5.7cm},clip]{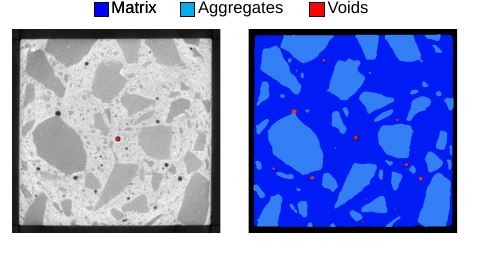}};
  \begin{scope}[x={(main.south east)},y={(main.north west)}]
                \node[anchor=north] at (0.23,.1){Grayscale}; 
\node[anchor=north] at (0.7,0.1){Segmented}; 
  \end{scope}
 \end{tikzpicture}
 }
                \end{minipage}
                     \caption{Comparison between the original and the segmented tomograms for one specimen: (a) 3D rendering and (b) a representative 2D cross-section (whose position is indicated in Fig.~\ref{fig:pf_segmentation3d} by the black plane).}
\label{fig:pf_segmentation}
\end{figure*}

The result of the segmentation process showing the identified mortar matrix, aggregates and pore space phases is compared to the original tomograms in Fig.~\ref{fig:pf_segmentation3d} for the entire volume of a specimen and in Fig.~\ref{fig:pf_segmentation2d} for a representative horizontal slice.

\begin{figure}
\centering
{\footnotesize
\subfloat[]{
\begin{tikzpicture}[baseline]
  \begin{axis}[
      axis line style = thick,
      width=5.5cm,
      height=6cm,
      enlargelimits=false,
      xmin=-.5, xmax=3.5,
      xtick={0,1,2,3},
     xticklabels={$2.0$, $4.0$, $8.0$, $11.25$},
     xticklabel style={text width=.6cm, align=right, rotate=90},
      ymin=0, ymax=100,
     ytick={0,20,40,60,80,100},
     yticklabels={0\%, 20\%, 40\%,60\%,80\%,100\%},
      xlabel={{\normalsize	 Sieve size [mm]}},
      ylabel={{\normalsize	Vol. fraction [\%]}},
      title={\normalsize Coarse aggregates},
      grid style={line width=.1pt, draw=gray!20, dashed},
      xmajorgrids,
      ymajorgrids,
      legend style={
					anchor=north west},
	legend pos=outer north east,
    ]{
      \addplot[gray!80, mark=*,mark options={thick, scale=0.75, solid}, dash dot] table [col sep=comma, x index=0, y index =1] {sieve_coarse.txt};
      \addplot[blue, mark=x,mark options={thick, scale=1.5, solid}, dash dot] table [col sep=comma, x index=0, y index =1] {sieve_comp.txt};
    \addplot[red, mark=x,mark options={thick, scale=1.5, solid}, dash dot] table [col sep=comma, x index=0, y index =2] {sieve_comp.txt};
\addplot[black, mark=x,mark options={thick, scale=1.5, solid}, dash dot] table [col sep=comma, x index=0, y index =3] {sieve_comp.txt};}
\legend{Ref. (Fig.~\ref{fig:pf_sieve_coarse}), Specimen \#1,Specimen \#2,Specimen \#3}
    \end{axis}
  \end{tikzpicture} \label{fig:pf_sieve_comp}}}\hspace{10mm}
\subfloat[]{\footnotesize
\begin{tikzpicture}[baseline]
\begin{axis}[
      xmajorgrids,
      ymajorgrids,
	      axis line style = thick,
      width=5.5cm,
      height=6cm,
  xmin=0.5, xmax=4.5,
      xtick={1,2,3,4},
       xticklabels={Ref., Spcm. \\ \#1, Spcm. \#2, Spcm. \#3},
   xticklabel style={text width=.8cm, align=center, rotate=90},
      ymin=0, ymax=100,
     ytick={0,20,40,60,80,100},
     yticklabels={0\%, 20\%, 40\%,60\%,80\%,100\%},
	 grid style={line width=.1pt, draw=gray!20, dashed},
      xmajorgrids,
      ymajorgrids,
      xtick align=inside,
every axis plot/.append style={
	ybar,
       xticklabels={Ref., Spcm. \\ \#1, Spcm. \#2, Spcm. \#3},
   xtick=data,
          bar width=.5,
 	bar shift=1pt,         
          fill,
        },
title={\normalsize Resolved aggregates},
 ylabel={{\normalsize	Vol. fraction [\%]}},
]
\addplot[gray!80, mark=none] coordinates {(1,34.4)};
\addplot[blue!40, mark=none] coordinates {(2,28.95)};
\addplot[red!40, mark=none] coordinates {(3,28.59)};
\addplot[black!80, mark=none] coordinates {(4,31.66)};
\end{axis}
\end{tikzpicture}
\vphantom{\begin{tikzpicture}[baseline]
  \begin{axis}[
      axis line style = thick,
      width=5.5cm,
      height=6cm,
      enlargelimits=false,
      xmin=-.5, xmax=3.5,
      xtick={0,1,2,3},
     xticklabels={$2.0$, $4.0$, $8.0$, $11.25$},
     xticklabel style={text width=.6cm, align=right, rotate=90},
      ymin=0, ymax=100,
     ytick={0,20,40,60,80,100},
     yticklabels={0\%, 20\%, 40\%,60\%,80\%,100\%},
      xlabel={{\normalsize	 Sieve size [mm]}},
      ylabel={{\normalsize	Vol. fraction [\%]}},
      title={\normalsize Coarse aggregates},
      grid style={line width=.1pt, draw=gray!20, dashed},
      xmajorgrids,
      ymajorgrids,
      legend style={
					anchor=north west},
	legend pos=outer north east,
    ]{
      \addplot[gray!80, mark=*,mark options={thick, scale=0.75, solid}, dash dot] table [col sep=comma, x index=0, y index =1] {sieve_coarse.txt};
      \addplot[blue, mark=x,mark options={thick, scale=1.5, solid}, dash dot] table [col sep=comma, x index=0, y index =1] {sieve_comp.txt};
    \addplot[red, mark=x,mark options={thick, scale=1.5, solid}, dash dot] table [col sep=comma, x index=0, y index =2] {sieve_comp.txt};
\addplot[black, mark=x,mark options={thick, scale=1.5, solid}, dash dot] table [col sep=comma, x index=0, y index =3] {sieve_comp.txt};}
\legend{Reference, Specimen \#1,Specimen \#2,Specimen \#3}
    \end{axis}
  \end{tikzpicture}
}
\label{fig:pf_tot_vol}
}
  \caption{Comparison between the reference aggregate sieve size distribution
  and that obtained from the segmented aggregates in the tomograms of the three specimens, considering only the explicitly modeled aggregate fraction (sieve size larger than 1 mm). The reference distribution is the one from the mix design in Tab.~\ref{tab:pf_conc_vol_frac} and Fig.~\ref{fig:pf_sieve_coarse}. (a) Particle size distribution curves and (b) total volume fraction of the resolved aggregates.}
\label{fig:pf_comp_segm}\end{figure}
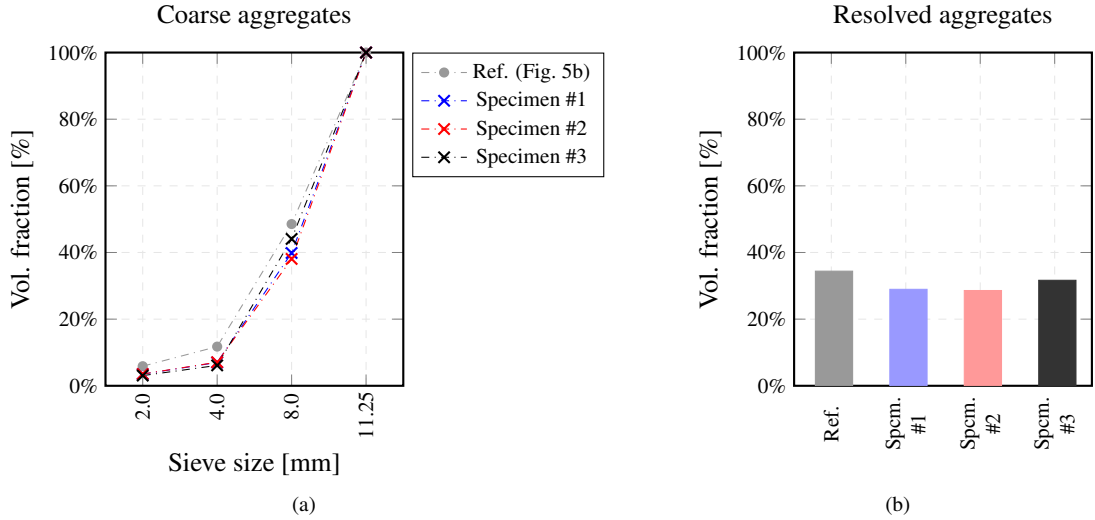

To validate the segmentation procedure, we compare the experimental and the segmented particle size distributions and the total volume fraction of the aggregates with $\varnothing \ge$ 1 mm (Fig.~\ref{fig:pf_comp_segm}). As shown in Fig.~\ref{fig:pf_sieve_comp}, the agreement between the two particle size distributions is quite close, with limited deviations from the reference one corresponding to the mix design. For the total volume fraction of the resolved aggregates, the segmented value marginally underestimates the reference one (Fig.~\ref{fig:pf_tot_vol}). We remark that a small variation in the aggregate volume is inevitable, due to different effects \cite{Michele:2016}. First, the heterogeneous nature of both aggregates and cementitious matrix can lead to strong differences in the X-ray attenuation values, which, along with the finite spatial resolution of the tomograms, may lead to a mislabeling of some voxels  \cite{Michele:2016}. This is particularly relevant for the smallest fraction of aggregates, where we indeed observe the largest discrepancy between experimental and segmentation-derived values (Fig.~\ref{fig:pf_sieve_comp}). Secondly, the identification of the aggregate edges, despite the baryte, is not always straightforward due to partial volume artifacts \cite{Ketcham_2019}. Both effects are  enhanced by the intrinsic noise level of the tomograms.

The segmented tomograms 
are used to obtain a mesh for the numerical simulations. This task is not trivial and can be performed in different ways \cite{Pietro:2018}. The specific approach followed here is detailed in Sect.~\ref{sct:pf_mesh}.

\subsection{Failure mechanisms}
\label{sct:pf_exp_test}
\begin{figure*}[htb]
       \centering
                \includegraphics[width=0.5\textwidth]{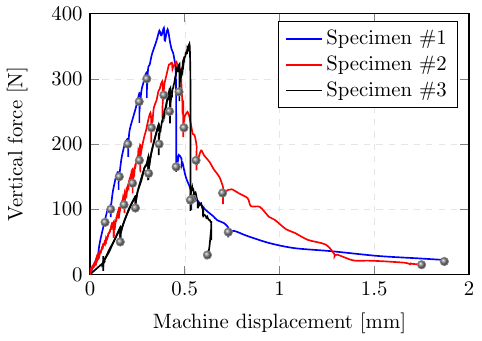}
                     \caption{Vertical force vs. machine displacement curves for the in-situ  WSTs.}
\label{fig:pf_WST_tests}
\end{figure*}

\begin{figure*}[!htb]
  \begin{minipage}{\textwidth}%
       \centering
       \subfloat[]{
                \includegraphics[height=4.5cm]{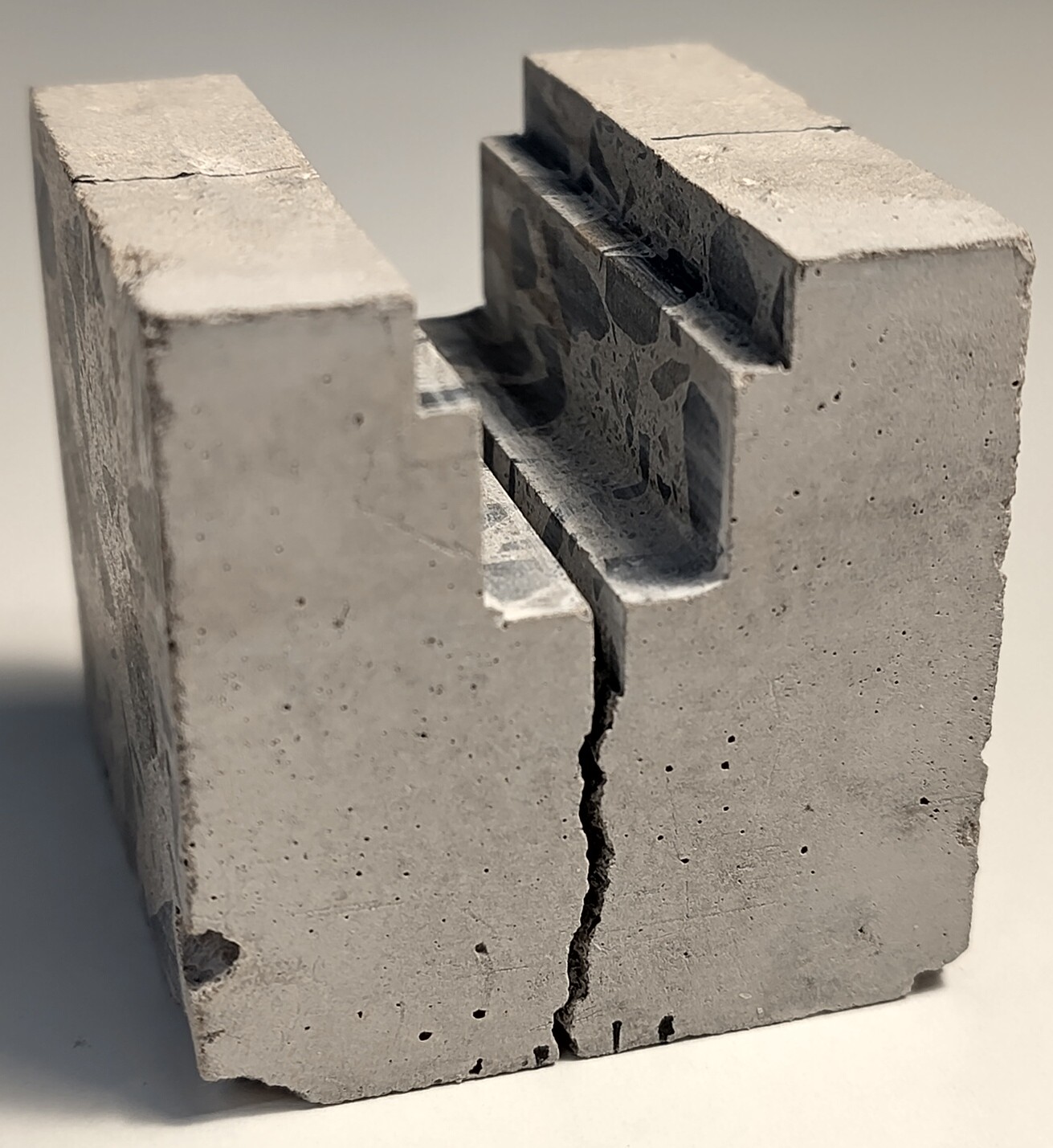}
               \label{fig:pf_expspec1}}
               \hspace{1cm}
               \subfloat[]{
                \includegraphics[height=4.5cm]{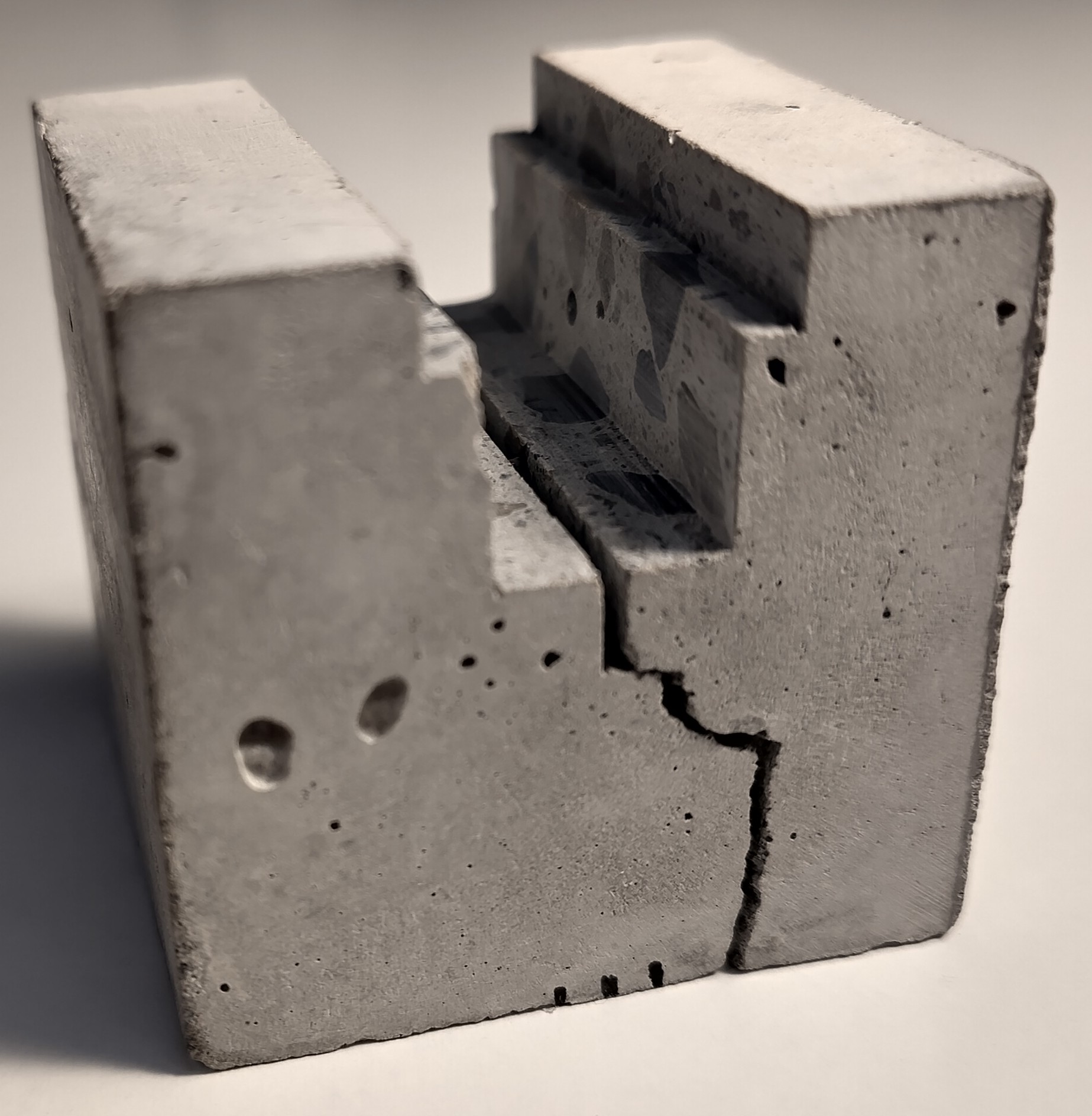}
               \label{fig:pf_expspec2}}
                \hspace{1cm}
                   \subfloat[]{
                \includegraphics[trim={0cm 0cm 0cm 2.6cm},clip,height=4.5cm]{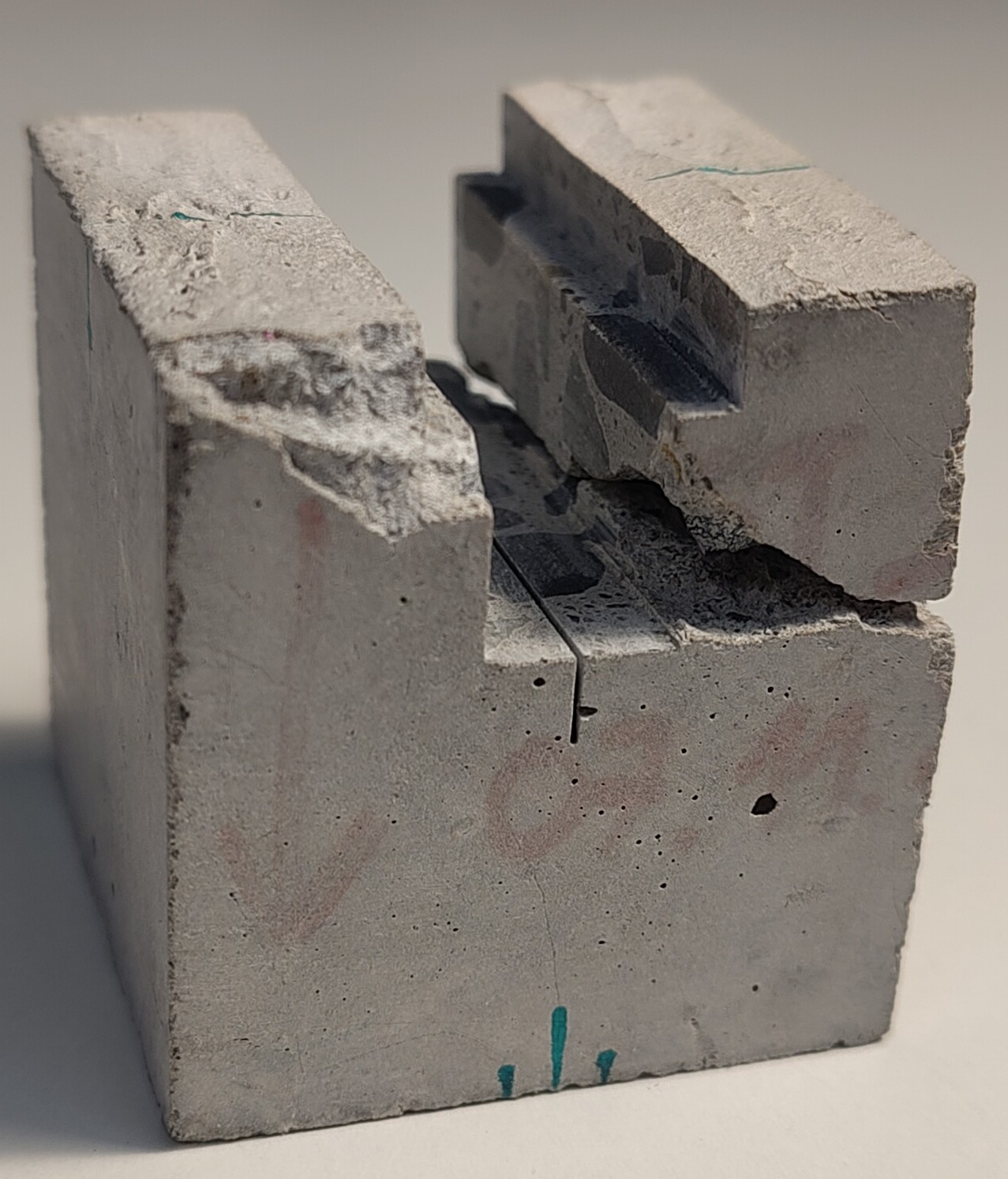}
               \label{fig:pf_expspec3}}
                \end{minipage}
                       \caption{Crack paths of the WSTs: (a) specimen $\#$1 (b) specimen $\#$2 and (c) specimen $\#$3.}
\label{fig:pf_test_cracks}
\end{figure*}

The three WSTs are performed using the load protocol and X-CT settings summarized in Tabs.~\ref{tab:pf_load_protocol} and \ref{tab:pf_adopted_CT_parameters}. The measured load-displacement plots are shown in Fig.~\ref{fig:pf_WST_tests}, where the load levels at which the tomograms are acquired are indicated with bullets ({\color{gray}$\bullet$}).  Since the displacement is held constant during the tomographic acquisition time (i.e., for about 25 minutes), the specimens experience a limited decrease in load due to relaxation. \textit{Post-mortem} photographs of the three specimens are shown in Fig.~\ref{fig:pf_test_cracks}. Although the peak loads show reasonably low variability ($\approx350\pm25$ N in the three cases), the crack patterns and post-peak behavior differ significantly, revealing the role of heterogeneous mesostructural arrangements and imperfect boundary conditions. 
Interestingly, the failure mechanism observed in specimen $\#$3 does not involve the stable propagation of a crack from the tip of the notch (Fig.~\ref{fig:pf_test_setup}), but rather the unstable propagation of a crack nucleating at the re-entrant corner of one of the two arms of the T-slot. In a homogeneous material, this mechanism is less likely to take place compared to the propagation of a crack from the sharp notch. However, here the presence of two large aggregates at the bottom of the arm creates a weak plane leading to unstable fracture right after the peak load (Fig.~\ref{fig:pf_WST_tests}). 

The heterogeneous mesostructure appears to have a strong influence also on the crack pattern of specimens $\#$1 (Fig.~\ref{fig:pf_expspec1}) and $\#$2 (Fig.~\ref{fig:pf_expspec2}).
Here the crack deviates from the perfectly vertical direction expected for a homogeneous specimen, as it tends to avoid aggregates and is attracted by pores. Also, we observe that the fracture mainly propagates within the mortar matrix (Fig.~\ref{fig:pf_expspec2}), confirming the assumption made in Sect.~\ref{sct:pf_tpb_calibration}. 

For the two tests on specimens $\#$1 and $\#$2, we acquire ten X-CT tomograms at different load levels in addition to two reference scans. Three of these acquisitions are performed during the post-peak branch where the majority of the crack propagation takes place, and a final one is performed at failure. Due to the unstable crack propagation after the peak, for specimen $\#$3 only two tomograms are obtained in the post-peak regime, leading to a total of two reference images and nine tomograms during this test. 

\section{DVC analysis} \label{S:pf_dvc_an}

This section discusses the DVC analyses, performed using the DVC module in Avizo \cite{Avizo}, and their main results.

\subsection{DVC approach}\label{S:adopted_DVC}

The estimation of the displacement vector field, $\bs u(\bs x), $ using DVC involves the  optimization of the cross-correlation between the reference tomogram, $f(\bs x)$, and that of the deformed specimen, $g(\bs x' )$, under the constraint of conservation of the total voxel greyscale values \cite{Buljac:2017}, i.e., 
\begin{equation}
    \label{eq:pf_dvc_Corr}
	   f(\bs x) = g(\bs x + \bs u(\bs x)).
\end{equation}
Ideally, we would seek the field $\bs u(\bs x)$ that satisfies Eq. \ref{eq:pf_dvc_Corr}.
However, the voxel-wise fulfillment of \eqref{eq:pf_dvc_Corr} leads to an ill-posed problem, since it entails the detection of vectorial displacements based on a scalar field \cite{Leclerc:2012}. Also, the optimization of the correlation between tomograms is further complicated by the presence of noise \cite{Buljac:2017,Buljac2018}. Hence, \eqref{eq:pf_dvc_Corr} requires a relaxation \cite{Leclerc:2012}, achieved by replacing it with the following minimization problem
\begin{equation}
\label{eq:pf_dvc_res}
	\bs u(\bs x) = \arg \min_{\bs{\tilde u}(\bs x)}  \left\{ r^2\left(\bs x,\,\bs{\tilde u}(\bs x)\right)\right\} \quad \text{with} \quad r\left(\bs x,\,\bs{\tilde u}(\bs x)\right)=  f\left(\bs x\right)-g\left(\bs x + \bs{\tilde u}(\bs x)\right)\,,
	\end{equation}
\noindent where $r\left(\bs x,\,\bs{\tilde u}(\bs x)\right)$ represents the voxel value residual field.
Also, a weak form of \eqref{eq:pf_dvc_res} is preferred leading to 
\begin{equation}
\label{eq:pf_dvc_relaxed}
	\bs u(\bs x)  = \arg \min_{\bs{\tilde u}(\bs x)}  \left\{ \int_\Omega r^2\left(\bs x,\,\bs{\tilde u}(\bs x)\right)\,d\Omega\right\}\,,
	\end{equation}
\noindent where $\Omega$ now represents the analyzed volume. The numerical solution of \eqref{eq:pf_dvc_relaxed} is  found after the parameterization of the displacement field $\bs u(\bs x)$.

 The adoption of a globally defined parameterization of the displacement field leads to the so-called \textit{global} DVC approach \cite{Roux:2008}. Conversely, a DVC approach is termed \textit{local} if residual minimization is independently performed on subvolumes of the domain \cite{Bay2008}. Local DVC does not guarantee a continuous displacement field, possibly leading to kinematically incompatible deformations between neighboring subvolumes that may require smoothing \cite{Buljac2018}. Moreover, local DVC often yields a coarser description of the displacement field than its global counterpart \cite{Leclerc:2012}. For these reasons we consider here only the global DVC approach. In particular, we adopt linear FE-based \textit{ansatz} functions defined on a global DVC mesh composed of tetrahedral elements.
 
 The solution of \eqref{eq:pf_dvc_relaxed} leads to a non-linear system of equations that is solved iteratively. Here we use a gradient descent scheme and the solution is considered converged when the $L_2$ norm of the displacement correction falls below a given tolerance $tol_{DVC}$.

\subsection{DVC uncertainty analysis}
\label{sct:pf_uncertainty_analysis}

For the DVC analysis, it is crucial to appropriately choose the size of the DVC support mesh. 
A coarser mesh reduces the uncertainty but limits the spatial resolution of the estimated displacement field, therefore, a compromise must be sought between uncertainty and resolution \cite{Buljac2018}. 
In the following, we perform a \textit{DVC uncertainty quantification} analysis. This procedure involves registering two different X-CT images differing at most by a rigid body motion. Since no deformation is involved, the residuals obtained after removing the rigid body motion correspond to the effects on the measurements of the noise in tomograms. Performing this procedure for  different sizes of the DVC support mesh provides an objective way to define a good trade-off between resolution and uncertainty in the measurements \cite{Buljac2018}.

\begin{figure}[!htb]
\centering
\footnotesize
\subfloat[]{
\centering
\hspace{0.6cm}\includegraphics[width=.3\textwidth]{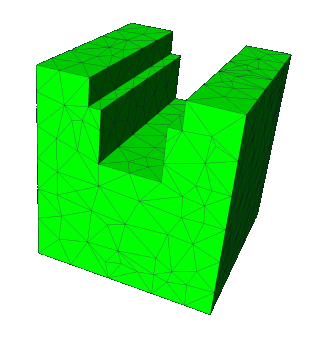}
       \label{fig:DVC_mesh}
}\hspace{2.48cm}
\subfloat[]{
\begin{tikzpicture}
  \draw[red, thick,-] (0.2,1.4) -- (0.6,1.4);
             \node[anchor=west,fill=white] at (0.7,1.4){$x_1$ direction};
  \draw[blue, thick,-] (0.2,1.1) -- (0.6,1.1);
             \node[anchor=west,fill=white] at (0.7,1.1){$x_2$ direction};
  \draw[green, thick,-] (0.2,0.8) -- (0.6,0.8);
             \node[anchor=west,fill=white] at (0.7,0.8){$x_3$ direction};
\begin{axis}[
y tick label style={
                /pgf/number format/fixed,
                /pgf/number format/fixed zerofill,
                /pgf/number format/precision=1},
scaled y ticks=base 10:1,
    width=0.38\textwidth,
     xmin = 60,
     xmax = 140,
    ymin= 0.05,
    ymax=0.25,
 ytick={ 0.5e-1,1e-1,1.5e-1,2e-1,2.5e-1 },
    legend style={draw=none, font=\small},
    legend cell align=left,
    legend pos=north east,
    every axis plot/.append style={thick},
    legend pos= outer north east,
     xlabel={\normalsize DVC mesh size [vox.]},
      ylabel={\normalsize Displacement st. dev. [vox.]},
        grid style={line width=.1pt, draw=gray!20, dashed},
      xmajorgrids,
      ymajorgrids,
      title style={anchor=south,yshift=0.5cm},
]
\addplot+[smooth, no marks, color=blue, solid] table[x=Cell Size, y=std(uy), col sep=comma] {spec4_uncertainty.csv};
\addplot+[smooth, no marks, color=red, solid] table[x=Cell Size, y=std(ux), col sep=comma] {spec4_uncertainty.csv};
\addplot+[smooth, no marks, color=green, solid] table[x=Cell Size, y=std(uz), col sep=comma] {spec4_uncertainty.csv};
\end{axis}
\node[anchor=west] at (.15,0.3){Specimen \#1};
\end{tikzpicture}
\label{fig:unc_1}
}

\subfloat[]{
\begin{tikzpicture}
  \draw[red, thick,-] (0.2,3.6) -- (0.6,3.6);
             \node[anchor=west,fill=white] at (0.7,3.6){$x_1$ direction};
  \draw[blue, thick,-] (0.2,3.3) -- (0.6,3.3);
             \node[anchor=west,fill=white] at (0.7,3.3){$x_2$ direction};
  \draw[green, thick,-] (0.2,3.0) -- (0.6,3.0);
             \node[anchor=west,fill=white] at (0.7,3.0){$x_3$ direction};
\begin{axis}[
y tick label style={
                /pgf/number format/fixed,
                /pgf/number format/fixed zerofill,
                /pgf/number format/precision=1},
scaled y ticks=base 10:1,
    width=0.38\textwidth,
    legend style={draw=none, font=\small},
    legend cell align=left,
    legend pos=north east,
     xmin = 60,
     xmax = 140,
    ymin= 0.05,
    ymax=0.25,
 ytick={ 0.5e-1,1e-1,1.5e-1,2e-1,2.5e-1 },
    every axis plot/.append style={thick},
    legend pos= outer north east,
     xlabel={\normalsize  DVC mesh size [vox.]},
     ylabel={\normalsize Displacement st. dev. [vox.]},
      grid style={line width=.1pt, draw=gray!20, dashed},
      xmajorgrids,
      ymajorgrids,
]
\addplot+[smooth, no marks, color=blue, solid] table[x=CellSize, y=std(uy), col sep=comma] {Spec3_uncertainty.csv};
\addplot+[smooth, no marks, color=red, solid] table[x=CellSize, y=stdux, col sep=comma] {Spec3_uncertainty.csv};
\addplot+[smooth, no marks, color=green, solid] table[x=CellSize, y=std(uz), col sep=comma] {Spec3_uncertainty.csv};
\end{axis}
\node[anchor=west] at (.15,0.3){Specimen \#2};
\end{tikzpicture}
\label{fig:unc_2}}\hspace{2cm}
\subfloat[]{
\begin{tikzpicture}
  \draw[red, thick,-] (0.2,1.4) -- (0.6,1.4);
             \node[anchor=west,fill=white] at (0.7,1.4){$x_1$ direction};
  \draw[blue, thick,-] (0.2,1.1) -- (0.6,1.1);
             \node[anchor=west,fill=white] at (0.7,1.1){$x_2$ direction};
  \draw[green, thick,-] (0.2,0.8) -- (0.6,0.8);
             \node[anchor=west,fill=white] at (0.7,0.8){$x_3$ direction};
\begin{axis}[
y tick label style={
                /pgf/number format/fixed,
                /pgf/number format/fixed zerofill,
                /pgf/number format/precision=1},
scaled y ticks=base 10:1,
    width=0.38\textwidth,
    legend style={draw=none, font=\small},
    legend cell align=left,
    legend pos=north east,
     xmin = 60,
     xmax = 140,
    ymin= 0.05,
    ymax=0.55,
 ytick={1.0e-1,2.0e-1,3.0e-1,4.0e-1,5.0e-1},
    every axis plot/.append style={thick},
    legend pos= outer north east,
     xlabel={\normalsize  DVC mesh size [vox.]},
     ylabel={\normalsize Displacement st. dev. [vox.]},
       grid style={line width=.1pt, draw=gray!20, dashed},
      xmajorgrids,
      ymajorgrids,
]
\addplot+[smooth, no marks, color=blue, solid] table[x=Size, y=std(uy), col sep=comma] {spec5_uncertainty.csv};
\addplot+[smooth, no marks, color=red, solid] table[x=Size, y=std(ux), col sep=comma] {spec5_uncertainty.csv};
\addplot+[smooth, no marks, color=green, solid] table[x=Size, y=std(uz), col sep=comma] {spec5_uncertainty.csv};
\end{axis}
\node[anchor=west] at (.15,0.3){Specimen \#3};
\end{tikzpicture}
\label{fig:unc_3}}
\caption{DVC uncertainty analysis: (a) example of DVC support mesh with average element size of 125 voxels and estimated standard deviation of the displacements for  specimen (b) $\#$1, (c) $\#$2 and (d) $\#$3. }
\label{fig:pf_uca}
\end{figure}
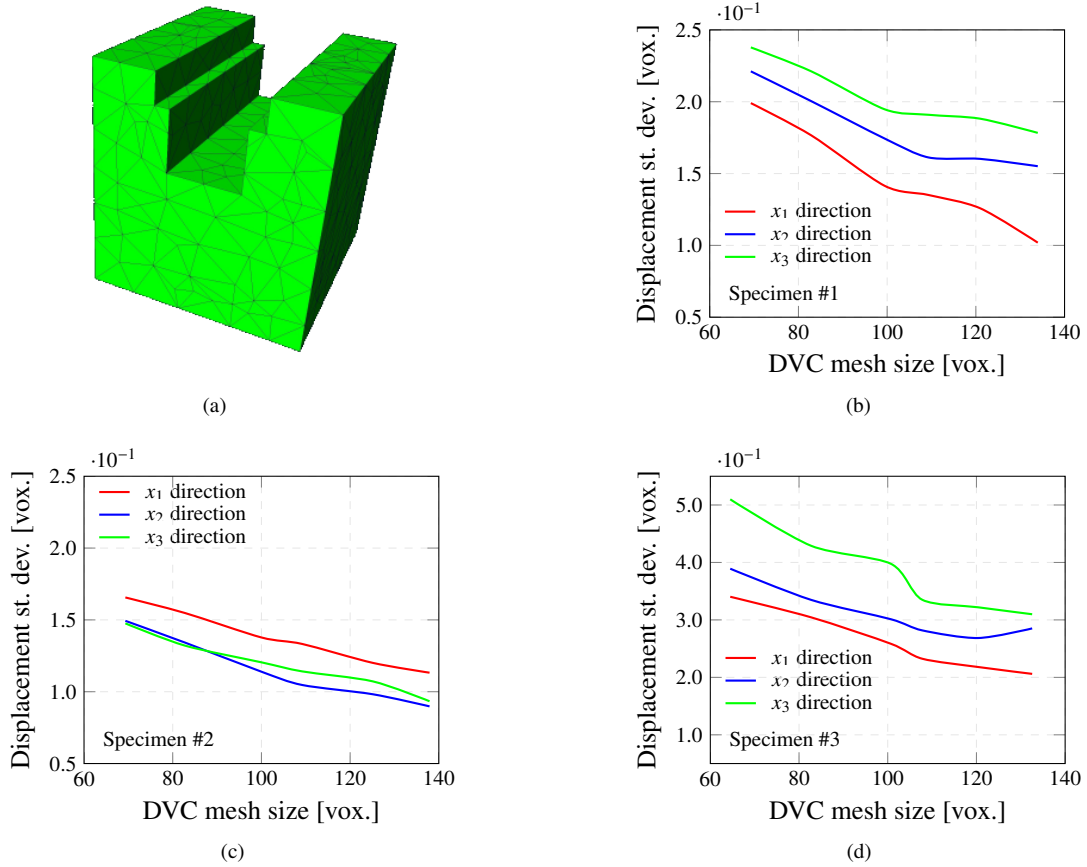

\begin{figure}
\centering
	\centering
\includegraphics[width=.3\textwidth]{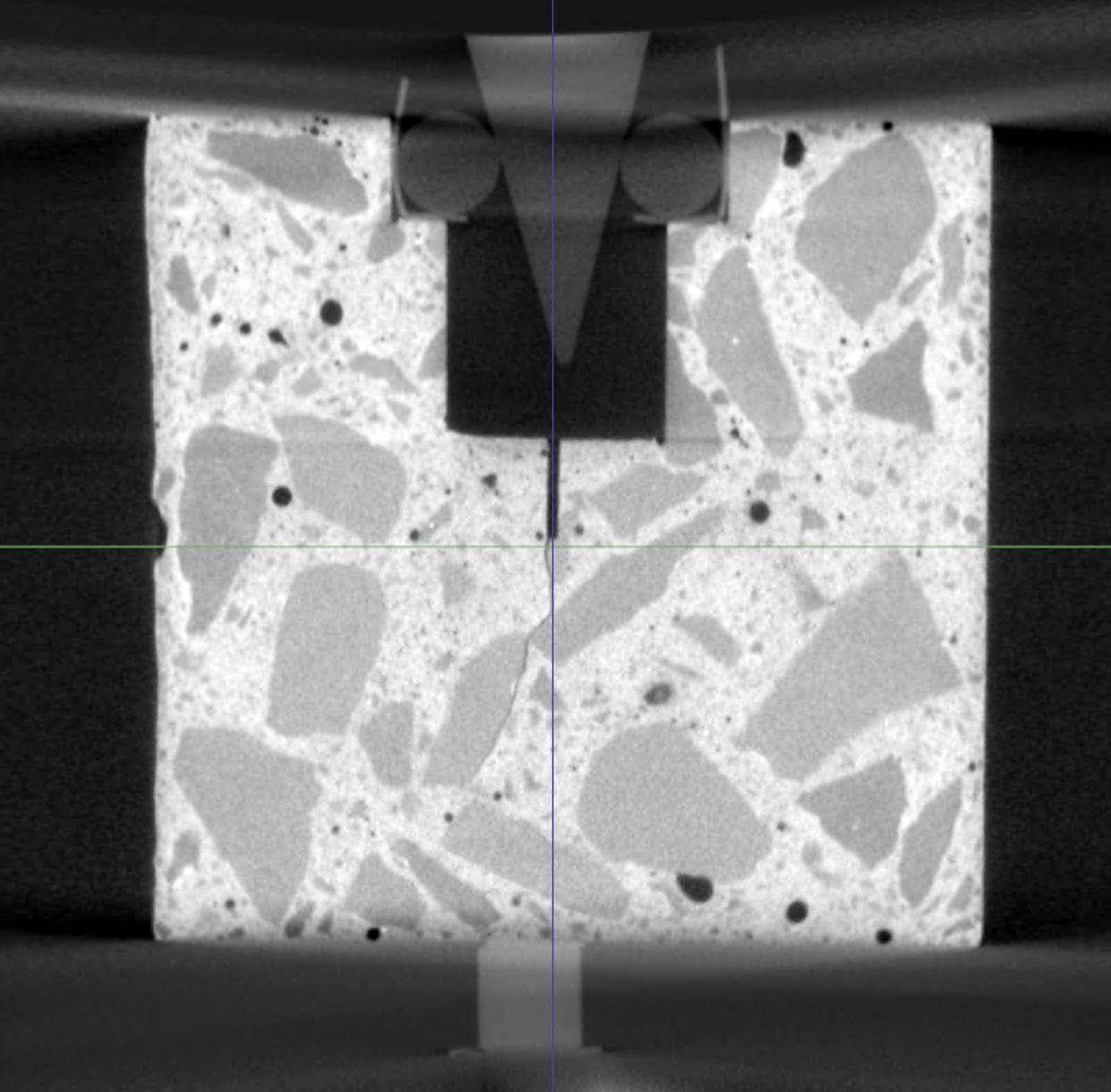};
       \label{fig:pf_artifacts}
 \caption{Artifacts at the top and bottom of specimen $\#$1, due to the X-ray cone beam geometry and shown in a representative 2D slice.  }
        \label{fig:pf_streaks_fig}
\end{figure}

First, we define different DVC support meshes with an average mesh size ranging from about 65 voxels to 140 voxels. We remark that the obtained meshes do not include the smallest features of the overall geometry such as the initial notch (Fig.~\ref{fig:DVC_mesh}). The description of such small features would require a fine mesh that, if uniform, would significantly increase the uncertainty in the displacement estimation. On the other hand, a local mesh refinement would lead to a spatially varying uncertainty with hardly predictable effects on the final results. To avoid cone beam geometry artifacts such as streaking and striation (Fig.~\ref{fig:pf_streaks_fig}) occurring close to the specimen surfaces that remain always orthogonal to the rotation axis and at the periphery of the radiographic field of view \cite{Barrett:2004,Schulze,Ruhrnschopf:2011,Ruhrnschopf2011b,Lifton2015}, we remove from the meshed volume about 1 mm of material from the top and bottom sides. 

Figs.~\ref{fig:unc_1}-\ref{fig:unc_3} show the results of the uncertainty analysis for the three tested specimens in terms of standard deviation of the obtained nodal displacements. As expected, we observe a reduction in the uncertainties while increasing the element size. The uncertainties corresponding to specimens $\#$1 and $\#$2 are lower compared to specimen $\#$3. This is due to artifacts created by motion of specimen $\#$3 during one of the two reference acquisitions. Nevertheless, for an average mesh size of $120-130$ voxels, the uncertainties reduce to less than $0.2$ voxels for specimens $\#$1 and $\#$2, and less than  $0.3$ voxels for specimen $\#$3. Lower uncertainties can be reached by increasing the mesh size. However, a size larger than about 140 voxels does not allow for a geometry conforming with the T-slot. 

Based on the obtained results, we adopt a mesh size of $\sim125$ voxels (similar to Fig.~\ref{fig:DVC_mesh}) for all the analyses presented in the following. Accounting for the voxel size (i.e., 50 $\mu$m), this results in a precision in the displacement estimation of about 10 $\mu$m for specimens $\#$1 and $\#$2 and 15 $\mu$m for specimen $\#$3.

\subsection{DVC results}
\label{sct:pf_dvc_result}

\begin{figure}[!htb]
    \centering
    \subfloat[]{
    \includegraphics[width=0.95\textwidth]{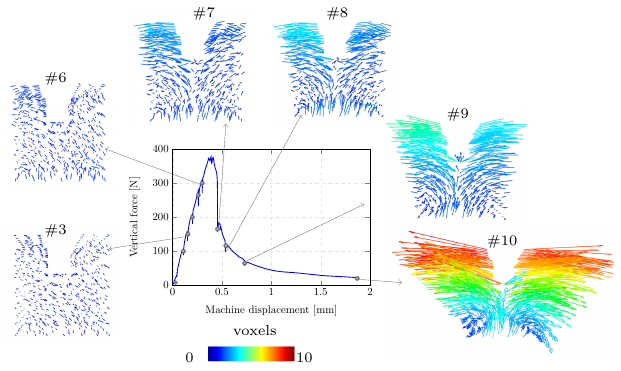}
    \label{fig:pf_dvc1_displ} } 
    
        \subfloat[]{ \includegraphics[height=4cm,valign=t]{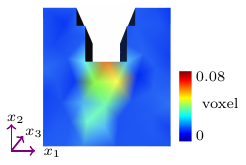} \label{fig:pf_dvc1_strain}} \hspace{1cm}
        \subfloat[]{ \adjustbox{valign=t}{\begin{tikzpicture}\vphantom{ \includegraphics[height=4cm,valign=t]{DVC_result1_strain.pdf}}
        \node [anchor=north west,inner sep=0] (residual2) at (0,0){\includegraphics[trim={.5cm .51cm .58cm .425cm},clip,height=3.5cm,valign=t]{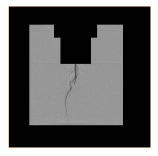}\label{fig:pf_dvc1_residual}};
        \draw [red, thick] (1.1,-1.3) rectangle (2.12,-3.5);
        \end{tikzpicture}} }\hspace{1cm}
        \subfloat[]{ \includegraphics[trim={1.0cm .53cm 1.095cm 1.08cm},clip,height=3.5cm,valign=t]{DVC_result1_res.pdf} \vphantom{ \includegraphics[height=4cm,valign=t]{DVC_result1_strain.pdf}}\label{fig:pf_dvc1_residual_zoom}}
    
  \caption{DVC analysis for specimen $\#$1: (a) displacement field measured at different load steps, magnified 75$\times$, (b) normal strains in $x_1$ direction at load step \#10, (c) residual field related to the central slice of the specimen at load step \#10 along with (d) the detail of the cracked region (in red in Fig.~\ref{fig:pf_dvc1_residual}).}
 \label{fig:pf_dvc1}
\end{figure}

\begin{figure}[!htb]
   \centering
   \subfloat[]{
    \includegraphics[width=0.95\textwidth]{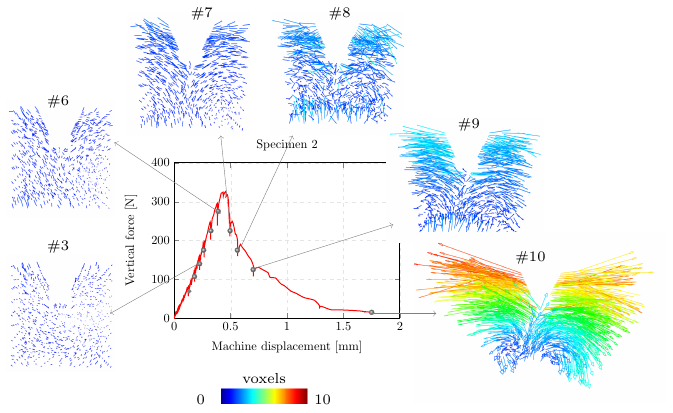}
     \label{fig:pf_dvc2_displ}} \hspace{0.1cm}
        \subfloat[]{ \includegraphics[height=4cm,valign=t]{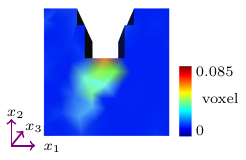}  \label{fig:pf_dvc2_strain}} \hspace{1cm}
        \subfloat[]{ \adjustbox{valign=t}{\begin{tikzpicture}
        \node [anchor=north west,inner sep=0] (residual2) at (0,0){\includegraphics[trim={.51cm .468cm .448cm .447cm},clip,height=3.5cm,valign=t]{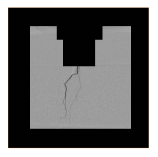}\vphantom{ \includegraphics[height=4cm,valign=t]{DVC_result2_strain.pdf}}\label{fig:pf_dvc2_residual}};  
        \draw [red, thick] (0.7,-1.25) rectangle (2.1,-3.5);
        \end{tikzpicture}} }\hspace{1cm}
        \subfloat[]{ \includegraphics[trim={.9cm .475cm 1.095cm 1.12cm},clip,height=3.5cm,valign=t]{DVC_result2_res.pdf}\vphantom{ \includegraphics[height=4cm,valign=t]{DVC_result2_strain.pdf}}  \label{fig:pf_dvc2_residual_zoom}}
    
  \caption{DVC analysis for specimen $\#$2: (a) displacement field measured at different load steps, magnified 75$\times$, (b) normal strains in $x_1$ direction at load step \#10 and (c) residual field related to the central slice of the specimen at load step \#10 along with (d) the detail of the cracked region (in red in Fig.~\ref{fig:pf_dvc2_residual}).}
 \label{fig:pf_dvc2}
\end{figure}

\begin{figure}[!htb]
    \centering
   \subfloat[]{
    \includegraphics[width=0.95\textwidth]{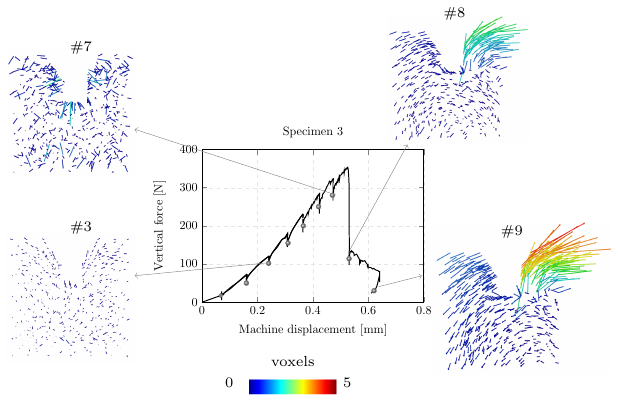}
    \label{fig:pf_dvc3_displ} } \hspace{0.1cm}
        \subfloat[]{ \includegraphics[height=4cm,valign=t]{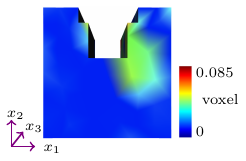} \label{fig:pf_dvc3_strain}} \hspace{1cm}
        \subfloat[]{ \adjustbox{valign=t}{\begin{tikzpicture}
        \node [anchor=north west,inner sep=0] (residual2) at (0,0){\includegraphics[trim={.443cm .538cm .555cm .43cm},clip,height=3.5cm,valign=t]{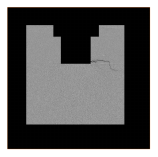}\vphantom{ \includegraphics[height=4cm,valign=t]{DVC_result3_strain.pdf}} \label{fig:pf_dvc3_residual} }; \draw [red, thick] (2,-1) rectangle (3.4,-1.9);
        \end{tikzpicture}} }\hspace{1cm}
        \subfloat[]{ \includegraphics[trim={1.45cm 1.25cm .6cm .9cm},clip,height=3.0cm,valign=t]{DVC_result3_res.pdf}\vphantom{ \includegraphics[height=4cm,valign=t]{DVC_result3_strain.pdf}} \label{fig:pf_dvc3_residual_zoom}}
    
  \caption{DVC analysis for specimen $\#$3: (a) displacement field measured at different load steps, magnified 75$\times$, (b) normal strains in $x_1$ direction at load step \#9 and (c) residual field related to the central slice of the specimen at load step \#9  along with (d) the detail of the cracked region (in red in Fig.~\ref{fig:pf_dvc3_residual}).}
 \label{fig:pf_dvc3}
\end{figure}

Figs.~\ref{fig:pf_dvc1}-\ref{fig:pf_dvc3} show the results of DVC analyses performed on the three specimens at different load levels along with the experimental load vs. vertical machine displacement curves, where the points at which the acquisitions are performed are highlighted with bullets (${\color{gray}\bs{\bullet}}$). Since the deformations remain small, the image registration is always performed between a deformed-state tomogram and the reference one. The tolerance for convergence of the DVC analyses is taken as $tol_{DVC}$ = 10$^{-3}$ voxels.

The displacement fields illustrated in the following are all corrected by subtracting the rigid body motion between the deformed and reference configurations. At each load step, the initial guess for the iterative DVC procedure is taken as the converged solution from the previous load step. For the first load step we initialize the solver with a zero displacement.

For specimens $\#$1 and $\#$2 (Figs.~\ref{fig:pf_dvc1}-\ref{fig:pf_dvc2}) the observed behavior is similar and involves a relatively limited deformation before the peak, with the lower part of the domain remaining almost undeformed. Close to the peak load, the displacements at the top of the T-slot start growing and increase steadily during the softening branch. There, the deformations of the lower part of the specimens grow as well, with a pattern that clearly highlights a splitting mechanism localized in the central region. This failure mechanism is further confirmed observing the normal strain measurements in $x_1$ direction at the last step of the test (step \#10 in Figs.~\ref{fig:pf_dvc1_displ} and \ref{fig:pf_dvc2_displ}) and the DVC residuals (Figs.~\ref{fig:pf_dvc1_residual},\subref*{fig:pf_dvc1_residual_zoom} and \ref{fig:pf_dvc2_residual},\subref*{fig:pf_dvc2_residual_zoom}). A localization of the strains is visible in the region with high residuals. The main difference between the results obtained for specimens $\#$1 and $\#$2 is a smoother post-peak behavior for the latter compared to the former. This can be justified considering that the crack pattern observed in Fig.~\ref{fig:pf_dvc2_residual_zoom} is quite complex with pronounced kinks and the presence of a main and wide crack along with a secondary and thinner crack. Conversely, the crack observed in Fig.~\ref{fig:pf_dvc1_residual_zoom} is less intricate and includes a single wide crack. These differences, linked to the different distributions of heterogeneities, lead to a lower energy dissipation in specimen $\#$1 than in specimen $\#$2, further demonstrating the role played by the mesostructure.
 
For specimen $\#$3, the displacement field illustrated in Fig.~\ref{fig:pf_dvc3_displ} in the pre-peak regime is similar to that of specimens $\#$1 and $\#$2. However, in the post-peak stage the displacement in $x_1$ direction evolves with a markedly asymmetric distribution, with the right side of the T-slot experiencing a larger lateral displacement than the left one (load steps $\#$8 and $\#$9 in Fig.~\ref{fig:pf_dvc3_displ}). Correspondingly, the observed failure behavior involves a crack separating the right T-slot arm from the remainder of the specimen. This mechanism likely involves a snap-back right after the load peak, that prevents observing the specimen evolution between a load level of $P\simeq380$ N (step $\#$7 in Fig.~\ref{fig:pf_dvc3_displ}) and $P\simeq100$ N (step $\#$8 in Fig.~\ref{fig:pf_dvc3_displ}).  The failure mechanism is confirmed also by Figs.~\ref{fig:pf_dvc3_strain} and \ref{fig:pf_dvc3_residual},\subref*{fig:pf_dvc3_residual_zoom} where the normal strain in $x_1$ direction and the DVC residuals at the last load step (\#9 in Fig.~\ref{fig:pf_dvc3_displ}) are shown. There we can observe a single crack separating the right arm of the T-slot, while neither crack nor strain concentration is present at or close to the notch.

\begin{figure}[!htb]
\centering
\begin{tikzpicture}
\begin{axis}[
    width=0.4\textwidth,
     ymin=0.01,
     ymax=0.038,
    legend style={draw=none, font=\small},
    legend cell align=left,
    legend pos=north east,
    every axis plot/.append style={thick},
    legend pos= outer north east,
    legend entries={Specimen 2 Uncertainty analysis,Specimen 2 DVC analysis, Specimen 1 Uncertainty analysis,Specimen 1 DVC analysis, Specimen 3 Uncertainty analysis,Specimen 3 DVC analysis },
     xlabel={Scan number},
      ylabel={(St. dev. of residuals)/range},
]
\addplot+[smooth, no marks, color=red, dotted] table[x=Step, y=Unc, col sep=comma] {Spec3_dvc.csv};
\addplot+[smooth, no marks, color=red, solid] table[x=Step, y=coeff, col sep=comma] {Spec3_dvc.csv};
\addplot+[smooth, no marks, color=blue, dashed] table[x=Step, y=Unc, col sep=comma] {spec4_dvc.csv};
\addplot+[smooth, no marks, color=blue, solid] table[x=Step, y=coeff, col sep=comma] {spec4_dvc.csv};
\addplot+[smooth, no marks, color=black, dashed] table[x=Step, y=Unc, col sep=comma] {spec5_dvc.csv};
\addplot+[smooth, no marks, color=black, solid] table[x=Step, y=coeff, col sep=comma] {spec5_dvc.csv};
\end{axis}
\end{tikzpicture}
\caption{Standard deviation of the DVC residuals, normalized by the dynamic range of the reference tomogram of each WST.}
\label{fig:pf_dvc_res}
\end{figure}
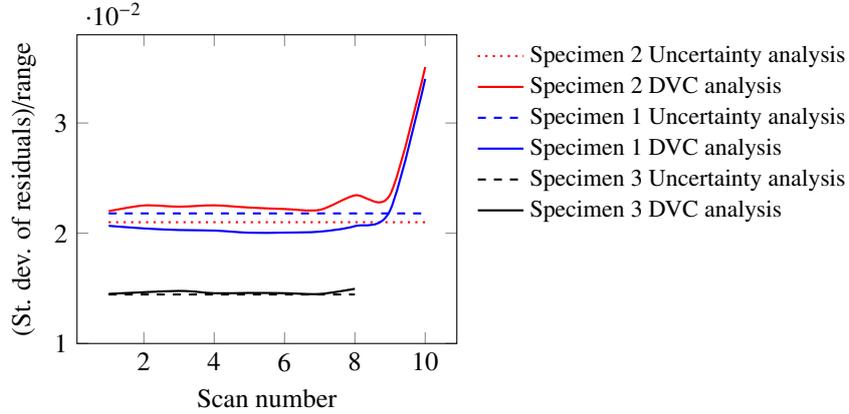

To validate the DVC analyses, we measure the standard deviation of the DVC residuals after each registration, normalized by the dynamic range of the reference image used to determine the displacements (Fig.~\ref{fig:pf_dvc_res}). Within the pre-peak regime, the obtained values are in agreement with the corresponding values obtained from the uncertainty analysis (Sect.~\ref{sct:pf_uncertainty_analysis}), confirming that both the measurement uncertainty and the noise level are preserved during the in-situ test. It is worth mentioning that the increase in the standard deviation of the residuals during the post-peak stage is not related to a deterioration of the measurement precision, rather to the development of cracks. Since the \textit{ansatz} functions used in the DVC analyses cannot reproduce any displacement jump, the displacement field at the cracks cannot be retrieved, leading to a local increase of the residuals. However, as observable in Figs.~\ref{fig:pf_dvc1_residual}-\ref{fig:pf_dvc3_residual}, the residuals are low away from the cracks.

\section{Phase-field computations} \label{S:pf_validation}
In this section, we discuss the simulation of the three WSTs using the phase-field model described in Sect.~\ref{S:pf_phase_field} and calibrated in Sect.~\ref{S:pf_mat_char}. 

\subsection{Spatial discretization and mesh generation}
\label{sct:pf_mesh}

The FE meshes used in the numerical computations are generated using the segmented images. Different methods exist to discretize the domain \cite{Lengsfeld:1998, Buljac:2017, Pietro:2018}. The two most commonly used involve either a \textit{voxel-based} discretization, whereby each voxel is treated as a hexahedral element \cite{Lengsfeld:1998} or the detection of the surface separating each phase followed by direct meshing of the enclosed volume with tetrahedral elements \cite{Pietro:2018}. Voxel-based meshing is the simplest approach, which, however, leads to a large number of degrees of freedom. Conversely, direct meshing with tetrahedral elements offers more flexibility in setting the size of the spatial discretization but it often leads to meshing errors, due to the approximations introduced during the identification of the separating surfaces by, e.g., the marching cubes algorithm \cite{Lorensen1998}. Artifacts such as non-watertight surfaces, overlapping of separate particles or presence of gaps in otherwise continuous subdomains are frequently present and their manual correction, if at all feasible, is extremely time consuming. 

\begin{figure}[!hbt]
\centering
    \includegraphics[width=.45\textwidth,valign=c]{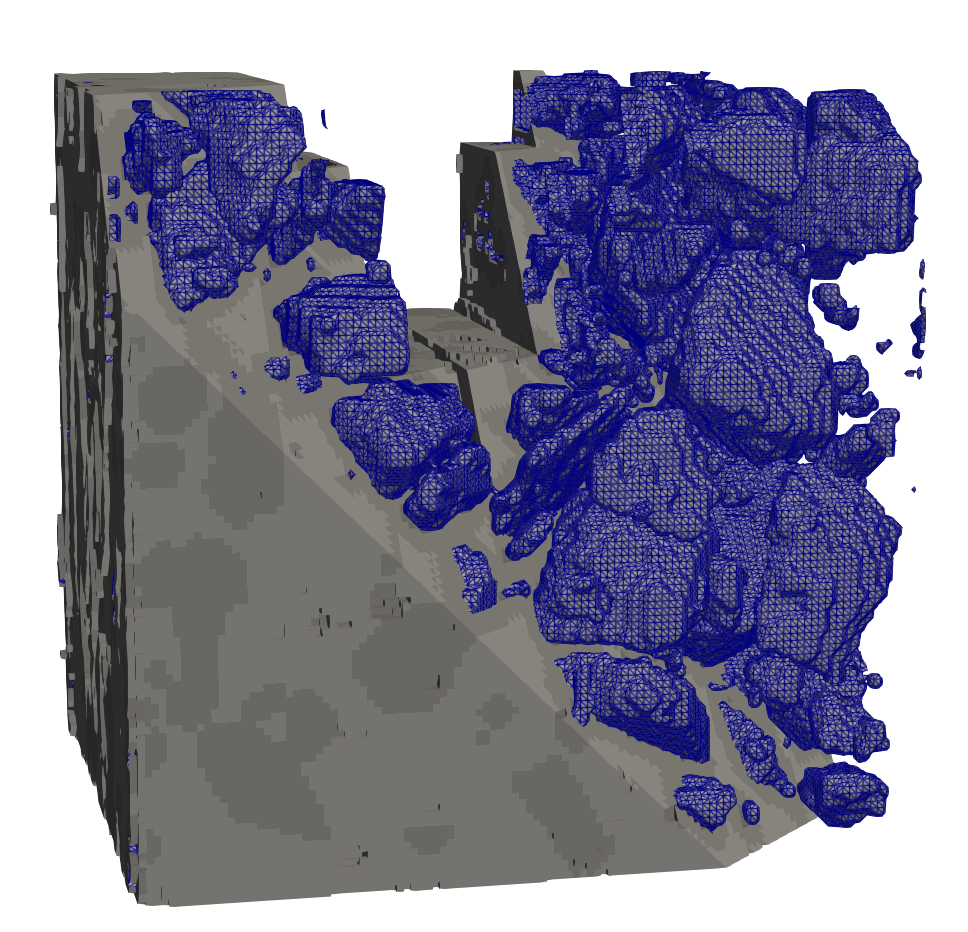}
        \caption{Example of FE mesh for a WST specimen. For better clarity, the FE mesh is shown only for the aggregates. } \label{fig:pf_meshed_agg}
\end{figure}

In this work, we generate voxel-based meshes using the library \texttt{deal.ii} \cite{Arndt:2021}. This step is crucial to obtain meshes compatible with a distributed memory architecture and to avoid copying the complete mesh in every core, which can lead to a rapid overload of the available memory. However, since each sample contains on average 512$\cdot$10$^6$ voxels, we first perform a
6$\times$6$\times$6$\times$ binning of the tomograms, reaching a voxel size of about $\sim$300$\times$300$\times$300 $\mu$m$^{3}$.
This downsampling serves only the purpose of reducing the global number of degrees of freedom.
A mesostructure phase label is then assigned to each resulting voxel, depending on the material corresponding to its centroid and a trilinear hexahedral FE mesh with size of $h\simeq$ 300 $\mu$m is defined. Once the elements belonging to aggregate and matrix phases are identified, the ITZ material is modeled by reassigning the label of the matrix elements within a width $\sim\ell$ from the aggregate boundaries as ITZ. This leads to an ITZ thickness larger than the realistic estimate of $\simeq$50 $\mu$m. However, we are here interested in the effects of the ITZ on the global behavior of concrete rather than on the local response of the ITZ phase itself.
Note that, in the phase-field approach, features with characteristic size smaller than the regularization length have a negligible effect on the response of the material. Therefore, with the above choice the ITZ has a sufficient size to influence the global response, and we adopt for the associated material parameters the upper bound of the reasonable range to approximately account for the choice of a larger thickness (see Sect.~\ref{sct:pf_el_prop} and Sect.~\ref{sct:pf_tpb_calibration}).
The final mesh obtained in this way has about $3.5 \cdot 10^7$ degrees of freedom. A representative example for a
WST specimen is shown in Fig. \ref{fig:pf_meshed_agg}.
 
The elements corresponding to the pores should ideally be removed from the domain. However, this would lead to an unstructured mesh, which would require the element connectivity matrix to be copied on each computational core, leading to an increase of the memory consumption. To avoid this, we keep the elements corresponding to the pores but we do not associate any degrees of freedom to them \cite{Arndt:2021}. This allows to preserve a structured mesh, while modeling the domain only with the necessary degrees of freedom. 

\subsection{Roles of the heterogeneous mesostructure and of the experimental boundary conditions}
\label{sct:pf_heter_role}

Before validating the calibrated phase-field model, we investigate the role of heterogeneities and boundary conditions on the numerically predicted fracture pattern. To this end, we use the phase-field model (Sect.~\ref{S:pf_phase_field}) to simulate the WST on specimen $\#$2, where, at first, we do not distinguish between mortar, aggregates and ITZ. Rather, we consider a homogeneous volume with mechanical characteristics set to the volume average of those calibrated in Sect.~\ref{S:pf_mat_char}. Note that this is not meant as a homogenization of the actual material and is only done to have some reasonable orders of magnitude for the homogeneous material properties. The adopted average material parameters are $E= 76.5$ GPa, $\nu = 0.17$, $G_f = 0.135$ N/mm and $f_t = 11.4$ MPa, while we use a regularization length $\ell= 1$ mm. We compare the results obtained using \texttt{vol/dev} and \texttt{spectral} splits.

\begin{figure}[!htb]
    \centering
        \begin{tikzpicture}
            \node[anchor=south west,inner sep=0] (main) at (0,0) {\includegraphics[width=.33\textwidth,trim={4.8cm 2.5cm 13cm 2.5cm},clip]{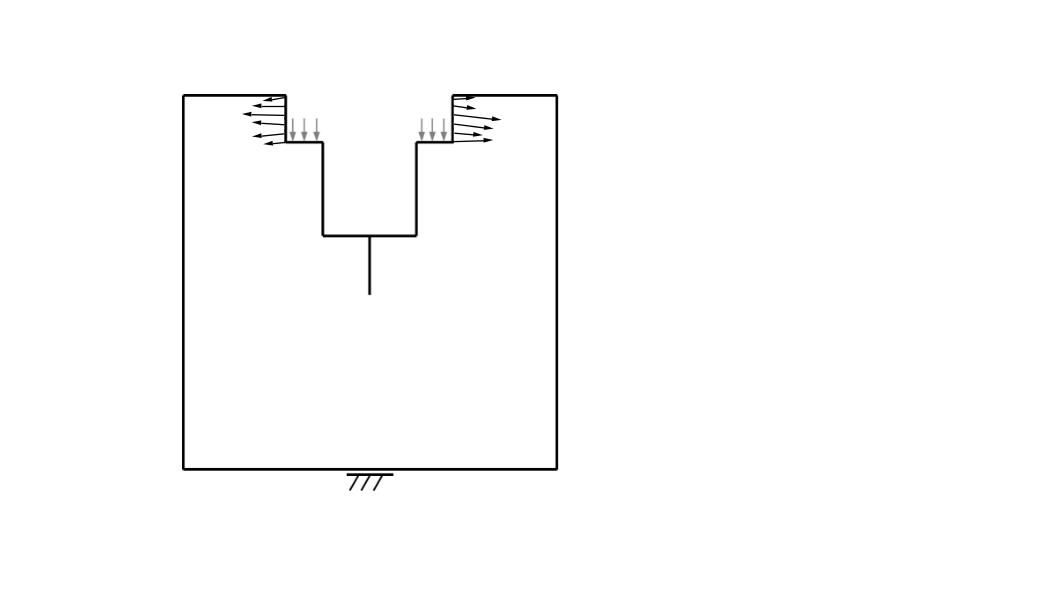}};
            \begin{scope}[x={(main.south east)},y={(main.north west)}]   
                \node[anchor=north] at (0.10,.98){$\bar{\bs u}(\bs{x})$};
                \node[anchor=north] at (0.9,.98){$\bar{\bs u}(\bs{x})$};
                \node[anchor=north] at (0.35,1.03){$\bar{\bs t}$};
                \node[anchor=north] at (0.65,1.03){$\bar{\bs t}$};
                \draw[thick,->] (1.15,0.05) -- (1.3,0.05);
                \node[anchor=north] at (1.3, 0.04){$x_1$};
                \draw[thick,->] (1.15,0.05) -- (1.15,0.20);
                \node[anchor=north] at (1.18, 0.26){$x_3$};
            \end{scope}
        \end{tikzpicture}
    \caption{Boundary conditions, illustrated qualitatively on a 2D cross-section along the $x_2$ axis. }\label{fig:pf_bc_homog}
\end{figure}

At each specimen cross-section along the $x_2$ direction, we use the boundary conditions qualitatively illustrated in Fig.~\ref{fig:pf_bc_homog}. We impose zero displacements in vertical (i.e., $x_3$) direction at the linear support (bottom bar in Fig.~\ref{fig:pf_bc_homog}), prescribed displacements  $\bs{\bar u}(\bs{x})$ on the upper part of the T-slot and tractions $\bs{\bar t}$ replicating the compressive forces exerted by the loading rollers on the specimen. Since only the vertical resultant force measured by the load cell is available, the tractions $\bar{\bs t}$ are considered uniformly distributed and are obtained by dividing the load cell readings by the horizontal area of the upper part of the T-slot (Fig.~\ref{fig:pf_bc_homog}). The imposed displacements $\bs{\bar u}(\bs{x})$ are obtained projecting the DVC displacements onto the FE mesh. This implies that the left and right parts of the T-slot are subjected to different imposed displacements, thus replicating the real situation. This is essential to investigate the role of the material heterogeneity while avoiding symmetric loading conditions that force the crack to propagate vertically.
 
  \begin{figure}[!htb]
\centering   
    \subfloat[]{
     \begin{tikzpicture}
             \node[anchor=south west,inner sep=0] (main) at (0,0) {\includegraphics[width=.36\textwidth]{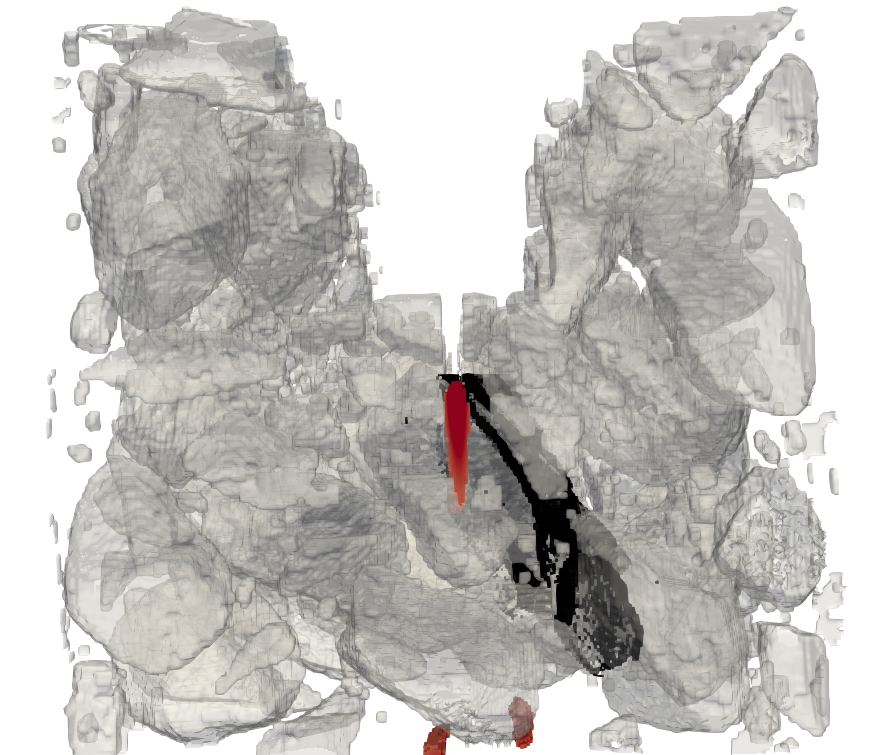}};
            \begin{scope}[x={(main.south east)},y={(main.north west)}]
  \end{scope}
  \end{tikzpicture}\label{fig:pf_homog_voldev}
   }\hspace{5mm}
 \subfloat[]{
    \begin{tikzpicture}
             \node[anchor=south west,inner sep=0] (main) at (0,0) {\includegraphics[width=.36\textwidth]{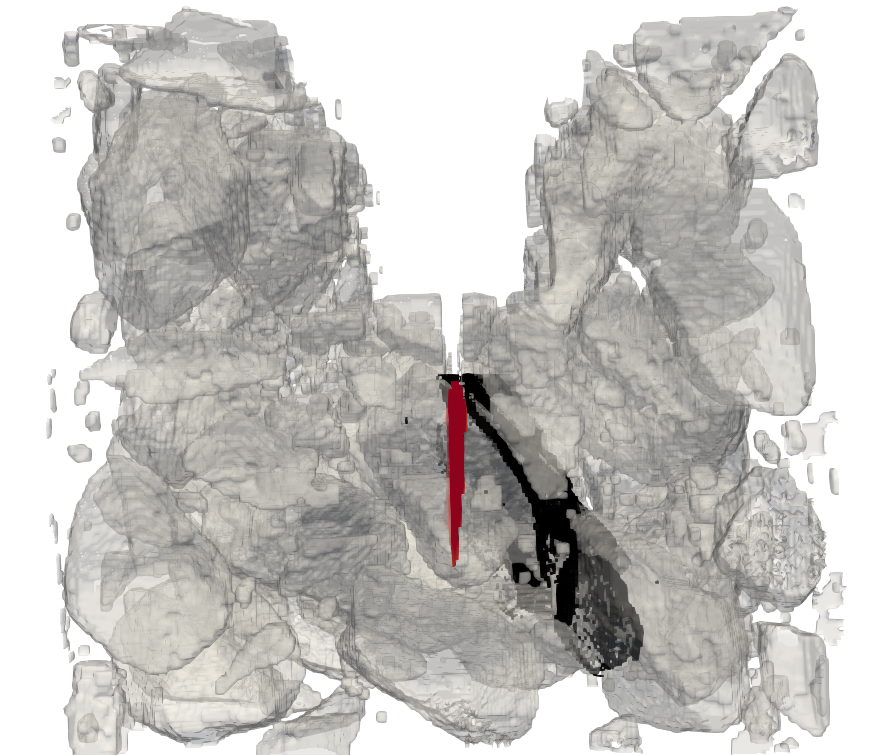}};
            \begin{scope}[x={(main.south east)},y={(main.north west)}]
  \end{scope}
 \end{tikzpicture}\label{fig:pf_homog_spect}
}
\caption{Comparison between the experimental crack pattern (in black) and (in red) the one predicted assuming a homogeneous material and the boundary conditions described schematically in Fig.~\ref{fig:pf_bc_homog}, for (a) the \texttt{vol/dev} and (b) the \texttt{spectral} splits, respectively. (The aggregates are semi-transparent, for illustrative purposes only).}
\label{fig:pf_ideal_bcs}
\end{figure}

To improve iterative convergence during the analyses, the displacement variations estimated between two subsequent tomograms are applied in increments by assuming a linear ramp. The total number of load steps is heuristically estimated for each test. Whenever the Newton-Raphson procedure which solves the displacement or the damage problem fails to converge, the number of increments between two consecutive X-CT acquisitions is increased, and the solution procedure is restarted from the last converged load step corresponding to a X-CT acquisition. 

The crack patterns at the final load step obtained with the \texttt{vol/dev} and \texttt{spectral} splits are illustrated in red in Figs.~\ref{fig:pf_homog_voldev} and \ref{fig:pf_homog_spect}, respectively. In the following, we identify the crack pattern as the set of points with $\alpha\ge$ 0.9. Although we consider in this case a specimen made of a homogeneous material, in Fig.~\ref{fig:pf_ideal_bcs} we include the outline of the aggregates for illustrative purposes only (since they are relevant for the experimental crack patterns). The results obtained with the \texttt{vol/dev} and the \texttt{spectral} splits are similar, with a sub-vertical crack emanating from the initial notch and no crack deviation in spite of the unsymmetric distribution of applied displacements. The main difference between the two results is the wider thickness and shorter length of the crack predicted by the \texttt{vol/dev} split (Fig.~\ref{fig:pf_homog_voldev}) compared to the  \texttt{spectral} one (Fig.~\ref{fig:pf_homog_spect}). This effect can be traced back to the higher residual stiffness $\eta_l$ used for the \texttt{vol/dev} split as discussed in Sect.~\ref{S:pf_num_aspects}. Experimentally, we obtain a more complex crack characterized by several kinks and the tendency to avoid the aggregates (in black in Fig.~\ref{fig:pf_homog_spect}). 

\begin{figure}[!hbt]
\centering
\subfloat[]{
 \centering
    \includegraphics[width=.35\textwidth,valign=c]{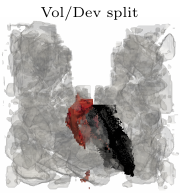}
\vphantom{\includegraphics[width=.3\textwidth,valign=c]{Adopted_mesh.png}}
 \label{fig:pf_heter_voldev}
    }\hspace{2.5cm}\subfloat[]{
     \centering
      \includegraphics[width=.35\textwidth,valign=c]{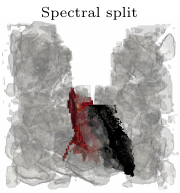}
      \vphantom{\includegraphics[width=.3\textwidth,valign=c]{Adopted_mesh.png}}
   \label{fig:pf_heter_spect}
}
        \caption{Phase-field modeling results obtained using the boundary conditions described in Fig.~\ref{fig:pf_bc_homog} for the heterogeneous mesostructure, using (a) the \texttt{vol/dev} and (c) the \texttt{spectral} splits, respectively.}
        \label{fig:pf_ideal_het_bcs}
\end{figure}

We explore now how the results change if the boundary conditions of Fig.~\ref{fig:pf_bc_homog} are complemented with the real heterogeneous geometry obtained with X-CT (Fig.~\ref{fig:pf_meshed_agg}). The regularization length is here kept as $\ell$ = 1 mm, since this is the minimum length that could be adopted given the available computational resources (Sect.~\ref{S:pf_PF})\footnote{With about 35 millions degrees of freedom and a heterogeneous domain, the computations with $\ell$ = 1 mm last for about 4 weeks. A reduction of the length scale parameter would lead to a much longer computation time (i.e., above 8 weeks). However, the results (not shown here) obtained performing different partial computations with $\ell$ varying from 2 mm to 0.7 mm do not indicate any major changes in the crack pattern. Therefore, we consider the predictions with $\ell$ = 1 mm reliable.}. 
The comparison between experimental (in black) and numerical (in red) crack pattern is presented in Fig.~\ref{fig:pf_ideal_het_bcs}. With respect to those shown in Fig.~\ref{fig:pf_ideal_bcs}, the numerical crack paths are here more intricate and tend to kink in order to avoid impinging into the aggregates for both the \texttt{vol/dev} (Fig.~\ref{fig:pf_heter_voldev}) and the \texttt{spectral} splits (Fig.~\ref{fig:pf_heter_spect}). Nevertheless, the agreement between numerical and experimental crack paths is not satisfactory.

\subsection{Boundary conditions, geometries and length scale adopted for validation}
\label{sct:pf_bc}

The results discussed in Sect.~\ref{sct:pf_heter_role} suggest that the improvement of the boundary conditions can be the key to improve the predictions. Following what proposed in \cite{Madi:2013, Nguyen2016_1}, where similar challenges are addressed, we proceed by considering a sub-volume of the domain completely included in the tested volume, while imposing at its boundary the experimentally obtained displacements. 

While the adoption of a sub-volume as computational domain precludes the numerical computation of the reaction forces, this procedure limits the effects of scattering, partial volume and cone beam X-CT artifacts affecting the outer surfaces of the specimens. This class of artifacts is more relevant in the regions where materials with very different attenuation coefficients are in contact (e.g., at concrete-air or concrete-PMMA interfaces).

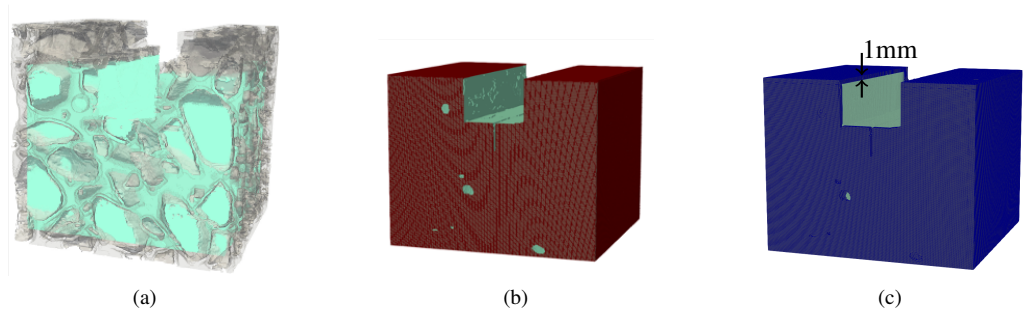
\begin{figure}
\centering
\subfloat[]{
    \centering
             {\includegraphics[width=.22\textwidth]{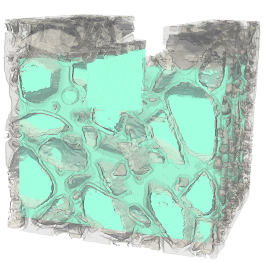}}
\label{fig:pf_comp_dom}
    }\hspace{1cm}
    \subfloat[]{
     \centering
     {\includegraphics[width=.22\textwidth]{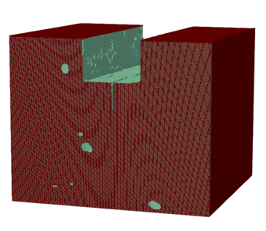}\label{fig:subv_BC}}
}\hspace{1cm}
\subfloat[]{
     \centering
    \begin{tikzpicture}
             \node[anchor=south west,inner sep=0] (main) at (0,0) {\includegraphics[width=.225\textwidth]{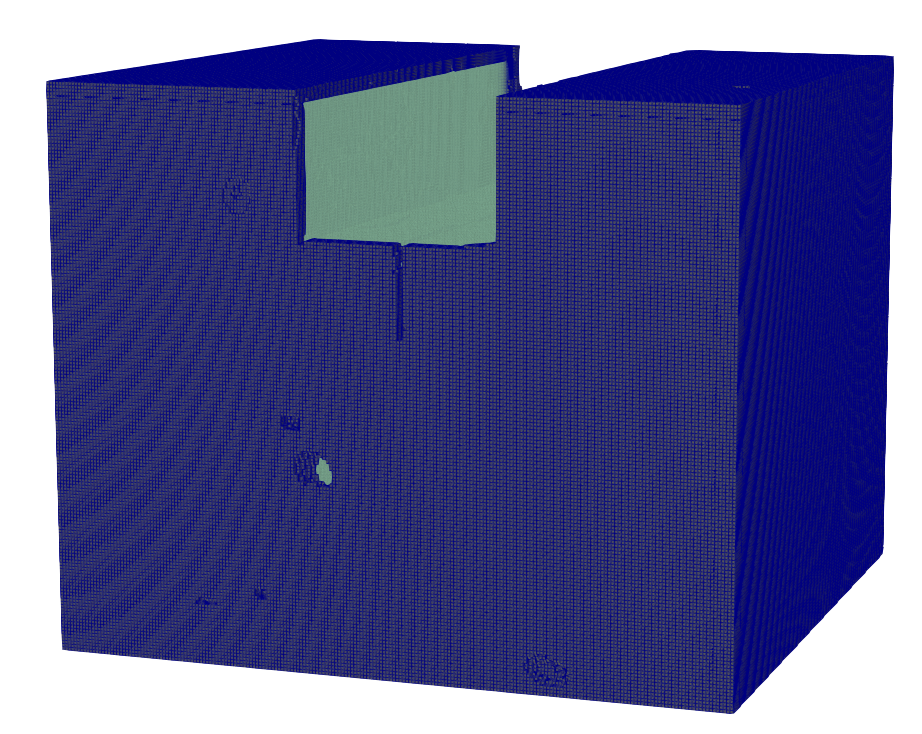}};
            \begin{scope}[x={(main.south east)},y={(main.north west)}]
             \draw[thick,->] (0.4,1) -- (0.4,0.9);
              \draw[thick,->] (0.4,0.8) -- (0.4,0.88);
               \node[anchor=north] at (0.5,1.1){1mm};
  \end{scope}
 \end{tikzpicture}\label{fig:pf_dmg_bc}
}
        \caption{ Computational sub-volume and boundary conditions: (a) comparison between the complete specimen (in gray) and the sub-volume used as computational domain (in cyan), (b) surfaces where non-homogeneous Dirichlet (in red) and homogeneous Neumann  (in cyan) boundary conditions are applied and (c) regions where $\alpha=0$ is imposed (in blue). }
        \label{fig:pf_aspects_fig}
\end{figure}

Since we expect that the crack propagates predominantly in the region of the specimens centered on the notch front, we define the computational domain by trimming $7$ mm from the top and $2$ mm from the bottom of the specimens. We eliminate a larger portion of the upper part of the specimens since these regions are mostly unloaded. Additionally, we reduce the lateral size of the specimens in directions $x_1$ and $x_2$ by $1$ mm from each side.
The final computational domain has outer size of about 38$\times$38$\times$29 mm$^3$ (width $\times$ thickness $\times$ height) and is compared to the complete specimen in Fig.~\ref{fig:pf_comp_dom}.  
Besides allowing the estimation of more accurate boundary conditions, the employment of  a sub-volume also yields a reduction in the number of degrees of freedom to about $2.5 \cdot 10^7$.

This reduction in turn enables the selection of a smaller regularization length $\ell$, which from now on is fixed to $\ell$ = 700 $\mu$m. As discussed in  Sect.~\ref{S:pf_PF}, the adopted phase-field model is length scale independent provided that $\ell$ is sufficiently small compared to the smallest size of the geometry. 
For our heterogeneous mesostructure, the presence of aggregates in contact leads to a vanishing smallest size (see also Sect.~\ref{S:eff_ell}), hence $\ell$ should be chosen as small as possible and its chosen value is governed by the minimum mesh size affordable with the available computational resources. Considering a minimum mesh size $h_{min}$ = 300 $\mu$m, we choose $\ell = $ 700 $\mu$m leading to $\ell / h\simeq$ 2.3.

Once the computational sub-domain is extracted, we apply non-homogeneous Dirichlet boundary conditions to the faces corresponding to the cutting planes (highlighted in red in  Fig.~\ref{fig:subv_BC}). These imposed displacements are obtained by projecting on the FE mesh the displacements obtained from the DVC analyses on the associated support mesh. Homogeneous Neumann boundary conditions are imposed to the remaining boundaries, namely the T-slot, notch and pores surfaces highlighted in cyan in Fig.~\ref{fig:subv_BC}.  We apply the same load stepping scheme illustrated in Sect.~\ref{sct:pf_heter_role}. To avoid spurious evolution of damage due to local stress concentrations arising from fluctuations of the applied displacements, the condition $\alpha=0$ is enforced in the regions within $1$ mm of the Dirichlet boundaries, highlighted in blue in Fig.~\ref{fig:pf_dmg_bc}.  

\subsection{Comparison between numerical and experimental results}

In this section we compare the experimental observations obtained for the three WST specimens, tested in-situ (Sects.~\ref{S:pf_WST_res}-\ref{S:pf_dvc_an}), with the numerical predictions considering the actual heterogeneous mesostructures and the experimental boundary conditions (Sect.~\ref{sct:pf_bc}). All the computations are performed with the parameters summarized in Tab.~\ref{tab:pf_final_pat} along with $\ell$ = 700 $\mu$m. 

Augmented reality (AR) renderings corresponding to the illustrated 3D computational results and with the experimental observations can be accessed at \url{https://ar.compmech.ethz.ch} or using the QR-code in
\ref{app:AR_render}. The renderings show the aggregates with sieve size $\varnothing>$1 mm as segmented in the
tomograms and described computationally, rendered in gray and semi-transparent.
The renderings also show, in black, the cracked regions obtained from the segmentation of the DVC residuals, obtained after image registration (Sect.~\ref{sct:pf_dvc_result}).
Finally, they show in red the regions where $\alpha\ge$0.9, as obtained from the phase-field computations.

\subsubsection{Test on specimen $\#$1}
\label{sct:pf_test1}

\begin{figure}[h]
\begin{minipage}{\textwidth}
     \centering
    \subfloat[]{
     \begin{overpic}[width=0.365\textwidth]{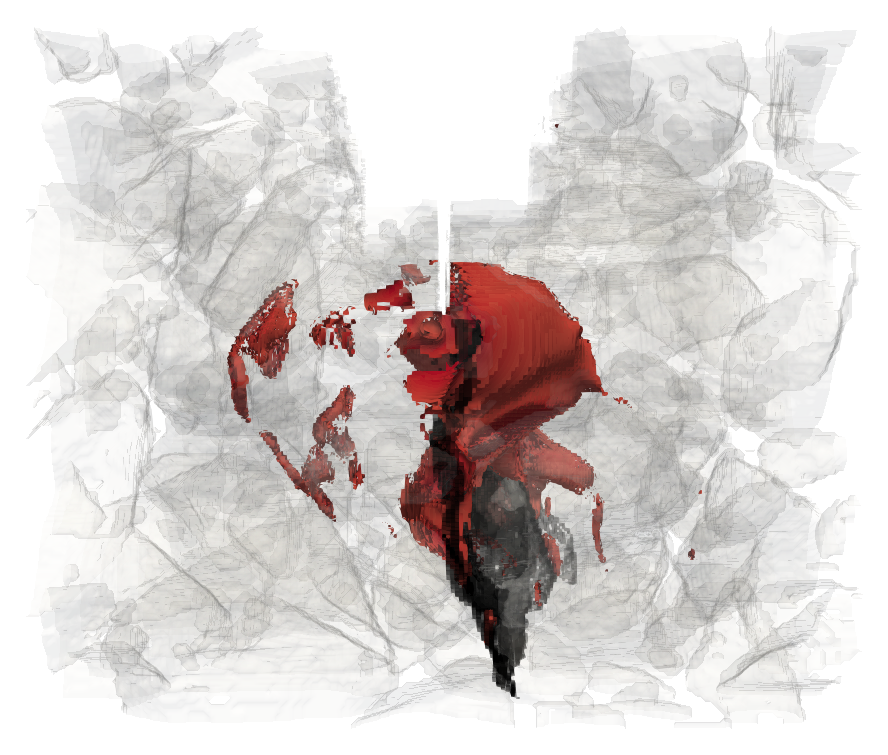}
    \end{overpic}}
    \hspace{10mm}
     \subfloat[]{
     \begin{overpic}[width=0.365\textwidth]{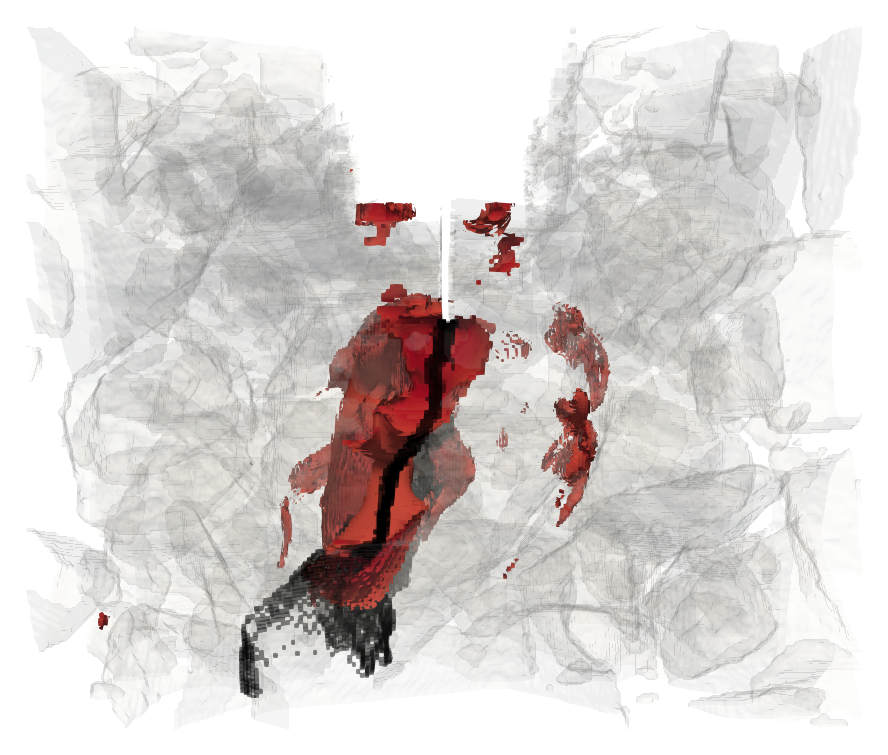}
    \end{overpic}}
\end{minipage}
   \begin{minipage}{\textwidth}
     \centering
    \subfloat[]{
     \begin{overpic}[width=0.365\textwidth]{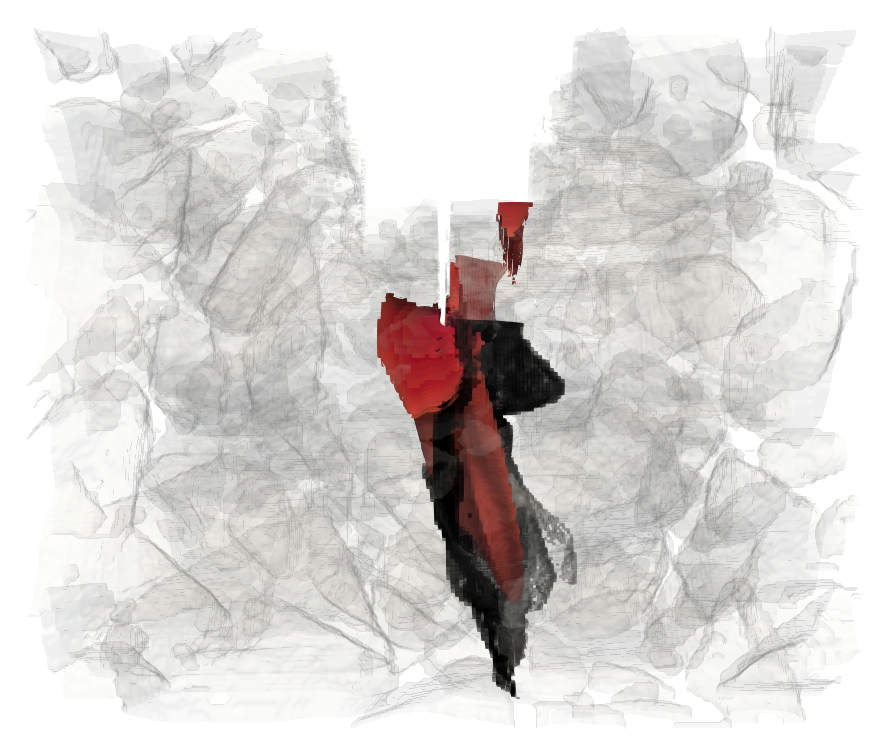}
    \end{overpic}}
    \hspace{10mm}
     \subfloat[]{
     \begin{overpic}[width=0.365\textwidth]{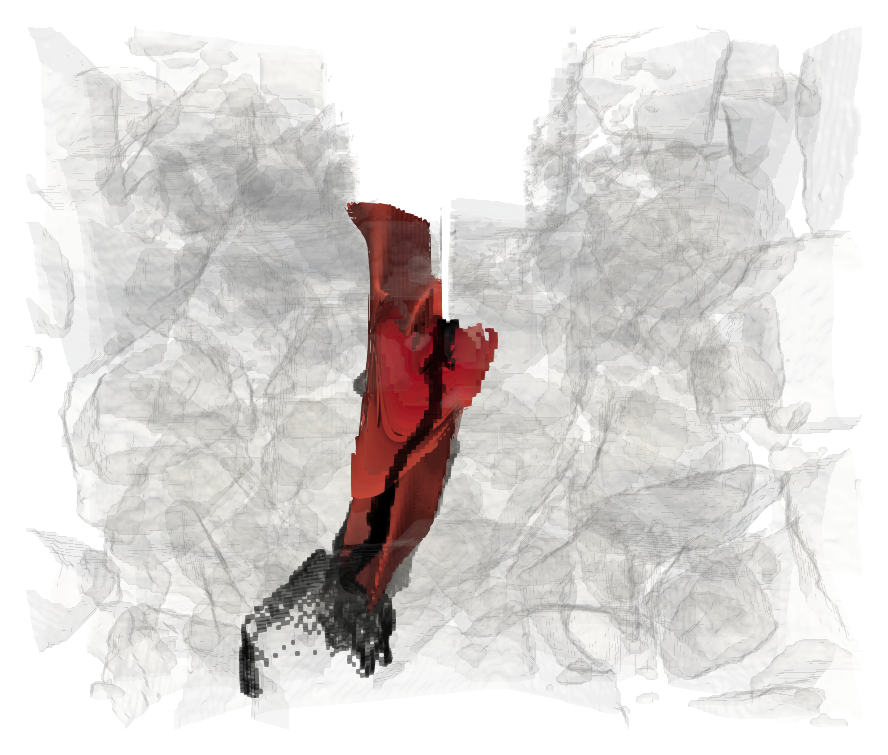}
    \end{overpic}}
\end{minipage}
    \caption{Comparison of the numerical (in red) and experimental (in black) crack patterns at the last load step of the WST on specimen $\#$1:  (a) front  and (b) back views of the crack pattern obtained using the \texttt{vol/dev} split, and (c) front and (d) back views obtained using the \texttt{spectral} split. The black rendered volume is
        the physical crack, as segmented from the tomograms, while the red volume corresponds to the $\alpha\geq$ 0.9 region obtained from the numerical computations.}
    \label{fig:pf_spec1_front}
\end{figure}

We start by comparing in Fig.~\ref{fig:pf_spec1_front} the experimentally resolved crack region at  failure for specimen $\#$1 with the corresponding numerical results, obtained using both \texttt{vol/dev} and \texttt{spectral} splits. The agreement is significantly better than in Fig.~\ref{fig:pf_ideal_het_bcs}. Due to some local stress concentrations triggered by the presence of pores and aggregates, the results obtained with the \texttt{vol/dev} split contain some further cracked regions away from the main crack, which appear to be due to spurious damage in compression. A possible strategy to avoid them could be adopting the so-called star-convex energy split \cite{Vicentini2024}, a modified version of the \texttt{vol/dev} split that allows to calibrate also the compressive strength of the material. The crack obtained with the \texttt{spectral} split tends to propagate through the aggregates rather than around them. This result can be better appreciated
with the AR 3D renderings, see \ref{app:AR_render}. Conversely, the \texttt{vol/dev} split leads the crack to kink and follow the shape of the aggregates, leading to a better agreement with the experimental evidence.  
 
\begin{figure}[!htb]
   \centering
    \includegraphics[width=0.75\textwidth]{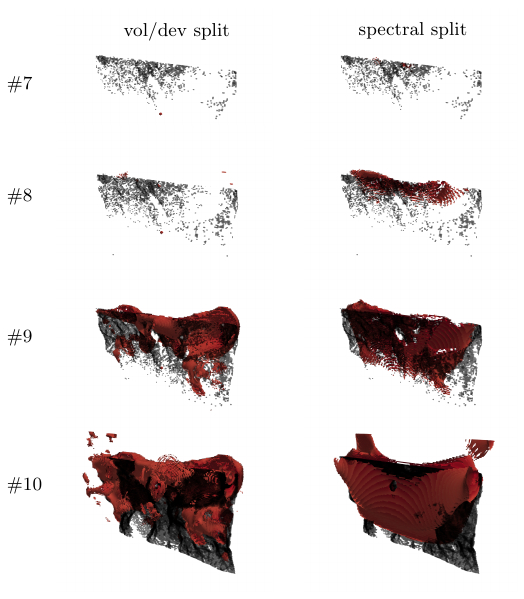}\label{fig:pf_evolution1_spec1}
        \caption{Damage evolution at subsequent load steps for the WST on specimen $\#$1.
        Similar rendering scheme as for Fig. \ref{fig:pf_spec1_front}.}
        \label{fig:pf_evolution_spec1}
\end{figure}

In Fig.~\ref{fig:pf_evolution_spec1} we compare the experimental results at different test stages with the numerical crack evolution, using both the \texttt{vol/dev} and the \texttt{spectral} splits.
In this comparison, care must be taken in interpreting both numerical and experimental results.  precision of the DVC measurements (Sect.~\ref{sct:pf_uncertainty_analysis}), the residuals become significant (i.e., higher than the background noise) when the displacement mismatch exceeds about 10-15 $\mu$m, which is thus the minimum detectable crack width. On the other hand, high residuals are only the signature of the crack presence, while no information is provided about the local crack opening value. Conversely, the damage variable $\alpha$ (in the limit for $\ell\to0$) is directly related to the value of the displacement jump across the phase-field support during the gradual crack opening (Sect.~\ref{S:pf_degradation_func}). Accounting for \eqref{eq:pf_cohesive_lorentz}-\eqref{eq:pf_cohesive_lorentz2}, $\alpha\geq$ 0.9 is reached only when the displacement jump corresponds to a fully formed crack with a width larger than $\delta(\alpha=0.9)\approx$ 40 $\mu$m. Hence, the finest portion of the experimental crack with a width in the range of approximately 15 $\mu$m to 40 $\mu$m is not represented by the numerical results in Fig.~\ref{fig:pf_evolution_spec1}. This is particularly evident for the steps corresponding to scans \#7 and \#8, where a portion of the lower crack section is not numerically reproduced. A similar observation can be made for the other propagation stages where the crack opening is limited, namely for the lower portion of the crack (steps \#9 and \#10 in Fig.~\ref{fig:pf_evolution_spec1}). Considering the aforementioned limitations, the comparison of the crack evolution in Fig.~\ref{fig:pf_evolution_spec1} confirms what observed for the final crack path in Fig.~\ref{fig:pf_spec1_front}, with the \texttt{vol/dev} split representing better the failure mechanism.

\begin{figure}[!htb]
      \centering
\subfloat[]{ \centering\includegraphics[scale=1.1]{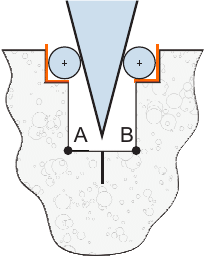} \label{fig:pf_codpts}}\hspace{4mm}
\subfloat[]{
\centering\includegraphics[scale=1.2]{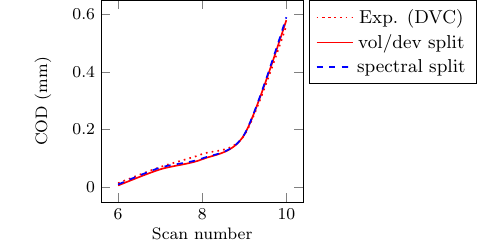} 
  \label{fig:pf_displ_spec1}}
\caption{ Points A and B marked in (a) denote the base of the T-slot in a 2D cross-section of the specimen. The experimental and numerical comparison of the COD  for WST on specimen $\#$1 is shown in (b).}\label{fig:cod_comp}
\end{figure}

As a global measure of the accuracy of the computations, we compare the experimental and numerical evolution of the COD. This is calculated at the base of the T-slot by subtracting the displacements averaged on both sides of the T-slot over the whole specimen width at the points \textsf{A} and \textsf{B} in Fig.~\ref{fig:pf_codpts} (illustrated with a 2D cross-section for clarity). The results are given in Fig.~\ref{fig:pf_displ_spec1} and demonstrate an excellent agreement. Note that the experimental COD related to the steps \#7 and \#8 in Fig.~\ref{fig:pf_evolution_spec1} is equal to 75 $\mu$m and 122 $\mu$m, respectively. These values, along with the limited magnitude of the splitting displacement, observable at the same load steps in Fig.~\ref{fig:pf_dvc1_displ}, are fully compatible with a crack width below 40 $\mu$m at the notch tip. 
               
\begin{figure}[!htb]
\centering
  \begin{tikzpicture}
             \node[anchor=south west,inner sep=0] (main) at (0,0) {\includegraphics[width=.35\textwidth]{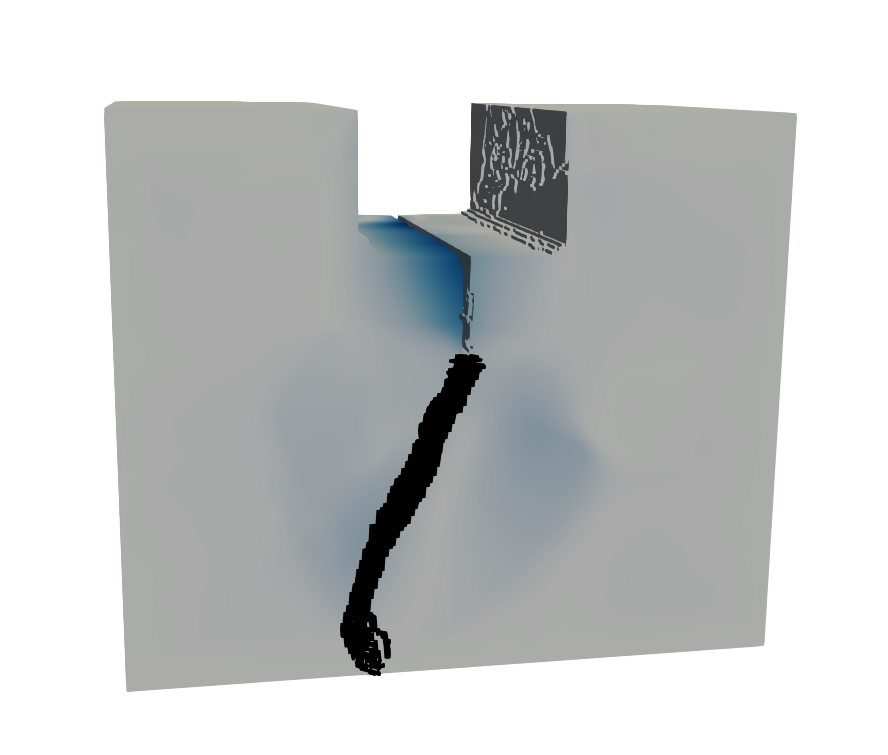}};
            \begin{scope}[x={(main.south east)},y={(main.north west)}]
             \node[anchor=south west] at (1,0.01) {\includegraphics[width=.12\textwidth]{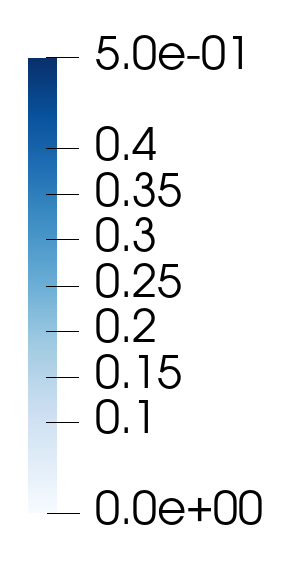}};
  \end{scope}
  \end{tikzpicture}
 \caption{Displacement mismatch (magnitude of the difference of the displacement vector, in mm) between numerical (using the \texttt{vol/dev} split) and experimental results at the last load step (\# 10) of the WST on specimen $\#$1, with experimental crack region shown in black.}
\label{fig:pf_disp_difference}
  \end{figure}

The local accuracy of the calibrated model is demonstrated in Fig.~\ref{fig:pf_disp_difference}, where we show the difference in magnitude between the experimental displacements and those obtained numerically with the \texttt{vol/dev} split at the final load step (\#10 in Fig.~\ref{fig:pf_evolution_spec1}). Given the different size of the FE and DVC support meshes, the results of the latter are projected onto the former. As expected, a mismatch close to the notch and to the crack is observed due to the inability of the DVC ansatz functions to reproduce displacement jumps at both locations (Sect.~\ref{S:pf_dvc_an}). In particular, a displacements mismatch close to the observed COD value is observed at the bottom of the T-slot (Figs.~\ref{fig:cod_comp} and \ref{fig:pf_disp_difference}), quantitatively confirming the accuracy of the analyses. As expected, the displacement mismatch is highly localized close to the discontinuities and vanishes away from them. 

Since the \texttt{vol/dev} split provides better results compared to the \texttt{spectral} one, unless otherwise specified, in the following we consider only the \texttt{vol/dev} split.

\subsubsection{Test on specimen $\#$2}
\label{sct:pf_test2}
\begin{figure}[!htb]
    \centering
    \subfloat[]{
    \begin{overpic}[width=0.365\textwidth]{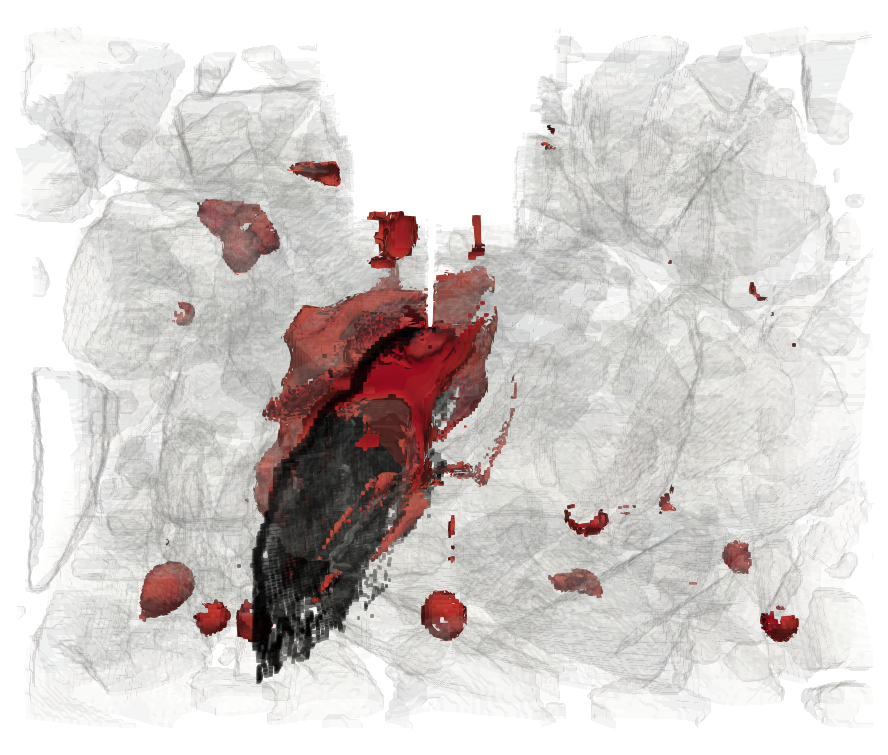}
    \end{overpic}}
    \hspace{10mm}
     \subfloat[]{
    \begin{overpic}[width=0.365\textwidth]{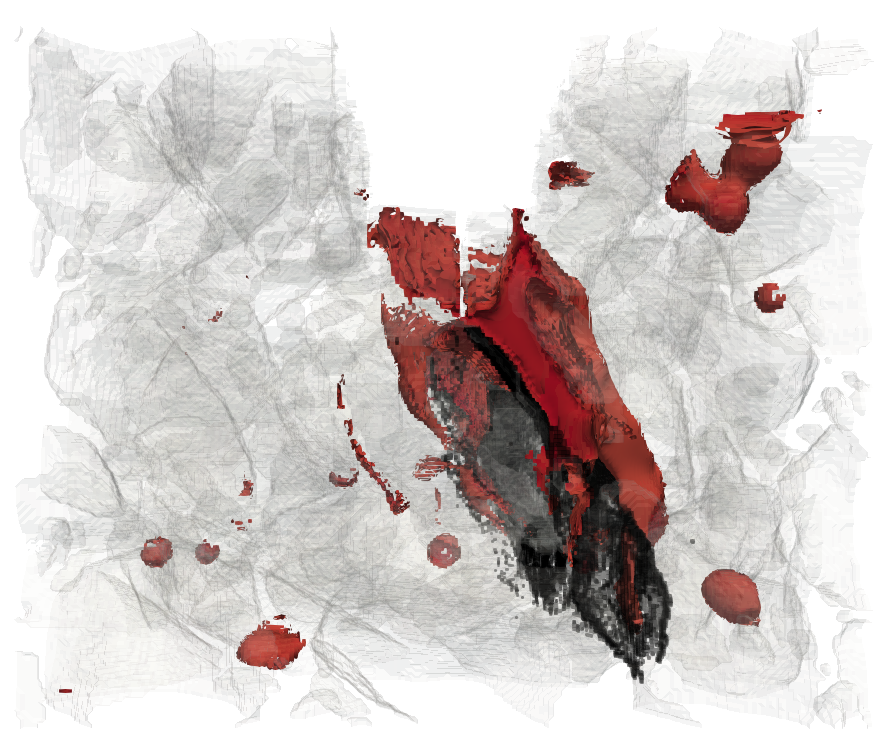}
    \end{overpic}}
    \caption{Comparison of the numerical and experimental crack patterns at the end of the WST on specimen $\#$2 modeled using the \texttt{vol/dev} split: (a) front view and (b) back view. The aggregates, as segmented from the tomogram, are rendered semi-transparent
    in gray. The crack region, also segmented from the tomogram, is rendered in black. The red-rendered region is the computed crack region.}
    \label{fig:pf_spec2}
\end{figure}

We now shift our attention to  specimen $\#$2 and, as before, we first compare the experimental and numerical crack patterns at the end of the test in Fig.~\ref{fig:pf_spec2}. The obtained results substantially support the observations of the previous example (Sect.~\ref{sct:pf_test1}), confirming the good agreement between numerical and experimental results. Also in this case, some spurious evolution of the damage variable is observed at pores or large aggregates (see Fig.~\ref{fig:pf_spec2} and the AR renders in \ref{app:AR_render}).

\begin{figure}[!htb]
    \centering
      \subfloat[]{\label{fig:pf_evolution_spec2_int1}
       \begin{tikzpicture}
      \node[anchor=west] at (1,0){\includegraphics[width=0.3\textwidth]{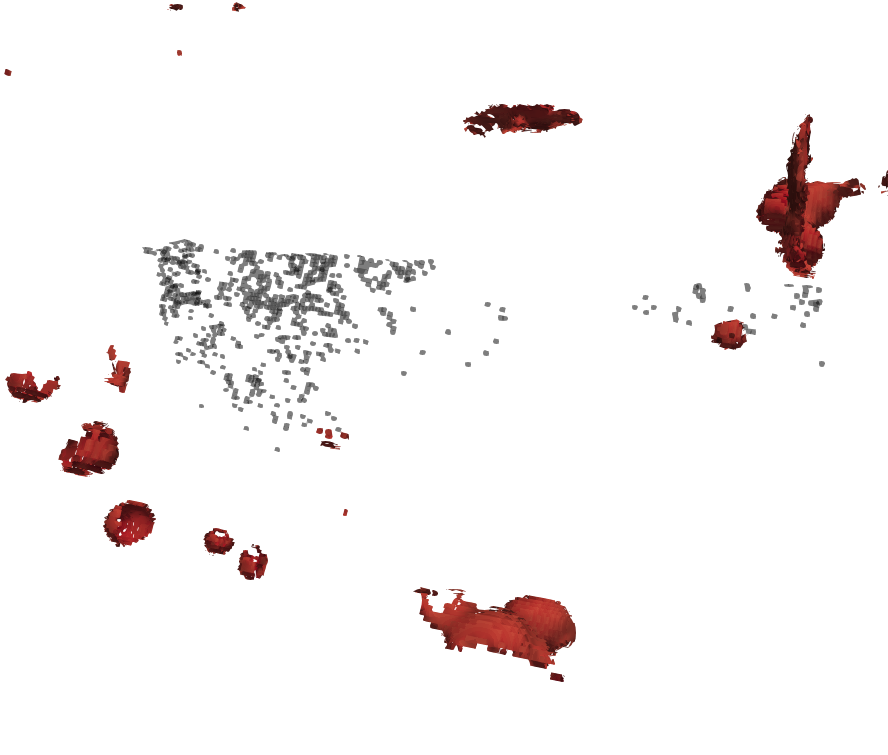}};
      \node[anchor=north west] at (0,0){\Large  \#8};
      \end{tikzpicture}
      }
      \hspace{15mm}
    \subfloat[]{\label{fig:pf_evolution_spec2_int2}
           \begin{tikzpicture}
      \node[anchor=west] at (1,0){\includegraphics[width=0.3\textwidth]{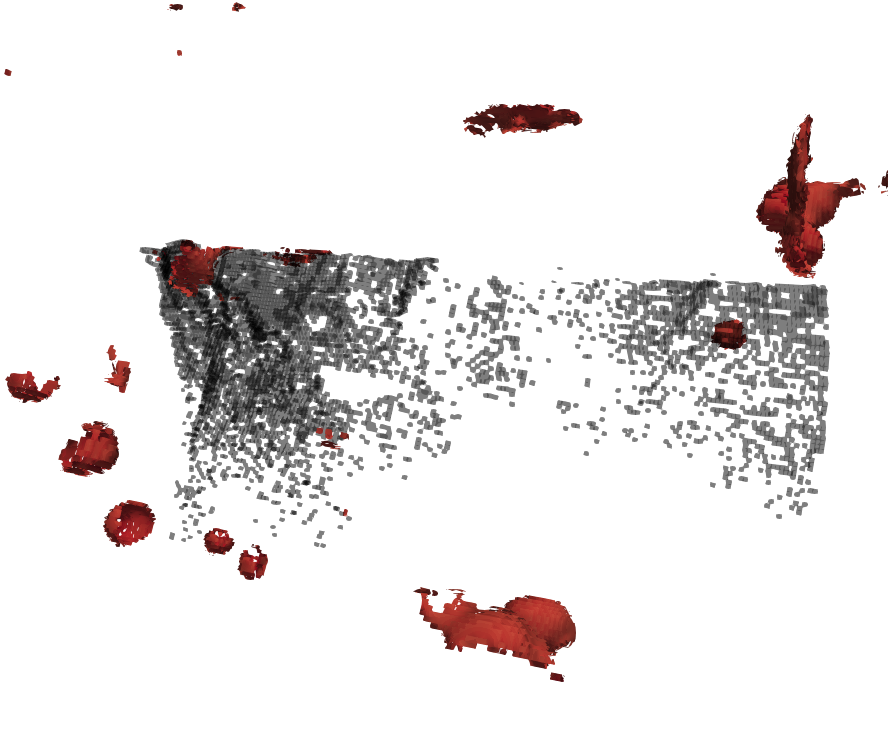}};
 \node[anchor=north west] at (0,0){\Large  \#9};
      \end{tikzpicture}
      }
    \hspace{8mm}
     \subfloat[]{\label{fig:pf_evolution_spec2_int3}
     \begin{tikzpicture}
      \node[anchor=west] at (1,0){\includegraphics[width=0.3\textwidth]{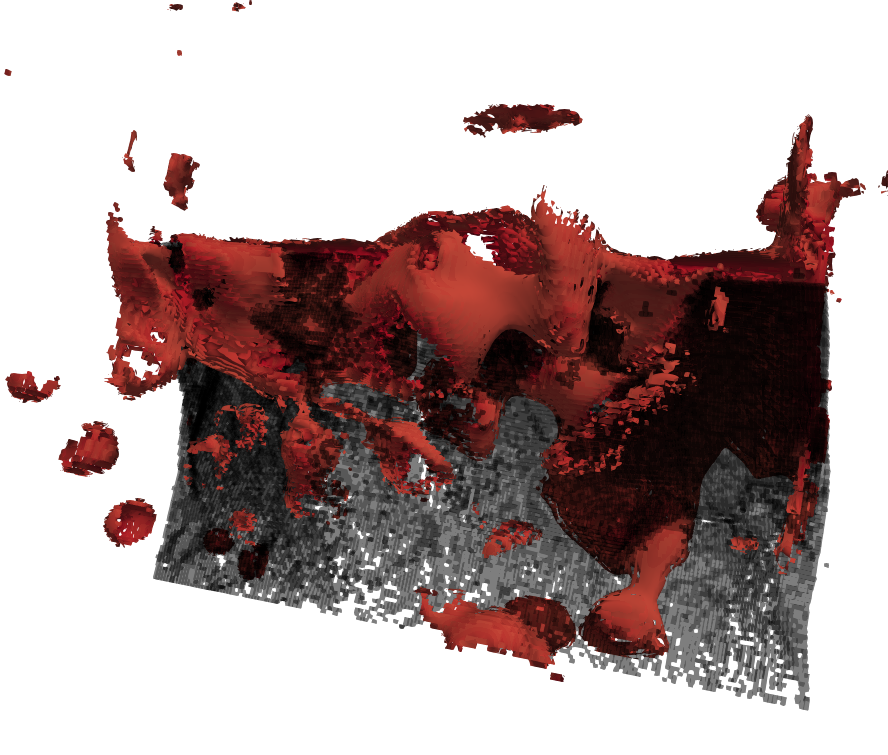}};
      \node[anchor=north west] at (0,0){\Large  \#10};
      \end{tikzpicture}
       }
        \caption{Damage evolution for the WST on specimen $\#$2, modeled with the \texttt{vol/dev} split, for the load steps (a) \#8, (b) \#9 and (c) \#10. The black-rendered region is the physical crack, as segmented from the tomograms, while the red
        region is the corresponding crack as obtained from the computation.}
        \label{fig:pf_evolution_spec2}
\end{figure}

As for specimen $\#$1, we now compare the experimental crack evolution with its numerical counterpart in Fig.~\ref{fig:pf_evolution_spec2}.  
We start here from load step \#8 since, before, neither experimental nor numerical cracks are visible. As observed for specimen $\#$1, at the initial stages of cracking the crack width is below 40 $\mu$m, hence, it is not properly represented by a value of $\alpha$ = 0.9 (Figs.~\ref{fig:pf_evolution_spec2_int1}-\ref{fig:pf_evolution_spec2_int2}). The onset of fracture is experimentally observed to take place at the upper-left corner, and this is also the case for the numerical predictions (Fig.~\ref{fig:pf_evolution_spec2_int2} and \ref{app:AR_render}). The evolution of the damage pattern occurs predominantly between the load steps \#9 and \#10, where a load drop of about 50\% is recorded (Fig.~\ref{fig:pf_dvc2}). Fig.~\ref{fig:pf_evolution_spec2_int3} confirms that, at the final load step, the numerical results describe a reasonably realistic crack pattern.

For this specimen, the results obtained using the \texttt{spectral} split are illustrated in \ref{s:spec_2}.

\subsubsection{Test on specimen $\#$3}

\begin{figure}[!hbt]
     \centering
    \subfloat[]{
      \includegraphics[width=0.37\textwidth]{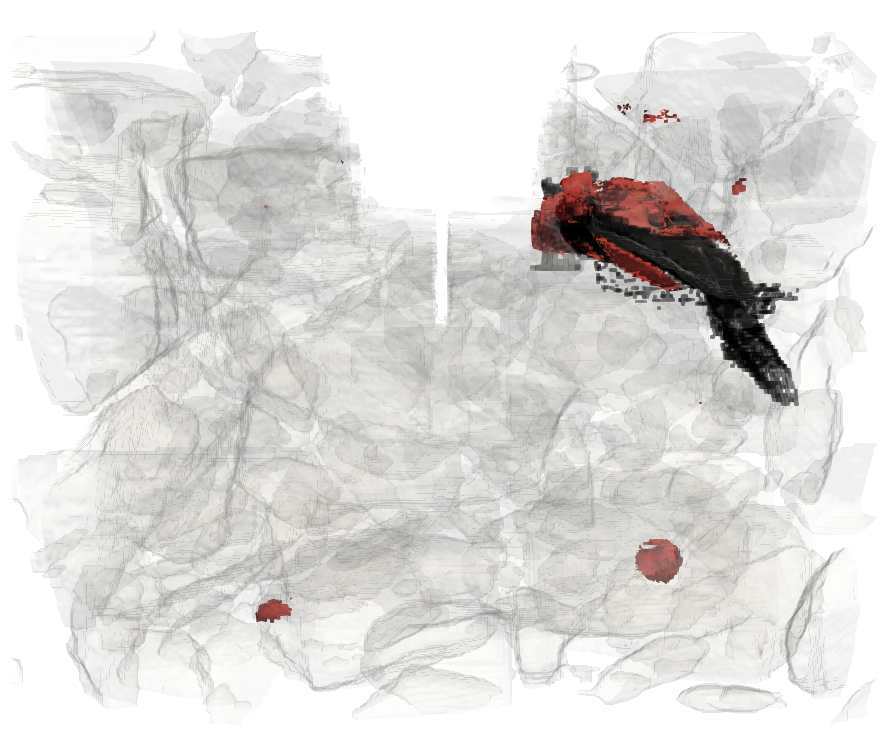}
   }
    \hspace{10mm}
     \subfloat[]{
     \includegraphics[width=0.37\textwidth]{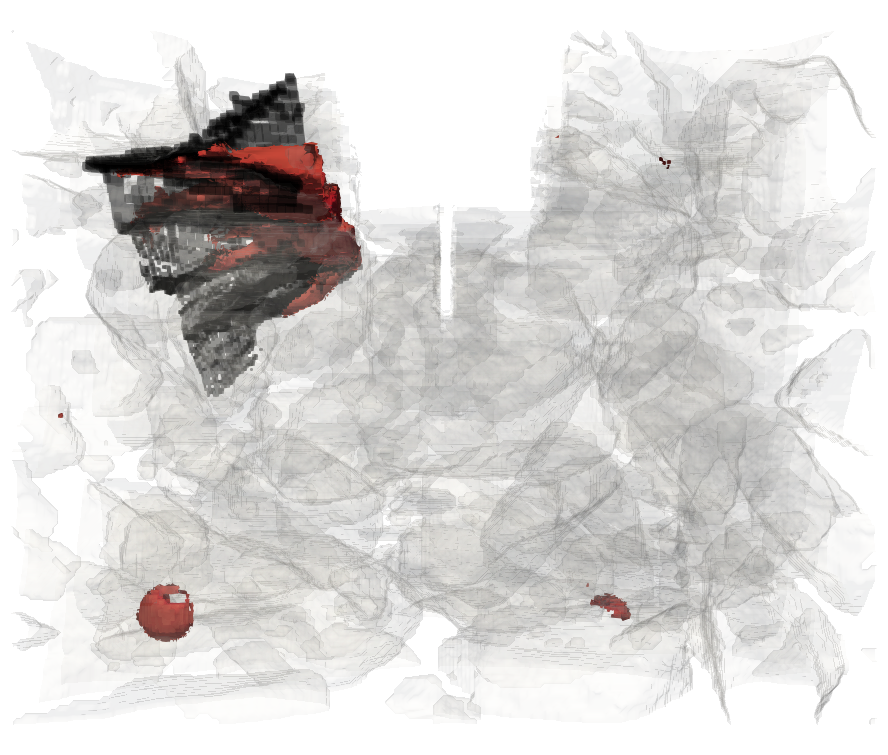}
    }
    \caption{Comparison of the numerical and experimental crack patterns at the final load step for specimen $\#$3: (a) front  and (b) back views of the crack pattern. The aggregates, segmented from the tomograms, are rendered semi-transparent in gray.
    The crack region also segmented from the tomogram is rendered in black.
    The computed crack region, rendered in red, is obtained using the \texttt{vol/dev} split. (Results using the \texttt{spectral} split are reported in \ref{s:spec_3})}
    \label{fig:pf_spec5}
\end{figure}

The comparison between numerical and experimental crack patterns at the last load step for specimen $\#$3 is shown in Fig.~\ref{fig:pf_spec5}. Also in this case a good agreement is observed in spite of the atypical failure mode discussed in Sect.~\ref{sct:pf_exp_test}. Note that the loading conditions during crack propagation are mainly of the opening (or mode-I) type, therefore different from those of specimens $\#$1 and $\#$2 where the results are influenced by the compressive load exerted by the loading rollers (Fig.~\ref{fig:pf_test_setup}) as well as by a significant shear at the crack kinking locations induced by aggregates and the ITZ. Nevertheless, the simulation is able to reproduce the experimental observations, demonstrating its predictive ability of the failure mechanism of the WST concrete specimens and of the influence exerted by the specific mesostructure.

The results obtained with the \texttt{spectral} split (Fig.~\ref{fig:pf_spec3_miehe} in \ref{s:spec_3}) are similar to those in Fig.~\ref{fig:pf_spec5}. This is due to the predominant mode-I loading conditions, which are known to lead to similar predictions when either the \texttt{vol/dev} or \texttt{spectral} splits are used \cite{Vicentini2024}.

\begin{figure}[htb]
    \centering
    \subfloat[]{\label{fig:spcm3_7}
    \begin{tikzpicture}
    \node[anchor=west] at (1,0){\includegraphics[width=0.35\textwidth]{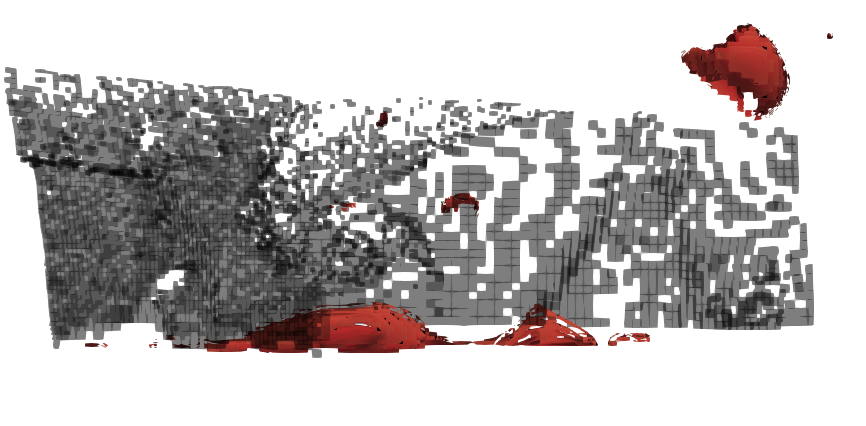}};
      \node[anchor=north west] at (0,0){\Large  \#8};
      \end{tikzpicture}
    }
    \hspace{10mm}
    \subfloat[]{\label{fig:spcm3_8}
    \begin{tikzpicture}
    \node[anchor=west] at (1,0){ \includegraphics[width=0.35\textwidth]{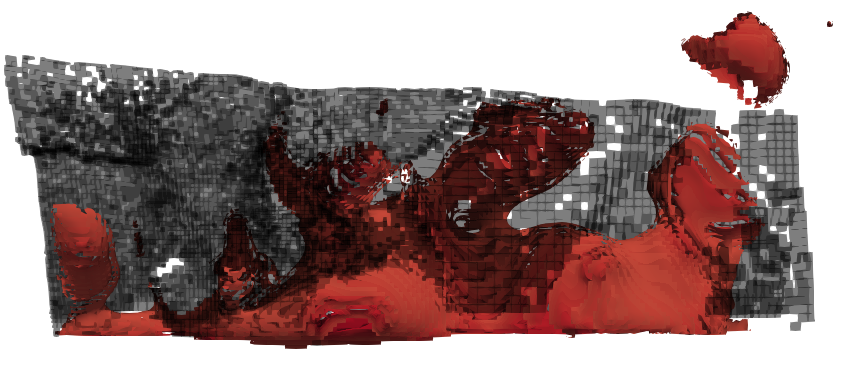}};
\node[anchor=north west] at (0,0){\Large  \#9};
      \end{tikzpicture}
     }
        \caption{Damage evolution for the WST on specimen $\#$3 for the load steps (a) \#8 and (b) \#9.
        The experimentally resolved crack region is rendered in black, while the numerical counterpart, obtained with \texttt{vol/dev} split,
        is rendered in red.}
        \label{fig:pf_evolution_spec3}
\end{figure}

We now compare the damage evolution through the unstable softening phase (steps \#8 and \#9 in Fig.~\ref{fig:pf_dvc3_displ}). The results are presented in Fig.~\ref{fig:pf_evolution_spec3} by showing the damage and crack patterns as observed from the top of the specimen. In Fig.~\ref{fig:spcm3_7} we can observe that at load step \#8 the experimental crack involves almost the whole width of the T-slot arm, however, the crack width is so limited that it cannot be represented numerically. As expected, the numerical predictions show a sudden damage evolution passing from the pre-peak phase, where damage is essentially zero and hence not shown,  to the post-peak stage (Fig.~\ref{fig:spcm3_8}) where damage evolves to $\alpha\geq0.9$ for a large width of the T-slot arm, consistent with the widening of the experimental crack.. This is consistent with the unstable propagation of the experimental crack, which coincides with a load drop of about 75\%. 

\section{Conclusions}\label{S:pf_conclusions}
In this contribution, we propose a model and a computational workflow, within the phase-field approach to fracture, to predict fracture of concrete. To this end, we account for the real concrete mesostructure, constituted by pores, a
homogenized mortar matrix, aggregates and an interfacial transition zone between the two. Since the mortar is itself a heterogeneous material with a composition similar to concrete, a cutoff size is selected below which aggregates and pores are embedded into
a homogenized matrix and above which they are explicitly resolved. The quasi-brittle nature of the phases is accounted for by adopting the phase-field formulation proposed by Lorentz \cite{Lorentz:2011}. The model  is calibrated by means of a set of physical and numerical experiments through which we characterize the elastic and fracture properties of the phases accounting for the aforementioned cutoff length. 
To validate the model, we perform a set of wedge splitting tests inside an X-ray tomograph (in-situ testing). The tomograms are analyzed using digital volume correlation, delivering the 3D evolution of both the displacement vector field and the crack pattern. In addition, the tomograms provide the real 3D mesoscale geometry of the specimens. They are meshed to obtain a realistic finite element discretization of the specimens, which is used to numerically reproduce the  wedge splitting tests driven by the real boundary conditions extracted from the measured displacements.
Numerical and experimental results yield the following main conclusions:
\begin{itemize}

\item[-] The real mesostructure and the real boundary conditions play a crucial role in determining the fracture response of the specimens. In particular, different aggregate distributions, along with the unavoidable imperfections in the loading conditions, may lead to significantly different failure mechanisms in nominally identical tests.

\item[-] The phase-field fracture model is able to correctly predict the experimental results, provided that a proper parameter calibration is performed and that sufficient information about the geometry and boundary conditions is available, which is enabled by in-situ testing and respective DVC analyses.

\item[-] The comparison between numerical and experimental results reveals a very good agreement in terms of local and global quantities, including the evolution, during the test, of the crack pattern and of the crack opening displacement as well as the 3D displacement fields.

\item[-] When dealing with highly heterogeneous materials such as concrete, the adoption of a sub-volume of the complete domain and the application of the experimental displacements to its outer surfaces greatly improves the accuracy of the numerical predictions. This comes at the cost of being unable to compute the reaction forces, an issue that remains open and will be addressed in future work. 

\end{itemize}

With the progress and gradual popularization of laboratory-scale X-ray computed tomography, digital volume correlation and high-performance computing, we believe that the tight synergy of modeling, computation, imaging and testing pursued in this work holds significant promise to advance the knowledge of the failure behavior of concrete and many other complex materials. 

\appendix

 \section{Optimization of X-CT settings}\label{app:pf_tuning_CT}
Before performing the in-situ WSTs, the X-CT settings are fine-tuned to find the best compromise between image quality and tomographic acquisition time. To this end, we analyze the quality of a set of tomograms of the studied concrete type, using different settings. The latter are varied while trying to maximize the tomographic signal-to-noise ratio (SNR) while keeping the acquisition time as low as possible. Note that the SNR is computed within (nominally homogeneous) portions of the aggregate phase. 
     
The five settings leading to the best results in terms of SNR are summarized in Tab.~\ref{tab:pf_CT_parameters}.
For each of them, a representative 2D cross-section (slice) of the reconstructed volume is shown in Fig.~\ref{fig:pf_WST_tests}.
The latter shows that, although the settings vary quite broadly (e.g., the X-ray source power ranges from 37.5 W to 90 W), all the images appear similar at visual inspection. However, the acquisition time ranges from 25 to 61 minutes with a SNR of about 30 dB. Prioritizing the reduction of time over a marginal improvement of the SNR, we adopt the setting highlighted in red in Tab.~\ref{tab:pf_CT_parameters}, corresponding to Fig.~\ref{fig:pf_setting2}.   

\begin{table}[htb]
    \footnotesize
    \renewcommand{\arraystretch}{1.2}
    \begin{adjustwidth}{-4cm}{-4cm}
        \centering
        \begin{tabular}{c c c c c c}
            \hline
            \multicolumn{6}{c}{ \textbf{X-CT Parameters}}\\
            \hline
            & Setting 1 & Setting 2 & Setting 3 & Setting 4  & Setting 5 (adopted)\\
            \hline
            Mode & \multicolumn{5}{c}{step-by-step rotation and imaging}\\
            {Detector mode} & No binning & No binning &No binning& No binning  &2$\times$2 binning  \\
            Detector pixel Nr. &{2564$\times$2024}  & {2564$\times$2024} & {2564$\times$2024} & {2564$\times$2024}  &  1268$\times$1012\\
            Voxel size  ($\mu$m) &~$25$ &~$25$ &~$25$ &~$25$   & $~$50\\
            Frame rate ($s^{-1}$) & 5 & 6 &6  &5  &10 \\
            Frame average &10 & 10  &7 &10  & 10 \\
            Source voltage (kV) & 150 & 150 & 150 &140   & 140   \\
            Source current ($\mu$A( &250 &450  & 600 &500   & 500  \\
            Source power (W) & 37.5 & 67.5 & 90  & 70 & 70  \\
            Radiographs \textit{per} tomogram & 1504 & 1504 & 1504 & 1206   & 1206\\
            Tomographic signal-to-noise ratio, SNR ($dB$) & 36 & 31 &30 & 32   & 28 \\
            Tomographic acquisition time (minutes) &61 & 51  &38  & 49  & 25 \\
            \hline
        \end{tabular}
    \end{adjustwidth}
    \begin{tikzpicture}[overlay]
		\draw[very thick,red](13.3,-0.1) -- (15.6,-0.1) -- (15.6,5.5) -- (13.3,5.5) -- cycle;
    \end{tikzpicture}
    \caption{X-CT parameters corresponding to different settings for in-situ WSTs. The \textit{no binning} detector mode
    means no downsampling of each acquired radiographic image (frame). The \textit{frame rate} indicates how many frames are acquired every
    second. The \textit{frame average} indicates how many frames are averaged to store on file a single radiograph.}
    \label{tab:pf_CT_parameters}
\end{table}
    
\begin{figure*}[!htb]
  \begin{minipage}{\textwidth}%
       \centering
       \subfloat[]{
                \includegraphics[width=0.22\textwidth]{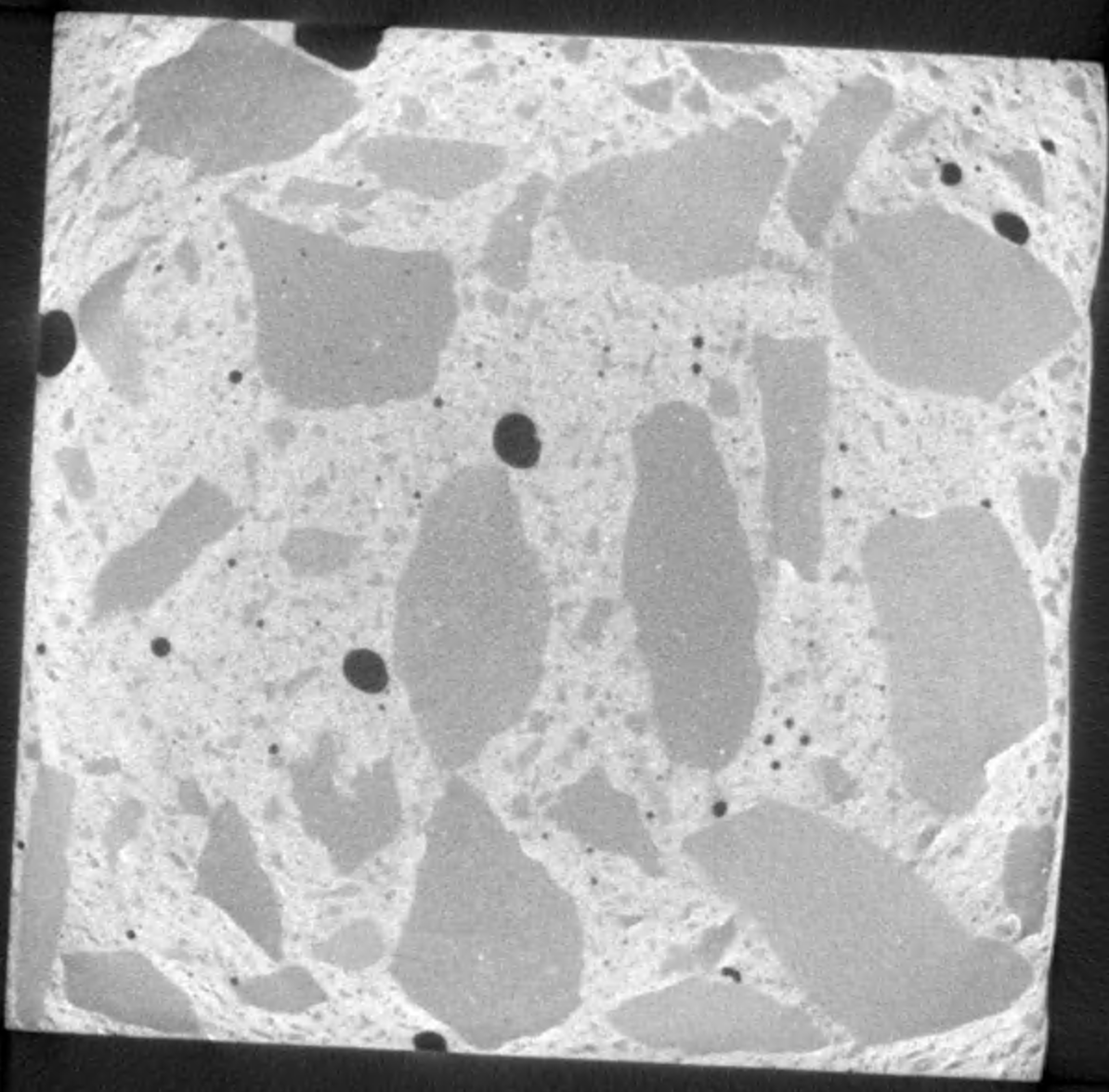}
               \label{fig:pf_setting1}}
               \hspace{0.5cm}
                    \subfloat[]{
                \includegraphics[width=0.22\textwidth]{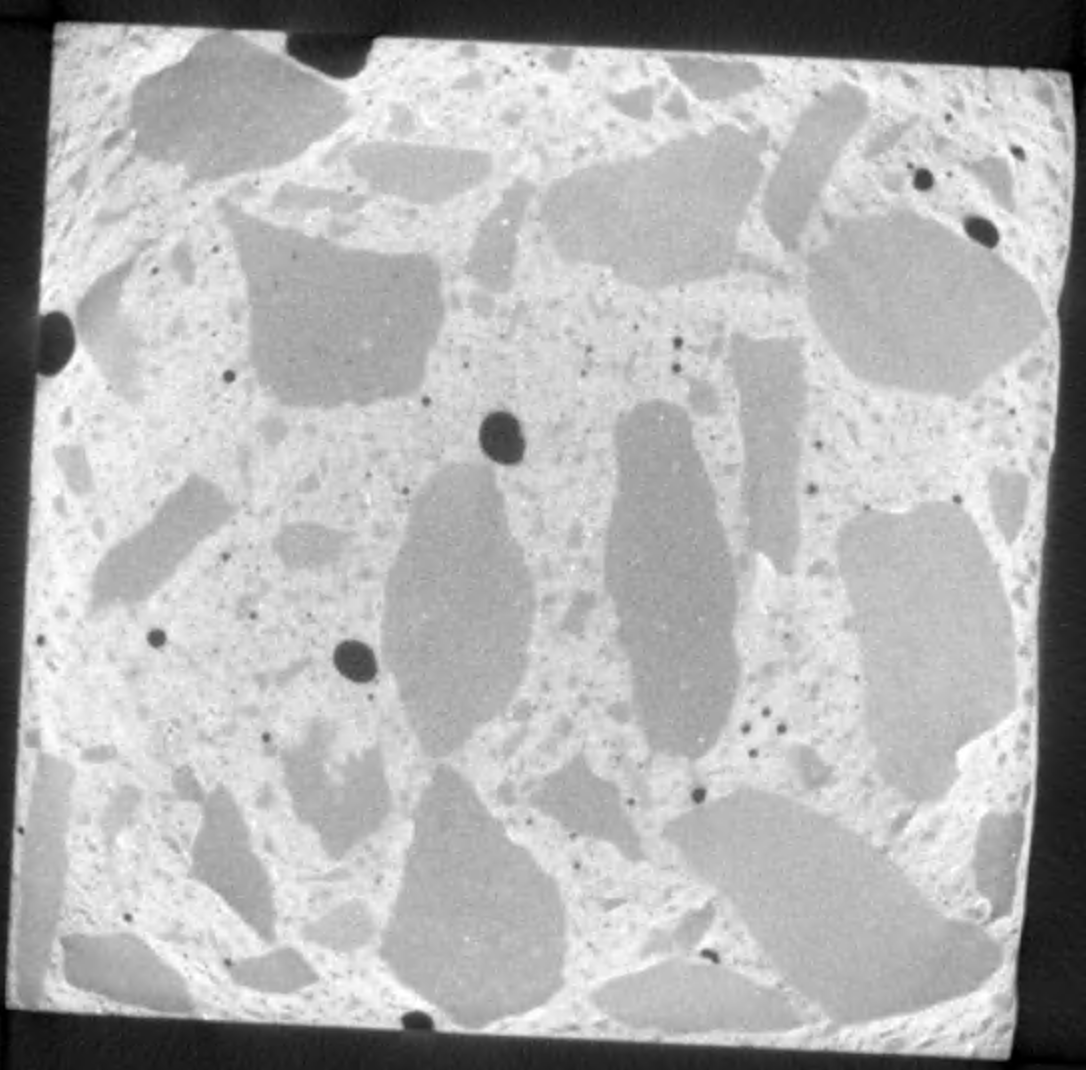}
               \label{fig:pf_setting2}}
               \hspace{0.5cm}
                \subfloat[]{
                \includegraphics[width=0.22\textwidth]{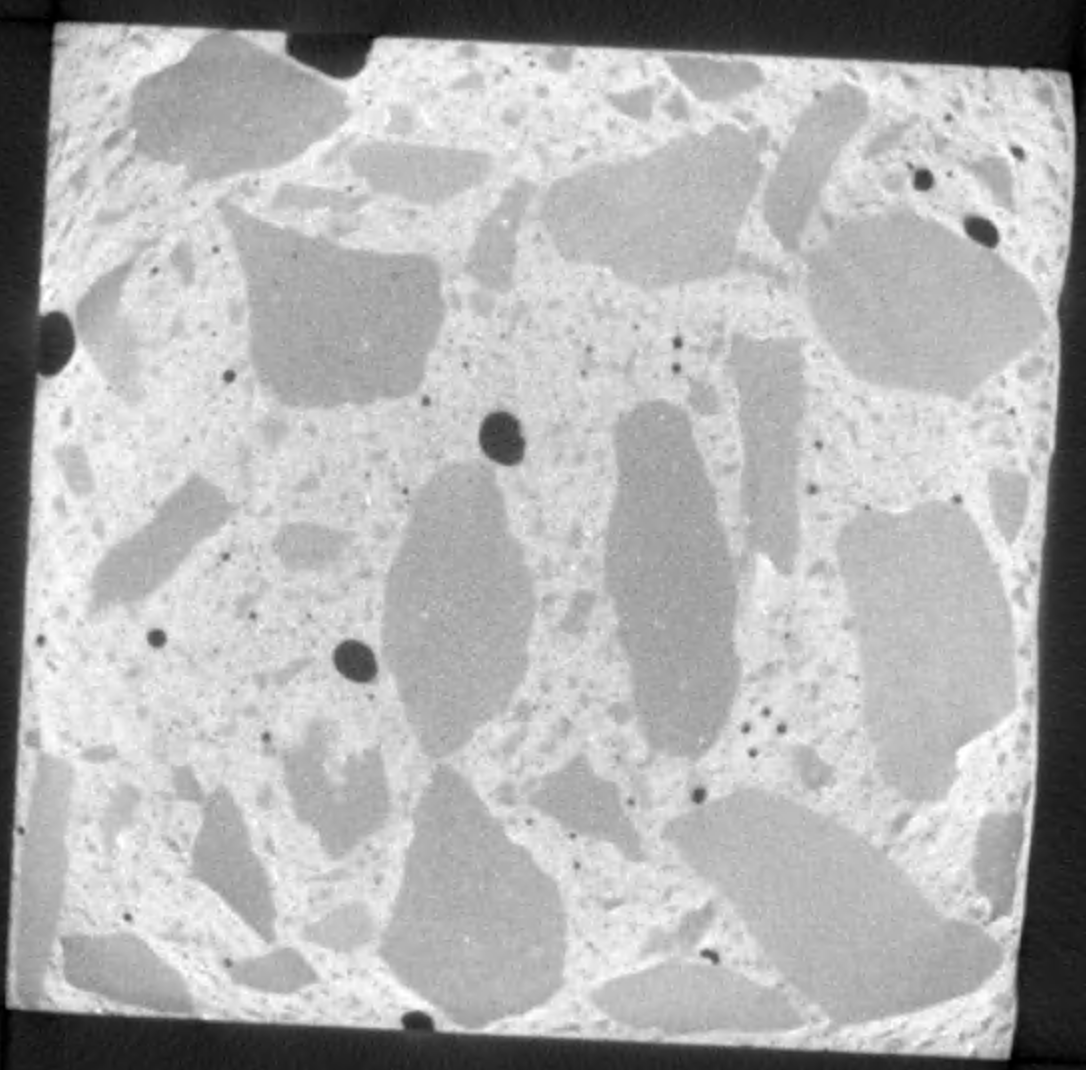}
               \label{fig:pf_setting2}}
                \end{minipage}
                  \begin{minipage}{\textwidth}%
       \centering
    
                    \subfloat[]{
                \includegraphics[width=0.22\textwidth]{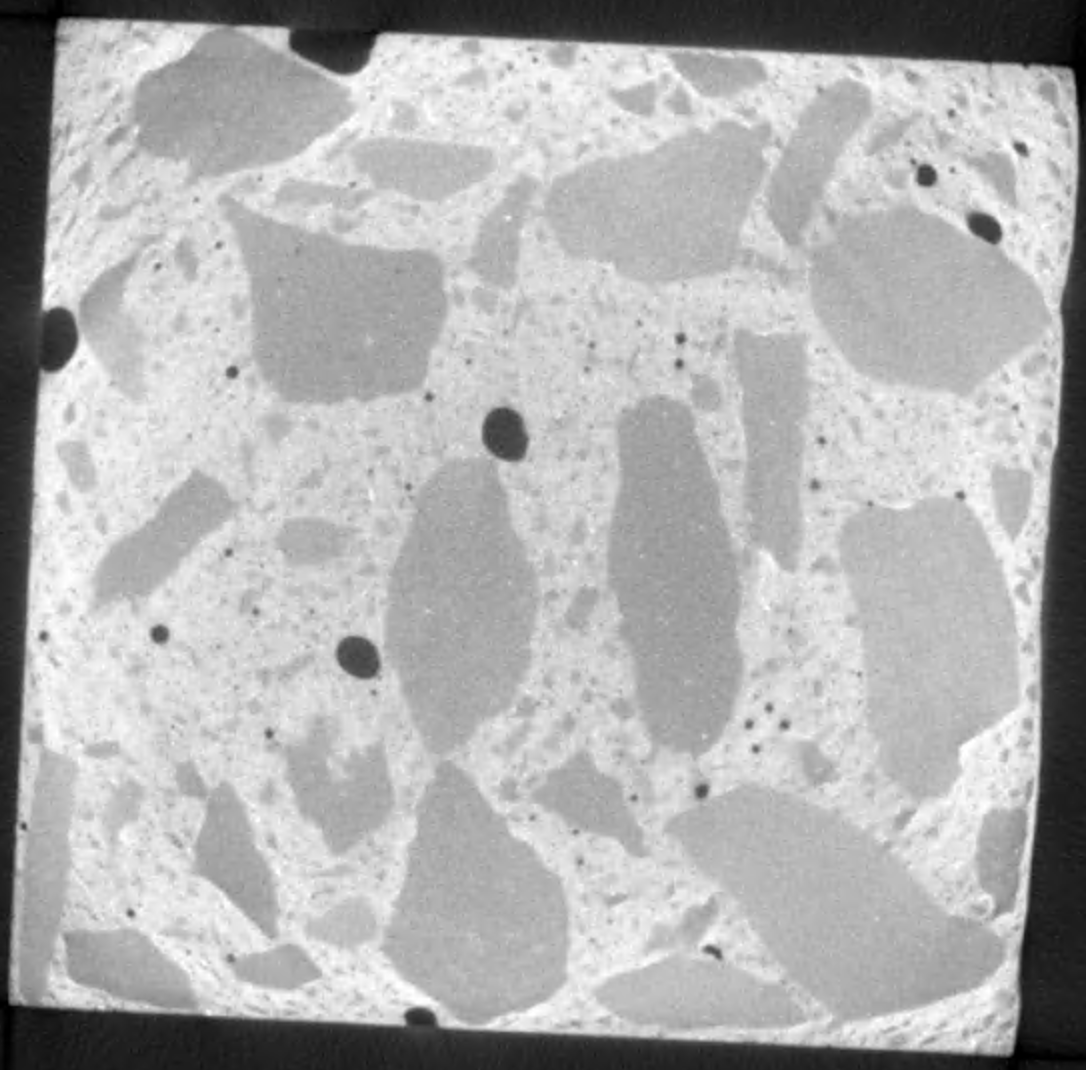}
               \label{fig:pf_setting2}}
               \hspace{0.4cm}
                \subfloat[]{
                \adjustbox{width=.22\linewidth,cfbox=red 1pt 3pt }{\includegraphics[width=0.22\textwidth]{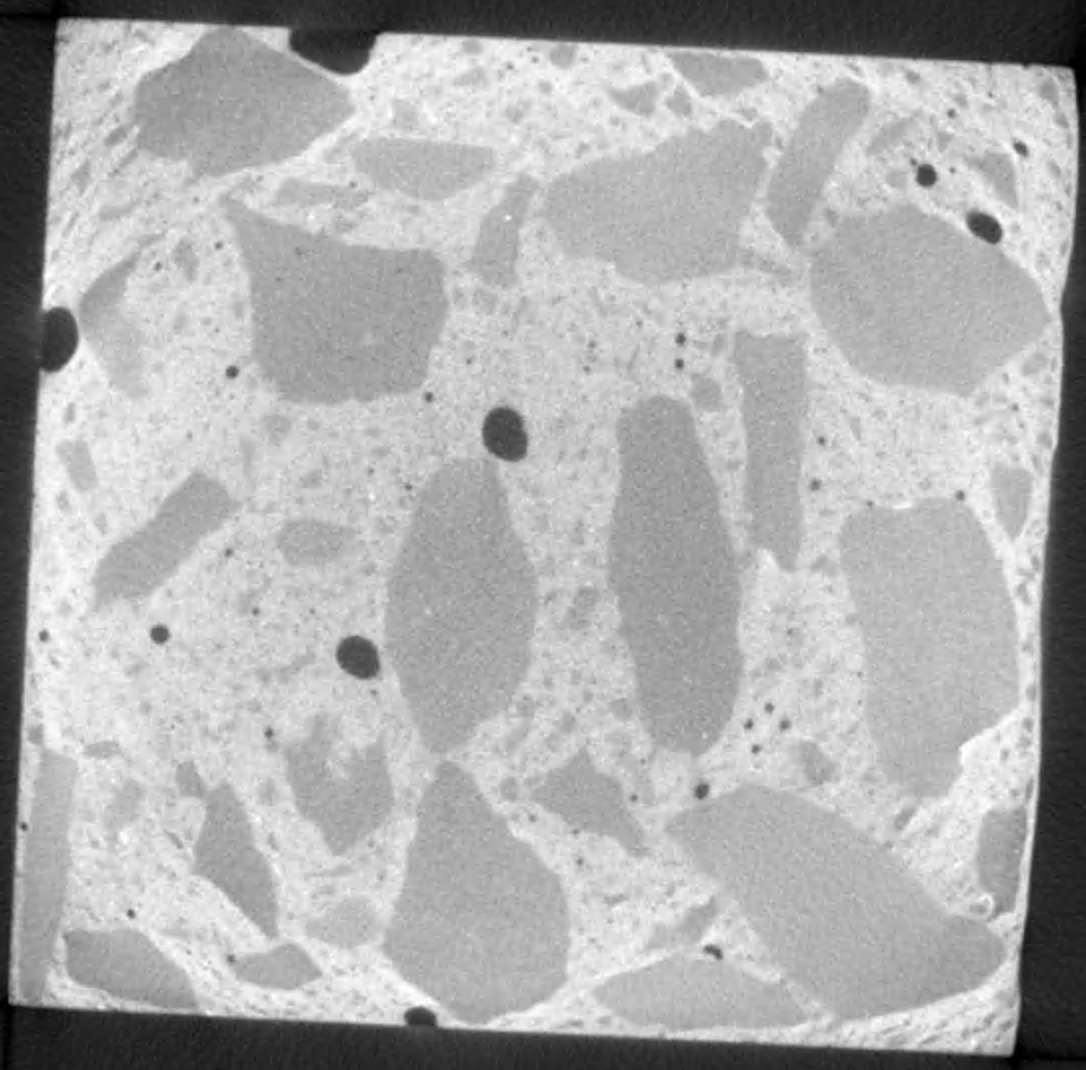}}
               \label{fig:pf_setting2}}
                \end{minipage}
                     \caption{X-CT measuring quality for (a) setting 1, (b) setting 2, (c) setting 3, (d) setting 4 and (e) setting 5 (adopted), defined in Tab.~\ref{tab:pf_CT_parameters}. The quality is assessed by comparing the SNR and by visual inspection of 2D cross-sections
                     (slices) from the correspondingly reconstructed 3D volumes (tomograms) of the same concrete specimen.
                     For each setting we show one slice at the same spatial position. The specimen cross-section has a lateral
                     size of 40 mm.}
\label{fig:pf_ctsetting_tests}
\end{figure*}

\section{Augmented reality (AR) renderings}\label{app:AR_render}

\begin{figure}[!hbt]
    \centering
        \includegraphics[width=4cm]{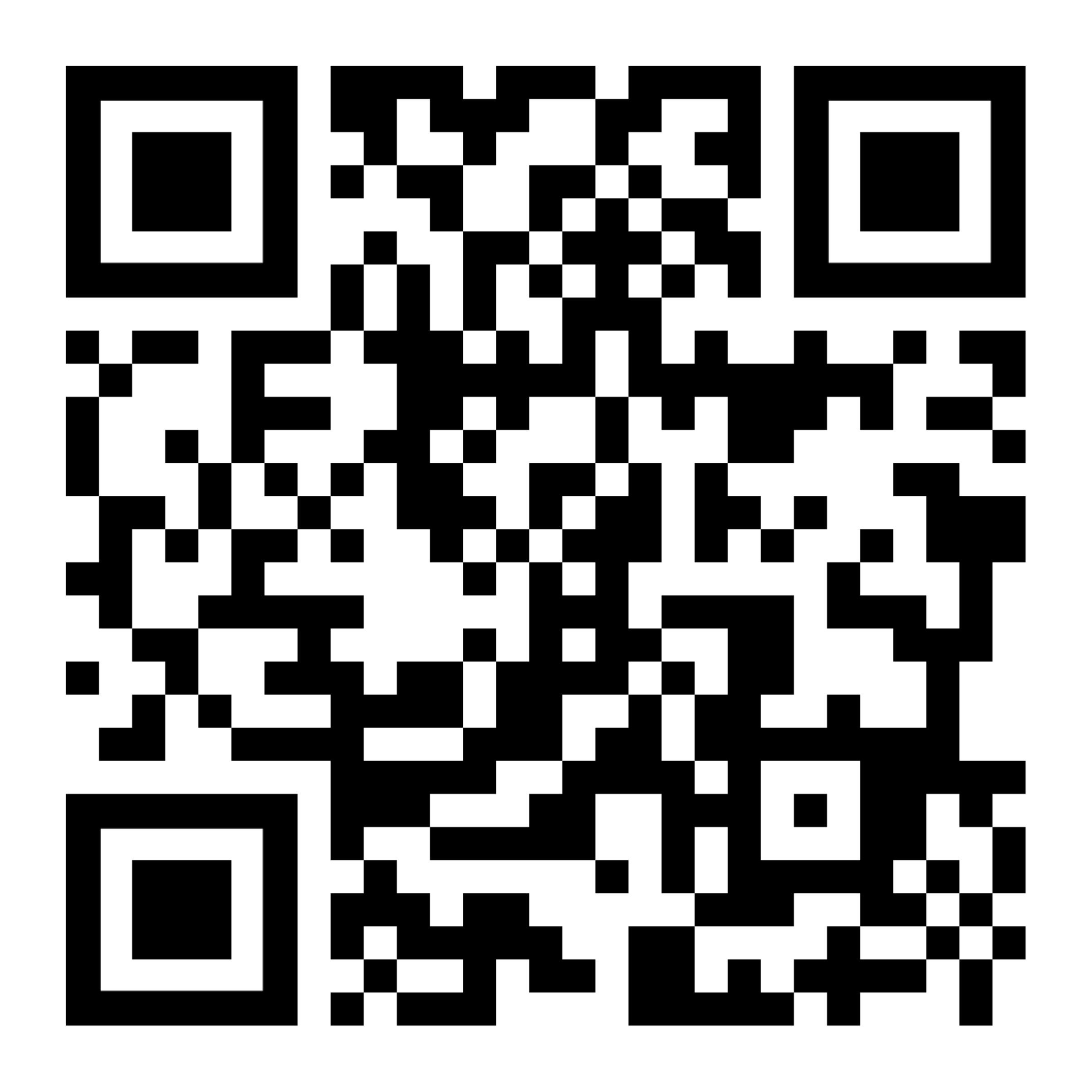}
    \caption{QR code for accessing the AR renderings of the 3D models.} \label{fig:pf_ar_qr}
\end{figure}

The AR 3D renderings are created following the procedure presented in \cite{mathur_2023_brief}, and can be viewed on/by any recent mobile phone or tablet or Internet browser.
The renderings are accessible at \url{https://ar.compmech.ethz.ch} or using the QR-code in Fig.~\ref{fig:pf_ar_qr}.
All scripts used for an automatic model and website generation can be found at \url{https://gitlab.ethz.ch/compmech/ar}.

\section{Results using \texttt{spectral} split}
\label{sct:pf_spect_split}

In the following we compare the experimental final crack pattern with the numerical predictions using the \texttt{spectral} split.

\subsection{Specimen $\#$2} \label{s:spec_2}
\begin{figure}[!htb]
    \centering
    \subfloat[]{
    \includegraphics[width=0.365\textwidth]{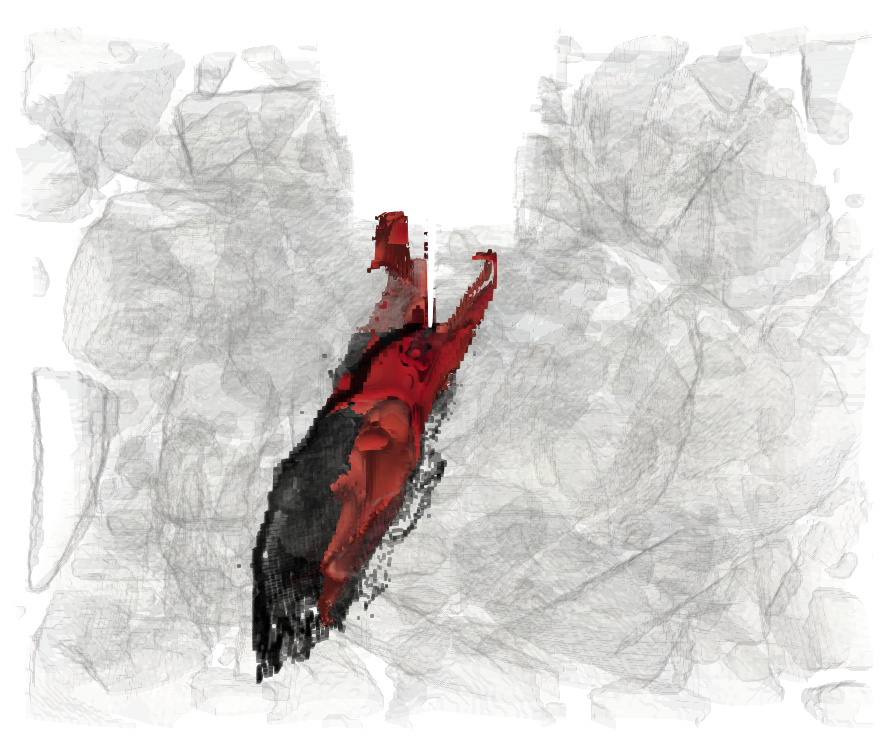}
}
    \hspace{10mm}
     \subfloat[]{
\includegraphics[width=0.365\textwidth]{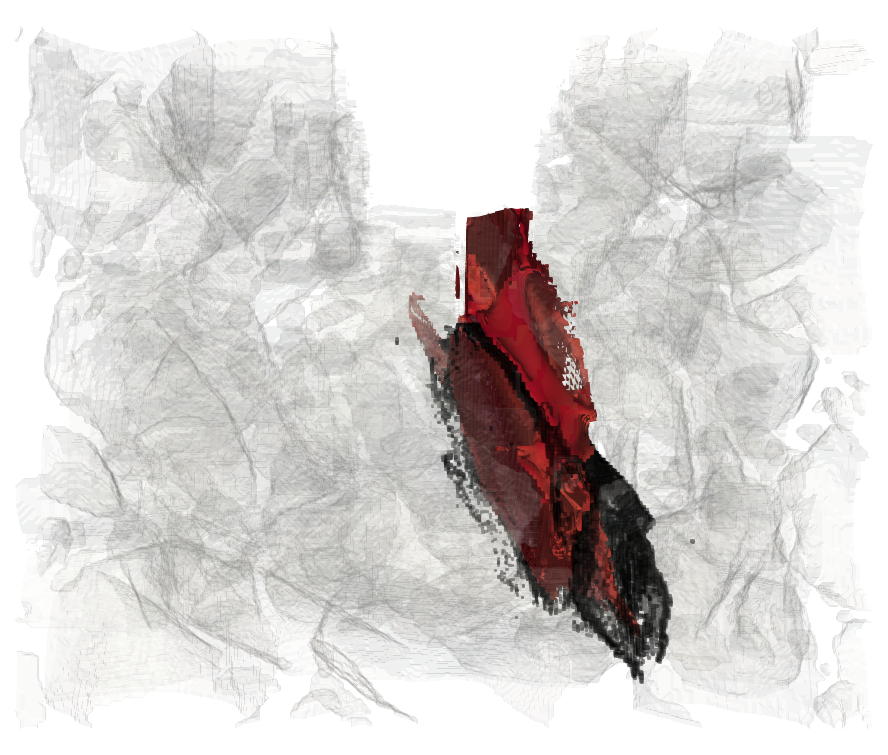}
}
    \caption{Comparison of the numerical and experimental crack patterns at the end of  WST on specimen $\#$2 modeled using the \texttt{spectral} split: (a) front view and (b) back view.}
    \label{fig:pf_spec2_miehe}
\end{figure}

Fig.~\ref{fig:pf_spec2_miehe} shows the comparison between experimental and numerical results obtained using the \texttt{spectral} split for specimen $\#$2 at the final load step.  
The obtained results support the observations outlined in Sect.~\ref{sct:pf_test1}. In particular, the crack obtained with the \texttt{spectral} split tends to impinge on the aggregates leading to a smoother pattern. These results can be better appreciated using the AR renders, which can be viewed using the QR code provided in \ref{app:AR_render}.

\subsection{Specimen $\#$3} \label{s:spec_3}

\begin{figure}[!htb]
    \centering
    \subfloat[]{
\includegraphics[width=0.36\textwidth]{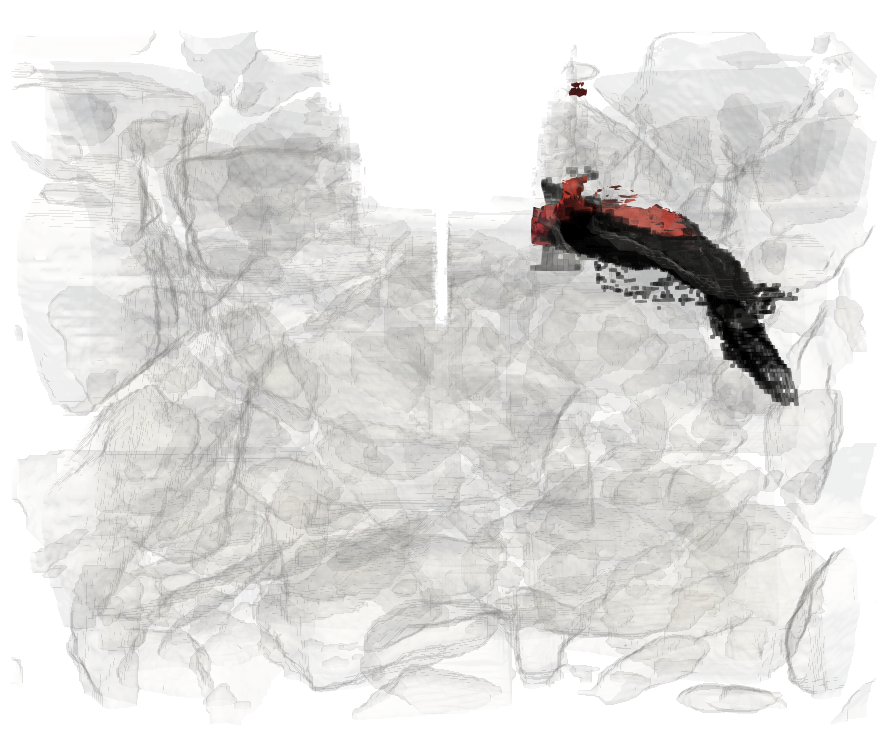}
}
    \hspace{10mm}
     \subfloat[]{
\includegraphics[width=0.36\textwidth]{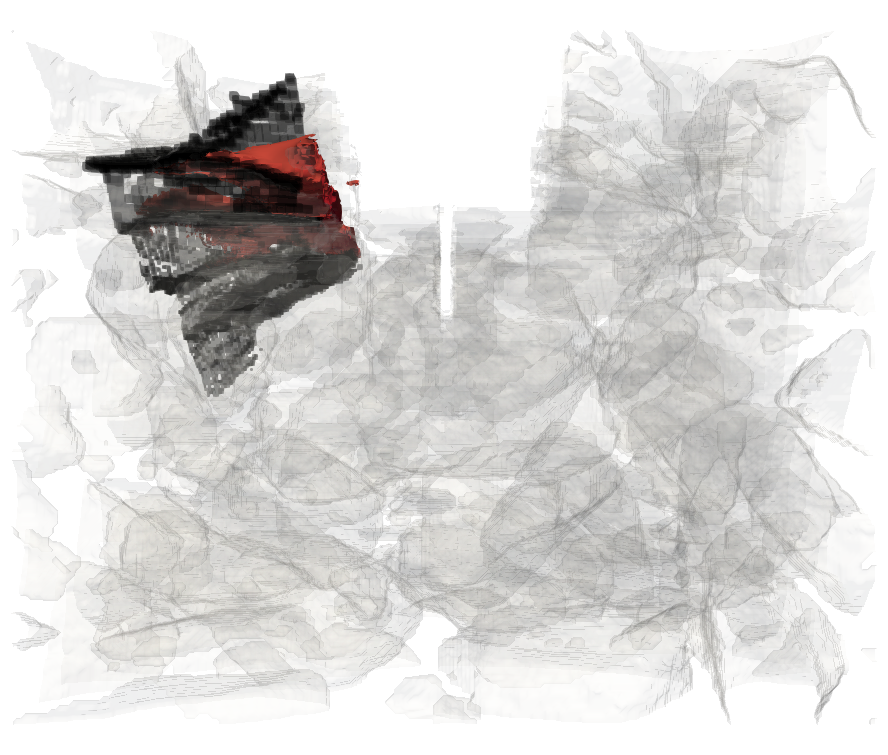}
}
      \caption{Comparison of the numerical and experimental crack patterns at the end of  WST on specimen $\#$3 modeled using the \texttt{spectral} split: (a) front view and (b) back view.}
    \label{fig:pf_spec3_miehe}
\end{figure}

Fig.~\ref{fig:pf_spec3_miehe} presents the comparison between experimental and numerical results obtained using the \texttt{spectral} split for specimen $\#$3 at the final load step.  Unlike for specimens $\#$1 and $\#2$, the predicted crack path is similar to that obtained from the \texttt{vol/dev} split (Fig.~\ref{fig:pf_spec2}) due to the predominantly mode-I loading conditions.

  \section*{Acknowledgements}
  We thank Veit Schönherr (ETH Zurich) for his invaluable support in designing and conducting the experiments and Marcel Käppeli (Empa)
  for his help by the elastic properties measurements.
  We also thank  Dr. Kamel Madi (3Dmagination Ltd.) for his assistance with the DVC analyses, RX Solutions for their support with the 3D reconstructions of tomograms and Deben UK for providing us with CAD images of the Deben open frame stage, which have been used in this article. AM, PC and LDL acknowledge funding from the Swiss National Science Foundation
through Grant No. 200021-219407 ‘Phase-field modeling of fracture and fatigue:
from rigorous theory to fast predictive simulations’.
 
\bibliographystyle{elsarticle-num-names}
\bibliography{ms.bib}

\begin{thebibliography}{131}
\expandafter\ifx\csname natexlab\endcsname\relax\def\natexlab#1{#1}\fi
\providecommand{\url}[1]{\texttt{#1}}
\providecommand{\href}[2]{#2}
\providecommand{\path}[1]{#1}
\providecommand{\DOIprefix}{doi:}
\providecommand{\ArXivprefix}{arXiv:}
\providecommand{\URLprefix}{URL: }
\providecommand{\Pubmedprefix}{pmid:}
\providecommand{\doi}[1]{\href{http://dx.doi.org/#1}{\path{#1}}}
\providecommand{\Pubmed}[1]{\href{pmid:#1}{\path{#1}}}
\providecommand{\bibinfo}[2]{#2}
\ifx\xfnm\relax \def\xfnm[#1]{\unskip,\space#1}\fi
\bibitem[{Carrara et~al.(2016)Carrara, Wu, Kruse, and
  De~Lorenzis}]{Pietro:2016}
\bibinfo{author}{P.~Carrara}, \bibinfo{author}{T.~Wu},
  \bibinfo{author}{R.~Kruse}, \bibinfo{author}{L.~De~Lorenzis},
\newblock \bibinfo{title}{Towards multiscale modeling of the interaction
  between transport and fracture in concrete},
\newblock \bibinfo{journal}{RILEM Technical Letters} \bibinfo{volume}{1}
  (\bibinfo{year}{2016}) \bibinfo{pages}{94}.
\bibitem[{Carrara et~al.(2018)Carrara, Kruse, Bentz, Lunardelli, Leusmann,
  Varady, and {De Lorenzis}}]{Pietro:2018}
\bibinfo{author}{P.~Carrara}, \bibinfo{author}{R.~Kruse},
  \bibinfo{author}{D.~Bentz}, \bibinfo{author}{M.~Lunardelli},
  \bibinfo{author}{T.~Leusmann}, \bibinfo{author}{P.~Varady},
  \bibinfo{author}{L.~{De Lorenzis}},
\newblock \bibinfo{title}{Improved mesoscale segmentation of concrete from 3d
  x-ray images using contrast enhancers},
\newblock \bibinfo{journal}{Cement and Concrete Composites}
  \bibinfo{volume}{93} (\bibinfo{year}{2018}) \bibinfo{pages}{30--42}.
\bibitem[{Yang et~al.(2017)Yang, Ren, Sharma, McDonald, Mostafavi, Vertyagina,
  and Marrow}]{Yang:2017}
\bibinfo{author}{Z.~Yang}, \bibinfo{author}{W.~Ren},
  \bibinfo{author}{R.~Sharma}, \bibinfo{author}{S.~McDonald},
  \bibinfo{author}{M.~Mostafavi}, \bibinfo{author}{Y.~Vertyagina},
  \bibinfo{author}{T.~Marrow},
\newblock \bibinfo{title}{In-situ x-ray computed tomography characterisation of
  3d fracture evolution and image-based numerical homogenisation of concrete},
\newblock \bibinfo{journal}{Cement and Concrete Composites}
  \bibinfo{volume}{75} (\bibinfo{year}{2017}) \bibinfo{pages}{74--83}.
\bibitem[{Thilakarathna et~al.(2020)Thilakarathna, {Kristombu Baduge}, Mendis,
  Vimonsatit, and Lee}]{Thilakarth:2020}
\bibinfo{author}{P.~Thilakarathna}, \bibinfo{author}{K.~{Kristombu Baduge}},
  \bibinfo{author}{P.~Mendis}, \bibinfo{author}{V.~Vimonsatit},
  \bibinfo{author}{H.~Lee},
\newblock \bibinfo{title}{Mesoscale modelling of concrete – a review of
  geometry generation, placing algorithms, constitutive relations and
  applications},
\newblock \bibinfo{journal}{Engineering Fracture Mechanics}
  \bibinfo{volume}{231} (\bibinfo{year}{2020}) \bibinfo{pages}{106974}.
\bibitem[{Jailin et~al.(2017)Jailin, Carpiuc, Kazymyrenko, Poncelet, Leclerc,
  Hild, and Roux}]{Jailin:2017}
\bibinfo{author}{C.~Jailin}, \bibinfo{author}{A.~Carpiuc},
  \bibinfo{author}{K.~Kazymyrenko}, \bibinfo{author}{M.~Poncelet},
  \bibinfo{author}{H.~Leclerc}, \bibinfo{author}{F.~Hild},
  \bibinfo{author}{S.~Roux},
\newblock \bibinfo{title}{Virtual hybrid test control of sinuous crack},
\newblock \bibinfo{journal}{Journal of the Mechanics and Physics of Solids}
  \bibinfo{volume}{102} (\bibinfo{year}{2017}) \bibinfo{pages}{239--256}.
\bibitem[{Carpiuc(2015)}]{Carpiuc:2015}
\bibinfo{author}{A.~Carpiuc}, \bibinfo{title}{Innovative tests for
  characterizing mixed-mode fracture of concrete: from pre-defined to
  interactive and hybrid tests}, Ph.D. thesis, Universit{\'e} Paris Saclay
  (COmUE), \bibinfo{year}{2015}.
\bibitem[{Grédiac and Hild(2012)}]{MichelG:2012}
\bibinfo{author}{M.~Grédiac}, \bibinfo{author}{F.~Hild},
  \bibinfo{title}{Full-Field Measurements and Identification in Solid
  Mechanics}, \bibinfo{year}{2012}.
\bibitem[{Malvar and Warren(1988)}]{Malvar}
\bibinfo{author}{L.~J. Malvar}, \bibinfo{author}{G.~E. Warren},
\newblock \bibinfo{title}{Fracture energy for three-point-bend tests on
  single-edge-notched beams},
\newblock \bibinfo{journal}{Experimental Mechanics} \bibinfo{volume}{28}
  (\bibinfo{year}{1988}) \bibinfo{pages}{266--272}.
\bibitem[{Carneiro(1943)}]{Carneiro:1943}
\bibinfo{author}{F.~Carneiro},
\newblock \bibinfo{title}{A new method to determine the tensile strength of
  concrete},
\newblock in: \bibinfo{booktitle}{Proceedings of the 5th meeting of the
  Brazilian Association for Technical Rules}, volume~\bibinfo{volume}{3},
  \bibinfo{year}{1943}, pp. \bibinfo{pages}{126--129}.
\bibitem[{Winkler(2001)}]{Winkler}
\bibinfo{author}{B.~J. Winkler}, \bibinfo{title}{Traglastuntersuchungen von
  unbewehrten und bewehrten Betonstrukturen auf der Grundlage eines objektiven
  Werkstoffgesetzes f{\"u}r Beton}, \bibinfo{publisher}{Innsbruck University
  Press}, \bibinfo{year}{2001}.
\bibitem[{Nooru-Mohamed et~al.(1993)Nooru-Mohamed, Schlangen, and {van
  Mier}}]{Nooru:1993}
\bibinfo{author}{M.~Nooru-Mohamed}, \bibinfo{author}{E.~Schlangen},
  \bibinfo{author}{J.~G. {van Mier}},
\newblock \bibinfo{title}{Experimental and numerical study on the behavior of
  concrete subjected to biaxial tension and shear},
\newblock \bibinfo{journal}{Advanced Cement Based Materials}
  \bibinfo{volume}{1} (\bibinfo{year}{1993}) \bibinfo{pages}{22--37}.
\bibitem[{Hoover et~al.(2013)Hoover, {P. Bažant}, Vorel, Wendner, and
  Hubler}]{Hoover:2013}
\bibinfo{author}{C.~G. Hoover}, \bibinfo{author}{Z.~{P. Bažant}},
  \bibinfo{author}{J.~Vorel}, \bibinfo{author}{R.~Wendner},
  \bibinfo{author}{M.~H. Hubler},
\newblock \bibinfo{title}{Comprehensive concrete fracture tests: Description
  and results},
\newblock \bibinfo{journal}{Engineering Fracture Mechanics}
  \bibinfo{volume}{114} (\bibinfo{year}{2013}) \bibinfo{pages}{92--103}.
\bibitem[{Sutton et~al.(1983)Sutton, Wolters, Peters, Ranson, and
  McNeill}]{Sutton:1983}
\bibinfo{author}{M.~Sutton}, \bibinfo{author}{W.~Wolters},
  \bibinfo{author}{W.~Peters}, \bibinfo{author}{W.~Ranson},
  \bibinfo{author}{S.~McNeill},
\newblock \bibinfo{title}{Determination of displacements using an improved
  digital correlation method},
\newblock \bibinfo{journal}{Image and Vision Computing} \bibinfo{volume}{1}
  (\bibinfo{year}{1983}) \bibinfo{pages}{133--139}.
\bibitem[{Pan et~al.(2009)Pan, Qian, Xie, and Asundi}]{Pan2009}
\bibinfo{author}{B.~Pan}, \bibinfo{author}{K.~Qian}, \bibinfo{author}{H.~Xie},
  \bibinfo{author}{A.~Asundi},
\newblock \bibinfo{title}{Two-dimensional digital image correlation for
  in-plane displacement and strain measurement: A review},
\newblock \bibinfo{journal}{Measurement Science and Technology}
  \bibinfo{volume}{20} (\bibinfo{year}{2009}).
\bibitem[{Hild and Roux(2006)}]{Hild:2006}
\bibinfo{author}{F.~Hild}, \bibinfo{author}{S.~Roux},
\newblock \bibinfo{title}{Digital image correlation: from displacement
  measurement to identification of elastic properties – a review},
\newblock \bibinfo{journal}{Strain} \bibinfo{volume}{42} (\bibinfo{year}{2006})
  \bibinfo{pages}{69--80}.
\bibitem[{Wu et~al.(2017)Wu, Carpiuc-Prisacari, Poncelet, and {De
  Lorenzis}}]{Wu:2017}
\bibinfo{author}{T.~Wu}, \bibinfo{author}{A.~Carpiuc-Prisacari},
  \bibinfo{author}{M.~Poncelet}, \bibinfo{author}{L.~{De Lorenzis}},
\newblock \bibinfo{title}{Phase-field simulation of interactive mixed-mode
  fracture tests on cement mortar with full-field displacement boundary
  conditions},
\newblock \bibinfo{journal}{Engineering Fracture Mechanics}
  \bibinfo{volume}{182} (\bibinfo{year}{2017}) \bibinfo{pages}{658--688}.
\bibitem[{Bu et~al.(2020)Bu, Hu, Yang, and Liu}]{Bu:2020}
\bibinfo{author}{J.~Bu}, \bibinfo{author}{L.~Hu}, \bibinfo{author}{H.~Yang},
  \bibinfo{author}{S.~Liu},
\newblock \bibinfo{title}{Experimental study on crack propagation of concrete
  under various loading rates with digital image correlation method},
\newblock \bibinfo{journal}{International Journal of Concrete Structures and
  Materials} \bibinfo{volume}{14} (\bibinfo{year}{2020}) \bibinfo{pages}{25}.
\bibitem[{Carpiuc-Prisacari et~al.(2017)Carpiuc-Prisacari, Poncelet,
  Kazymyrenko, Leclerc, and Hild}]{Carpuic:2017}
\bibinfo{author}{A.~Carpiuc-Prisacari}, \bibinfo{author}{M.~Poncelet},
  \bibinfo{author}{K.~Kazymyrenko}, \bibinfo{author}{H.~Leclerc},
  \bibinfo{author}{F.~Hild},
\newblock \bibinfo{title}{A complex mixed-mode crack propagation test performed
  with a 6-axis testing machine and full-field measurements},
\newblock \bibinfo{journal}{Engineering Fracture Mechanics}
  \bibinfo{volume}{176} (\bibinfo{year}{2017}) \bibinfo{pages}{1--22}.
\bibitem[{Nguyen et~al.(2016)Nguyen, Yvonnet, Bornert, Chateau, Sab, Romani,
  and Le~Roy}]{Nguyen2016_1}
\bibinfo{author}{T.~Nguyen}, \bibinfo{author}{J.~Yvonnet},
  \bibinfo{author}{M.~Bornert}, \bibinfo{author}{C.~Chateau},
  \bibinfo{author}{K.~Sab}, \bibinfo{author}{R.~Romani},
  \bibinfo{author}{R.~Le~Roy},
\newblock \bibinfo{title}{{On the choice of parameters in the phase field
  method for simulating crack initiation with experimental validation}},
\newblock \bibinfo{journal}{{International Journal of Fracture}}
  \bibinfo{volume}{197} (\bibinfo{year}{2016}) \bibinfo{pages}{213--226}.
\bibitem[{Buljac et~al.(2018)Buljac, Jailin, Mendoza, Neggers,
  Taillandier-Thomas, Bouterf, Smaniotto, Hild, and Roux}]{Buljac2018}
\bibinfo{author}{A.~Buljac}, \bibinfo{author}{C.~Jailin},
  \bibinfo{author}{A.~Mendoza}, \bibinfo{author}{J.~Neggers},
  \bibinfo{author}{T.~Taillandier-Thomas}, \bibinfo{author}{A.~Bouterf},
  \bibinfo{author}{B.~Smaniotto}, \bibinfo{author}{F.~Hild},
  \bibinfo{author}{S.~Roux},
\newblock \bibinfo{title}{Digital volume correlation: Review of progress and
  challenges},
\newblock \bibinfo{journal}{Experimental Mechanics} \bibinfo{volume}{58}
  (\bibinfo{year}{2018}) \bibinfo{pages}{661--708}.
\bibitem[{Bay et~al.(1999)Bay, Smith, Fyhrie, and Saad}]{Bay1999}
\bibinfo{author}{B.~K. Bay}, \bibinfo{author}{T.~S. Smith},
  \bibinfo{author}{D.~P. Fyhrie}, \bibinfo{author}{M.~Saad},
  \bibinfo{title}{Digital volume correlation: three-dimensional strain mapping
  using x-ray tomography}, \bibinfo{year}{1999}.
\bibitem[{Roux et~al.(2008)Roux, Hild, Viot, and Bernard}]{Roux:2008}
\bibinfo{author}{S.~Roux}, \bibinfo{author}{F.~Hild},
  \bibinfo{author}{P.~Viot}, \bibinfo{author}{D.~Bernard},
\newblock \bibinfo{title}{Three-dimensional image correlation from x-ray
  computed tomography of solid foam},
\newblock \bibinfo{journal}{Composites Part A: Applied Science and
  Manufacturing} \bibinfo{volume}{39} (\bibinfo{year}{2008})
  \bibinfo{pages}{1253--1265}. \bibinfo{note}{Full-field Measurements in
  Composites Testing and Analysis}.
\bibitem[{Buljac et~al.(2017)Buljac, Shakoor, Neggers, Bernacki, Bouchard,
  Helfen, Morgeneyer, and Hild}]{Buljac:2017}
\bibinfo{author}{A.~Buljac}, \bibinfo{author}{M.~Shakoor},
  \bibinfo{author}{J.~Neggers}, \bibinfo{author}{M.~Bernacki},
  \bibinfo{author}{P.-O. Bouchard}, \bibinfo{author}{L.~Helfen},
  \bibinfo{author}{T.~Morgeneyer}, \bibinfo{author}{F.~Hild},
\newblock \bibinfo{title}{Numerical validation framework for micromechanical
  simulations based on synchrotron 3d imaging},
\newblock \bibinfo{journal}{Computational Mechanics} \bibinfo{volume}{59}
  (\bibinfo{year}{2017}) \bibinfo{pages}{419--441}.
\bibitem[{Madi et~al.(2013)Madi, Tozzi, Zhang, Tong, Cossey, Au, Hollis, and
  Hild}]{Madi:2013}
\bibinfo{author}{K.~Madi}, \bibinfo{author}{G.~Tozzi},
  \bibinfo{author}{Q.~Zhang}, \bibinfo{author}{J.~Tong},
  \bibinfo{author}{A.~Cossey}, \bibinfo{author}{A.~Au},
  \bibinfo{author}{D.~Hollis}, \bibinfo{author}{F.~Hild},
\newblock \bibinfo{title}{Computation of full-field displacements in a scaffold
  implant using digital volume correlation and finite element analysis},
\newblock \bibinfo{journal}{Medical Engineering \& Physics}
  \bibinfo{volume}{35} (\bibinfo{year}{2013}) \bibinfo{pages}{1298--1312}.
\bibitem[{Hai et~al.(2024)Hai, Zhang, Wriggers, jie Huang, ying Zhuang, and
  lang Xu}]{Hai:2024}
\bibinfo{author}{L.~Hai}, \bibinfo{author}{H.~Zhang},
  \bibinfo{author}{P.~Wriggers}, \bibinfo{author}{Y.~jie Huang},
  \bibinfo{author}{X.~ying Zhuang}, \bibinfo{author}{S.~lang Xu},
\newblock \bibinfo{title}{3d concrete fracture simulations using an explicit
  phase field model},
\newblock \bibinfo{journal}{International Journal of Mechanical Sciences}
  \bibinfo{volume}{265} (\bibinfo{year}{2024}) \bibinfo{pages}{108907}.
\bibitem[{Lorentz(2017)}]{Lorentz:2017}
\bibinfo{author}{E.~Lorentz},
\newblock \bibinfo{title}{A nonlocal damage model for plain concrete consistent
  with cohesive fracture},
\newblock \bibinfo{journal}{International Journal of Fracture}
  \bibinfo{volume}{207} (\bibinfo{year}{2017}).
\bibitem[{Bazant and Oh(1983)}]{Bazant:1983}
\bibinfo{author}{Z.~Bazant}, \bibinfo{author}{B.~Oh},
\newblock \bibinfo{title}{Crack band theory for fracture of concrete},
\newblock \bibinfo{journal}{Matériaux et Constructions} \bibinfo{volume}{16}
  (\bibinfo{year}{1983}) \bibinfo{pages}{155--177}.
\bibitem[{Bocca et~al.(1991)Bocca, Carpinteri, and Valente}]{Bocca:1991}
\bibinfo{author}{P.~Bocca}, \bibinfo{author}{A.~Carpinteri},
  \bibinfo{author}{S.~Valente},
\newblock \bibinfo{title}{Mixed mode fracture of concrete},
\newblock \bibinfo{journal}{International Journal of Solids and Structures}
  \bibinfo{volume}{27} (\bibinfo{year}{1991}) \bibinfo{pages}{1139--1153}.
\bibitem[{Comi and Perego(2001)}]{Comi:2001}
\bibinfo{author}{C.~Comi}, \bibinfo{author}{U.~Perego},
\newblock \bibinfo{title}{Fracture energy based bi-dissipative damage model for
  concrete},
\newblock \bibinfo{journal}{International Journal of Solids and Structures}
  \bibinfo{volume}{38} (\bibinfo{year}{2001}) \bibinfo{pages}{6427--6454}.
\bibitem[{Schröder et~al.(2022)Schröder, Pise, Brands, Gebuhr, and
  Anders}]{Schroeder:2022}
\bibinfo{author}{J.~Schröder}, \bibinfo{author}{M.~Pise},
  \bibinfo{author}{D.~Brands}, \bibinfo{author}{G.~Gebuhr},
  \bibinfo{author}{S.~Anders},
\newblock \bibinfo{title}{Phase-field modeling of fracture in high performance
  concrete during low-cycle fatigue: Numerical calibration and experimental
  validation},
\newblock \bibinfo{journal}{Computer Methods in Applied Mechanics and
  Engineering} \bibinfo{volume}{398} (\bibinfo{year}{2022})
  \bibinfo{pages}{115181}.
\bibitem[{Bazant et~al.(1987)Bazant, Pan, and Pijaudier-Cabot}]{Bazant:1987}
\bibinfo{author}{Z.~Bazant}, \bibinfo{author}{J.~Pan},
  \bibinfo{author}{G.~Pijaudier-Cabot},
\newblock \bibinfo{title}{Softening in reinforced concrete beams and frames},
\newblock \bibinfo{journal}{Journal of Structural Engineering-asce - J STRUCT
  ENG-ASCE} \bibinfo{volume}{113} (\bibinfo{year}{1987}).
\bibitem[{López et~al.(2007)López, Carol, and Aguado}]{Lopez:2007}
\bibinfo{author}{C.~López}, \bibinfo{author}{I.~Carol},
  \bibinfo{author}{A.~Aguado},
\newblock \bibinfo{title}{Meso-structural study of concrete fracture using
  interface elements. i: Numerical model and tensile behavior},
\newblock \bibinfo{journal}{Materials and Structures/Materiaux et
  Constructions} \bibinfo{volume}{41} (\bibinfo{year}{2007})
  \bibinfo{pages}{583--599}.
\bibitem[{Li et~al.(2021)Li, Yang, Li, and Wu}]{Li:2021}
\bibinfo{author}{H.~Li}, \bibinfo{author}{Z.~Yang}, \bibinfo{author}{B.-B. Li},
  \bibinfo{author}{J.-Y. Wu},
\newblock \bibinfo{title}{A phase-field regularized cohesive zone model for
  quasi-brittle materials with spatially varying fracture properties},
\newblock \bibinfo{journal}{Engineering Fracture Mechanics}
  \bibinfo{volume}{256} (\bibinfo{year}{2021}) \bibinfo{pages}{107977}.
\bibitem[{Su et~al.(2010)Su, Yang, and Liu}]{Su:2010}
\bibinfo{author}{X.~Su}, \bibinfo{author}{Z.~Yang}, \bibinfo{author}{G.~Liu},
\newblock \bibinfo{title}{Monte carlo simulation of complex cohesive fracture
  in random heterogeneous quasi-brittle materials: A 3d study},
\newblock \bibinfo{journal}{International Journal of Solids and Structures}
  \bibinfo{volume}{47} (\bibinfo{year}{2010}) \bibinfo{pages}{2336--2345}.
\bibitem[{Yang and {Frank Xu}(2008)}]{Yang:2008}
\bibinfo{author}{Z.~Yang}, \bibinfo{author}{X.~{Frank Xu}},
\newblock \bibinfo{title}{A heterogeneous cohesive model for quasi-brittle
  materials considering spatially varying random fracture properties},
\newblock \bibinfo{journal}{Computer Methods in Applied Mechanics and
  Engineering} \bibinfo{volume}{197} (\bibinfo{year}{2008})
  \bibinfo{pages}{4027--4039}.
\bibitem[{Baxter et~al.(2001)Baxter, Hossain, and Graham}]{Baxter:2001}
\bibinfo{author}{S.~Baxter}, \bibinfo{author}{M.~Hossain},
  \bibinfo{author}{L.~Graham},
\newblock \bibinfo{title}{Micromechanics based random material property fields
  for particulate reinforced composites},
\newblock \bibinfo{journal}{International Journal of Solids and Structures}
  \bibinfo{volume}{38} (\bibinfo{year}{2001}) \bibinfo{pages}{9209--9220}.
\bibitem[{Wriggers and Moftah(2006)}]{Wriggers:2006}
\bibinfo{author}{P.~Wriggers}, \bibinfo{author}{S.~Moftah},
\newblock \bibinfo{title}{Mesoscale models for concrete: Homogenisation and
  damage behaviour},
\newblock \bibinfo{journal}{Finite Elements in Analysis and Design}
  \bibinfo{volume}{42} (\bibinfo{year}{2006}) \bibinfo{pages}{623--636}.
  \bibinfo{note}{The Seventeenth Annual Robert J. Melosh Competition}.
\bibitem[{Zhang et~al.(2017)Zhang, Song, Liu, Wu, and Song}]{Zhang:2017}
\bibinfo{author}{Z.~Zhang}, \bibinfo{author}{X.~Song},
  \bibinfo{author}{Y.~Liu}, \bibinfo{author}{D.~Wu}, \bibinfo{author}{C.~Song},
\newblock \bibinfo{title}{Three-dimensional mesoscale modelling of concrete
  composites by using random walking algorithm},
\newblock \bibinfo{journal}{Composites Science and Technology}
  \bibinfo{volume}{149} (\bibinfo{year}{2017}) \bibinfo{pages}{235--245}.
\bibitem[{López et~al.(2007)López, Carol, and Aguado}]{Lopez:20071}
\bibinfo{author}{C.~López}, \bibinfo{author}{I.~Carol},
  \bibinfo{author}{A.~Aguado},
\newblock \bibinfo{title}{Meso-structural study of concrete fracture using
  interface elements. ii: Compression, biaxial and brazilian test},
\newblock \bibinfo{journal}{Materials and Structures/Materiaux et
  Constructions} \bibinfo{volume}{41} (\bibinfo{year}{2007})
  \bibinfo{pages}{601--620}.
\bibitem[{Withers et~al.(2021)Withers, Bouman, Carmignato, Cnudde, Grimaldi,
  Hagen, Maire, Manley, Du~Plessis, and Stock}]{Withers:2021}
\bibinfo{author}{P.~J. Withers}, \bibinfo{author}{C.~Bouman},
  \bibinfo{author}{S.~Carmignato}, \bibinfo{author}{V.~Cnudde},
  \bibinfo{author}{D.~Grimaldi}, \bibinfo{author}{C.~K. Hagen},
  \bibinfo{author}{E.~Maire}, \bibinfo{author}{M.~Manley},
  \bibinfo{author}{A.~Du~Plessis}, \bibinfo{author}{S.~R. Stock},
\newblock \bibinfo{title}{X-ray computed tomography},
\newblock \bibinfo{journal}{Nature Reviews Methods Primers} \bibinfo{volume}{1}
  (\bibinfo{year}{2021}) \bibinfo{pages}{18}.
\bibitem[{Nguyen et~al.(2015)Nguyen, Yvonnet, Zhu, Bornert, and
  Chateau}]{Nguyen2015_1}
\bibinfo{author}{T.~T. Nguyen}, \bibinfo{author}{J.~Yvonnet},
  \bibinfo{author}{Q.~Z. Zhu}, \bibinfo{author}{M.~Bornert},
  \bibinfo{author}{C.~Chateau},
\newblock \bibinfo{title}{A phase field method to simulate crack nucleation and
  propagation in strongly heterogeneous materials from direct imaging of their
  microstructure},
\newblock \bibinfo{journal}{Engineering Fracture Mechanics}
  \bibinfo{volume}{139} (\bibinfo{year}{2015}) \bibinfo{pages}{18--39}.
\bibitem[{Nguyen et~al.(2016)Nguyen, Yvonnet, Zhu, Bornert, and
  Chateau}]{Nguyen2016}
\bibinfo{author}{T.~T. Nguyen}, \bibinfo{author}{J.~Yvonnet},
  \bibinfo{author}{Q.~Z. Zhu}, \bibinfo{author}{M.~Bornert},
  \bibinfo{author}{C.~Chateau},
\newblock \bibinfo{title}{A phase-field method for computational modeling of
  interfacial damage interacting with crack propagation in realistic
  microstructures obtained by microtomography},
\newblock \bibinfo{journal}{Computer Methods in Applied Mechanics and
  Engineering} \bibinfo{volume}{312} (\bibinfo{year}{2016})
  \bibinfo{pages}{567--595}.
\bibitem[{Ren et~al.(2015)Ren, Yang, Sharma, Zhang, and Withers}]{ren2015two}
\bibinfo{author}{W.~Ren}, \bibinfo{author}{Z.~Yang},
  \bibinfo{author}{R.~Sharma}, \bibinfo{author}{C.~Zhang},
  \bibinfo{author}{P.~J. Withers},
\newblock \bibinfo{title}{Two-dimensional x-ray ct image based meso-scale
  fracture modelling of concrete},
\newblock \bibinfo{journal}{Engineering Fracture Mechanics}
  \bibinfo{volume}{133} (\bibinfo{year}{2015}) \bibinfo{pages}{24--39}.
\bibitem[{Yang et~al.(2019)Yang, Li, and Wu}]{Yang:2019}
\bibinfo{author}{Z.~Yang}, \bibinfo{author}{B.-B. Li}, \bibinfo{author}{J.-Y.
  Wu},
\newblock \bibinfo{title}{X-ray computed tomography images based phase-field
  modeling of mesoscopic failure in concrete},
\newblock \bibinfo{journal}{Engineering Fracture Mechanics}
  \bibinfo{volume}{208} (\bibinfo{year}{2019}).
\bibitem[{Huang et~al.(2015)Huang, Yang, Ren, Liu, and Zhang}]{Huang:2015}
\bibinfo{author}{Y.~Huang}, \bibinfo{author}{Z.~Yang},
  \bibinfo{author}{W.~Ren}, \bibinfo{author}{G.~Liu},
  \bibinfo{author}{C.~Zhang},
\newblock \bibinfo{title}{3d meso-scale fracture modelling and validation of
  concrete based on in-situ x-ray computed tomography images using damage
  plasticity model},
\newblock \bibinfo{journal}{International Journal of Solids and Structures}
  \bibinfo{volume}{67-68} (\bibinfo{year}{2015}) \bibinfo{pages}{340--352}.
\bibitem[{Ren et~al.(2018)Ren, Yang, Sharma, McDonald, and Mummery}]{Ren:2018}
\bibinfo{author}{W.~Ren}, \bibinfo{author}{Z.~Yang},
  \bibinfo{author}{R.~Sharma}, \bibinfo{author}{S.~McDonald},
  \bibinfo{author}{P.~Mummery},
\newblock \bibinfo{title}{Three-dimensional in situ xct characterisation and fe
  modelling of cracking in concrete},
\newblock \bibinfo{journal}{Complexity} \bibinfo{volume}{2018}
  (\bibinfo{year}{2018}) \bibinfo{pages}{1--11}.
\bibitem[{Nguyen et~al.(2016)Nguyen, Yvonnet, Bornert, and
  Chateau}]{Nguyen:2016a}
\bibinfo{author}{T.~Nguyen}, \bibinfo{author}{J.~Yvonnet},
  \bibinfo{author}{M.~Bornert}, \bibinfo{author}{C.~Chateau},
\newblock \bibinfo{title}{Initiation and propagation of complex 3d networks of
  cracks in heterogeneous quasi-brittle materials: Direct comparison between in
  situ testing-microct experiments and phase field simulations},
\newblock \bibinfo{journal}{Journal of the Mechanics and Physics of Solids}
  \bibinfo{volume}{95} (\bibinfo{year}{2016}) \bibinfo{pages}{320--350}.
\bibitem[{Tsitova et~al.(2021)Tsitova, Bernachy-Barbe, Bary, Dandachli,
  Bourcier, Smaniotto, and Hild}]{Tsitova:2021}
\bibinfo{author}{A.~Tsitova}, \bibinfo{author}{F.~Bernachy-Barbe},
  \bibinfo{author}{B.~Bary}, \bibinfo{author}{S.~Dandachli},
  \bibinfo{author}{C.~Bourcier}, \bibinfo{author}{B.~Smaniotto},
  \bibinfo{author}{F.~Hild},
\newblock \bibinfo{title}{Damage quantification via digital volume correlation
  with heterogeneous mechanical regularization: Application to an in situ
  meso-flexural test on mortar},
\newblock \bibinfo{journal}{Experimental Mechanics}  (\bibinfo{year}{2021})
  \bibinfo{pages}{1--17}.
\bibitem[{Tsitova et~al.(2022)Tsitova, Bernachy-Barbe, Bary, Bourcier, and
  Hild}]{Tsitova:2022}
\bibinfo{author}{A.~Tsitova}, \bibinfo{author}{F.~Bernachy-Barbe},
  \bibinfo{author}{B.~Bary}, \bibinfo{author}{C.~Bourcier},
  \bibinfo{author}{F.~Hild},
\newblock \bibinfo{title}{Identification of microscale fracture models for
  mortar with in-situ tests},
\newblock \bibinfo{journal}{International Journal of Mechanical Sciences}
  \bibinfo{volume}{242} (\bibinfo{year}{2022}).
\bibitem[{Ren et~al.(2014)Ren, Yang, and Sharma}]{Ren:2014}
\bibinfo{author}{W.~Ren}, \bibinfo{author}{Z.~Yang},
  \bibinfo{author}{R.~Sharma},
\newblock \bibinfo{title}{3d meso-scale image-based fracture modelling of
  concrete using cohesive elements},
\newblock \bibinfo{year}{2014}.
\bibitem[{Yang et~al.(2013)Yang, Ren, Mostafavi, Mcdonald, and
  Marrow}]{Yang:2013}
\bibinfo{author}{Z.~Yang}, \bibinfo{author}{W.~Ren},
  \bibinfo{author}{M.~Mostafavi}, \bibinfo{author}{S.~A. Mcdonald},
  \bibinfo{author}{T.~J. Marrow},
\newblock \bibinfo{title}{Characterisation of 3d fracture evolution in concrete
  using in-situ x-ray computed tomography testing and digital volume
  correlation},
\newblock in: \bibinfo{booktitle}{VIII international conference on fracture
  mechanics of concrete and concrete structures},
  \bibinfo{organization}{Toledo, Spain CIMNE}, \bibinfo{year}{2013}, pp.
  \bibinfo{pages}{1--7}.
\bibitem[{Poinard et~al.(2011)Poinard, Piotrowska, Marin, Malecot, and
  Daudeville}]{Poinrad:2011}
\bibinfo{author}{C.~Poinard}, \bibinfo{author}{E.~Piotrowska},
  \bibinfo{author}{P.~Marin}, \bibinfo{author}{Y.~Malecot},
  \bibinfo{author}{L.~Daudeville},
\newblock \bibinfo{title}{Mesoscopic scale modeling of concrete under triaxial
  loading using x-ray tomographic images},
\newblock \bibinfo{year}{2011}, pp. \bibinfo{pages}{117--129}.
\bibitem[{Stamati et~al.(2018)Stamati, Roubin, Andò, and
  Malecot}]{Stamati:2018}
\bibinfo{author}{O.~Stamati}, \bibinfo{author}{E.~Roubin},
  \bibinfo{author}{E.~Andò}, \bibinfo{author}{Y.~Malecot},
\newblock \bibinfo{title}{Phase segmentation of concrete x-ray tomographic
  images at meso-scale: Validation with neutron tomography},
\newblock \bibinfo{journal}{Cement and Concrete Composites}
  \bibinfo{volume}{88} (\bibinfo{year}{2018}) \bibinfo{pages}{8--16}.
\bibitem[{Stamati et~al.(2021)Stamati, Roubin, Andò, Malecot, and
  Charrier}]{Stamati_2021}
\bibinfo{author}{O.~Stamati}, \bibinfo{author}{E.~Roubin},
  \bibinfo{author}{E.~Andò}, \bibinfo{author}{Y.~Malecot},
  \bibinfo{author}{P.~Charrier},
\newblock \bibinfo{title}{Fracturing process of micro-concrete under uniaxial
  and triaxial compression: {Insights} from in-situ {X}-ray mechanical tests},
\newblock \bibinfo{journal}{Cement and Concrete Research} \bibinfo{volume}{149}
  (\bibinfo{year}{2021}) \bibinfo{pages}{106578}.
\bibitem[{Ambati et~al.(2014)Ambati, Gerasimov, and De~Lorenzis}]{Ambati:2014}
\bibinfo{author}{M.~Ambati}, \bibinfo{author}{T.~Gerasimov},
  \bibinfo{author}{L.~De~Lorenzis},
\newblock \bibinfo{title}{A review on phase-field models of brittle fracture
  and a new fast hybrid formulation},
\newblock \bibinfo{journal}{Computational Mechanics} \bibinfo{volume}{55}
  (\bibinfo{year}{2014}).
\bibitem[{Wu et~al.(2021)Wu, Rosi{\'c}, {De Lorenzis}, and Matthies}]{Wu:2021}
\bibinfo{author}{T.~Wu}, \bibinfo{author}{B.~Rosi{\'c}},
  \bibinfo{author}{L.~{De Lorenzis}}, \bibinfo{author}{H.~Matthies},
\newblock \bibinfo{title}{Parameter identification for phase-field modeling of
  fracture: a bayesian approach with sampling-free update},
\newblock \bibinfo{journal}{Computational mechanics} \bibinfo{volume}{67}
  (\bibinfo{year}{2021}) \bibinfo{pages}{435--453}.
\bibitem[{Li et~al.(2019)Li, Chu, Gao, and Liu}]{Li:2019}
\bibinfo{author}{X.~Li}, \bibinfo{author}{D.~Chu}, \bibinfo{author}{Y.~Gao},
  \bibinfo{author}{Z.~Liu},
\newblock \bibinfo{title}{Numerical study on crack propagation in linear
  elastic multiphase composite materials using phase field method},
\newblock \bibinfo{journal}{Engineering Computations} \bibinfo{volume}{36}
  (\bibinfo{year}{2019}) \bibinfo{pages}{307--333}.
\bibitem[{Bourdin et~al.(2000)Bourdin, Francfort, and Marigo}]{Bourdin:2000}
\bibinfo{author}{B.~Bourdin}, \bibinfo{author}{G.~Francfort},
  \bibinfo{author}{J.-J. Marigo},
\newblock \bibinfo{title}{Numerical experiments in revisited brittle fracture},
\newblock \bibinfo{journal}{Journal of the Mechanics and Physics of Solids}
  \bibinfo{volume}{48} (\bibinfo{year}{2000}) \bibinfo{pages}{797--826}.
\bibitem[{Francfort and Marigo(1998)}]{Francfort:1998}
\bibinfo{author}{G.~Francfort}, \bibinfo{author}{J.-J. Marigo},
\newblock \bibinfo{title}{Revisiting brittle fracture as an energy minimization
  problem},
\newblock \bibinfo{journal}{Journal of the Mechanics and Physics of Solids}
  \bibinfo{volume}{46} (\bibinfo{year}{1998}) \bibinfo{pages}{1319--1342}.
\bibitem[{Ambati et~al.(2015)Ambati, Gerasimov, and De~Lorenzis}]{Ambati:2015}
\bibinfo{author}{M.~Ambati}, \bibinfo{author}{T.~Gerasimov},
  \bibinfo{author}{L.~De~Lorenzis},
\newblock \bibinfo{title}{Phase-field modeling of ductile fracture},
\newblock \bibinfo{journal}{Computational Mechanics}  (\bibinfo{year}{2015}).
\bibitem[{Alessi et~al.(2018)Alessi, Marigo, Maurini, and Vidoli}]{Alessi2018}
\bibinfo{author}{R.~Alessi}, \bibinfo{author}{J.~J. Marigo},
  \bibinfo{author}{C.~Maurini}, \bibinfo{author}{S.~Vidoli},
\newblock \bibinfo{title}{{Coupling damage and plasticity for a phase-field
  regularisation of brittle, cohesive and ductile fracture: One-dimensional
  examples}},
\newblock \bibinfo{journal}{International Journal of Mechanical Sciences}
  \bibinfo{volume}{149} (\bibinfo{year}{2018}) \bibinfo{pages}{559--576}.
\bibitem[{Carrara et~al.(2020)Carrara, Ambati, Alessi, and {De
  Lorenzis}}]{Carrara:2020}
\bibinfo{author}{P.~Carrara}, \bibinfo{author}{M.~Ambati},
  \bibinfo{author}{R.~Alessi}, \bibinfo{author}{L.~{De Lorenzis}},
\newblock \bibinfo{title}{A framework to model the fatigue behavior of brittle
  materials based on a variational phase-field approach},
\newblock \bibinfo{journal}{Computer Methods in Applied Mechanics and
  Engineering} \bibinfo{volume}{361} (\bibinfo{year}{2020})
  \bibinfo{pages}{112731}.
\bibitem[{Mesgarnejad et~al.(2019)Mesgarnejad, Imanian, and
  Karma}]{Mesgarnejad2019}
\bibinfo{author}{A.~Mesgarnejad}, \bibinfo{author}{A.~Imanian},
  \bibinfo{author}{A.~Karma},
\newblock \bibinfo{title}{{Phase-field models for fatigue crack growth}},
\newblock \bibinfo{journal}{Theoretical and Applied Fracture Mechanics}
  \bibinfo{volume}{103} (\bibinfo{year}{2019}) \bibinfo{pages}{102282}.
\bibitem[{Heinzmann et~al.(2024)Heinzmann, Carrara, Ambati, Mirzaei, and {De
  Lorenzis}}]{Heinzmann2024}
\bibinfo{author}{J.~Heinzmann}, \bibinfo{author}{P.~Carrara},
  \bibinfo{author}{M.~Ambati}, \bibinfo{author}{A.~M. Mirzaei},
  \bibinfo{author}{L.~{De Lorenzis}}, \bibinfo{title}{{An adaptive acceleration
  scheme for phase-field fatigue computations}}, \bibinfo{number}{September},
  \bibinfo{publisher}{Springer Berlin Heidelberg}, \bibinfo{year}{2024}.
\bibitem[{Nagaraja et~al.(2023)Nagaraja, Römer, Matthies, and {De
  Lorenzis}}]{Nagaraja:2023}
\bibinfo{author}{S.~Nagaraja}, \bibinfo{author}{U.~Römer},
  \bibinfo{author}{H.~G. Matthies}, \bibinfo{author}{L.~{De Lorenzis}},
\newblock \bibinfo{title}{Deterministic and stochastic phase-field modeling of
  anisotropic brittle fracture},
\newblock \bibinfo{journal}{Computer Methods in Applied Mechanics and
  Engineering} \bibinfo{volume}{408} (\bibinfo{year}{2023})
  \bibinfo{pages}{115960}.
\bibitem[{Verhoosel and de~Borst(2013)}]{Verhoosel:2013}
\bibinfo{author}{C.~V. Verhoosel}, \bibinfo{author}{R.~de~Borst},
\newblock \bibinfo{title}{A phase-field model for cohesive fracture},
\newblock \bibinfo{journal}{International Journal for Numerical Methods in
  Engineering} \bibinfo{volume}{96} (\bibinfo{year}{2013})
  \bibinfo{pages}{43--62}.
\bibitem[{Wu(2017)}]{WU2017}
\bibinfo{author}{J.-Y. Wu},
\newblock \bibinfo{title}{A unified phase-field theory for the mechanics of
  damage and quasi-brittle failure},
\newblock \bibinfo{journal}{Journal of the Mechanics and Physics of Solids}
  \bibinfo{volume}{103} (\bibinfo{year}{2017}) \bibinfo{pages}{72--99}.
\bibitem[{Chen and {de Borst}(2022)}]{Chen:2022}
\bibinfo{author}{L.~Chen}, \bibinfo{author}{R.~{de Borst}},
\newblock \bibinfo{title}{Phase-field regularised cohesive zone model for
  interface modelling},
\newblock \bibinfo{journal}{Theoretical and Applied Fracture Mechanics}
  \bibinfo{volume}{122} (\bibinfo{year}{2022}) \bibinfo{pages}{103630}.
\bibitem[{Chen and {de Borst}(2021)}]{Chen2021}
\bibinfo{author}{L.~Chen}, \bibinfo{author}{R.~{de Borst}},
\newblock \bibinfo{title}{Phase-field modelling of cohesive fracture},
\newblock \bibinfo{journal}{European Journal of Mechanics - A/Solids}
  \bibinfo{volume}{90} (\bibinfo{year}{2021}) \bibinfo{pages}{104343}.
\bibitem[{Conti et~al.(2016)Conti, Focardi, and Iaurlano}]{Conti:2016}
\bibinfo{author}{S.~Conti}, \bibinfo{author}{M.~Focardi},
  \bibinfo{author}{F.~Iaurlano},
\newblock \bibinfo{title}{Phase field approximation of cohesive fracture
  models},
\newblock \bibinfo{journal}{Annales de l'Institut Henri Poincaré C, Analyse
  non linéaire} \bibinfo{volume}{33} (\bibinfo{year}{2016})
  \bibinfo{pages}{1033--1067}.
\bibitem[{Lorentz et~al.(2011)Lorentz, Cuvilliez, and
  Kazymyrenko}]{Lorentz:2011}
\bibinfo{author}{E.~Lorentz}, \bibinfo{author}{S.~Cuvilliez},
  \bibinfo{author}{K.~Kazymyrenko},
\newblock \bibinfo{title}{Convergence of a gradient damage model toward a
  cohesive zone model},
\newblock \bibinfo{journal}{Comptes Rendus Mécanique} \bibinfo{volume}{339}
  (\bibinfo{year}{2011}) \bibinfo{pages}{20--26}.
\bibitem[{Brühwiler and Wittmann(1990)}]{Bruhweiler:1990}
\bibinfo{author}{E.~Brühwiler}, \bibinfo{author}{F.~Wittmann},
\newblock \bibinfo{title}{The wedge splitting test, a new method of performing
  stable fracture mechanics tests},
\newblock \bibinfo{journal}{Engineering Fracture Mechanics}
  \bibinfo{volume}{35} (\bibinfo{year}{1990}) \bibinfo{pages}{117--125}.
  \bibinfo{note}{Special Issue Fracture and Damage of Concrete and Rock}.
\bibitem[{Neuner et~al.(2022)Neuner, Smaniotto, and Hofstetter}]{Neuner:2022}
\bibinfo{author}{M.~Neuner}, \bibinfo{author}{S.~Smaniotto},
  \bibinfo{author}{G.~Hofstetter},
\newblock \bibinfo{title}{A modified wedge splitting test for susceptible
  quasi-brittle materials},
\newblock \bibinfo{journal}{Construction and Building Materials}
  \bibinfo{volume}{326} (\bibinfo{year}{2022}) \bibinfo{pages}{126733}.
\bibitem[{Yang et~al.(2020)Yang, Qsymah, Peng, Margetts, and
  Sharma}]{Yang:2020}
\bibinfo{author}{Z.~Yang}, \bibinfo{author}{A.~Qsymah},
  \bibinfo{author}{Y.~Peng}, \bibinfo{author}{L.~Margetts},
  \bibinfo{author}{R.~Sharma},
\newblock \bibinfo{title}{4d characterisation of damage and fracture mechanisms
  of ultra high performance fibre reinforced concrete by in-situ micro x-ray
  computed tomography tests},
\newblock \bibinfo{journal}{Cement and Concrete Composites}
  \bibinfo{volume}{106} (\bibinfo{year}{2020}) \bibinfo{pages}{103473}.
\bibitem[{Wang et~al.(2021)Wang, Li, Jivkov, Li, and Engelberg}]{Wang:2021}
\bibinfo{author}{J.~Wang}, \bibinfo{author}{X.~Li}, \bibinfo{author}{A.~P.
  Jivkov}, \bibinfo{author}{Q.~Li}, \bibinfo{author}{D.~L. Engelberg},
\newblock \bibinfo{title}{Interfacial transition zones in concrete meso-scale
  models – balancing physical realism and computational efficiency},
\newblock \bibinfo{journal}{Construction and Building Materials}
  \bibinfo{volume}{293} (\bibinfo{year}{2021}) \bibinfo{pages}{123332}.
\bibitem[{Gerasimov and {De Lorenzis}(2019)}]{Gerasimov:2019}
\bibinfo{author}{T.~Gerasimov}, \bibinfo{author}{L.~{De Lorenzis}},
\newblock \bibinfo{title}{On penalization in variational phase-field models of
  brittle fracture},
\newblock \bibinfo{journal}{Computer Methods in Applied Mechanics and
  Engineering} \bibinfo{volume}{354} (\bibinfo{year}{2019})
  \bibinfo{pages}{990--1026}.
\bibitem[{Amor et~al.(2009)Amor, Marigo, and Maurini}]{Amor:2009}
\bibinfo{author}{H.~Amor}, \bibinfo{author}{J.-J. Marigo},
  \bibinfo{author}{C.~Maurini},
\newblock \bibinfo{title}{Regularized formulation of the variational brittle
  fracture with unilateral contact: Numerical experiments},
\newblock \bibinfo{journal}{Journal of the Mechanics and Physics of Solids}
  \bibinfo{volume}{57} (\bibinfo{year}{2009}) \bibinfo{pages}{1209--1229}.
\bibitem[{Miehe et~al.(2010)Miehe, Welschinger, and Hofacker}]{Miehe:2010}
\bibinfo{author}{C.~Miehe}, \bibinfo{author}{F.~Welschinger},
  \bibinfo{author}{M.~Hofacker},
\newblock \bibinfo{title}{Thermodynamically consistent phase‐field models of
  fracture: Variational principles and multi‐field fe implementations},
\newblock \bibinfo{journal}{International Journal for Numerical Methods in
  Engineering} \bibinfo{volume}{83} (\bibinfo{year}{2010}) \bibinfo{pages}{1273
  -- 1311}.
\bibitem[{Vicentini et~al.(2024)Vicentini, Zolesi, Carrara, Maurini, and {De
  Lorenzis}}]{Vicentini2024}
\bibinfo{author}{F.~Vicentini}, \bibinfo{author}{C.~Zolesi},
  \bibinfo{author}{P.~Carrara}, \bibinfo{author}{C.~Maurini},
  \bibinfo{author}{L.~{De Lorenzis}},
\newblock \bibinfo{title}{{On the energy decomposition in variational
  phase-field models for brittle fracture under multi-axial stress states}},
\newblock \bibinfo{journal}{International Journal of Fracture}
  (\bibinfo{year}{2024}).
\bibitem[{De~Lorenzis and Maurini(2021)}]{DeLorenzis:2021}
\bibinfo{author}{L.~De~Lorenzis}, \bibinfo{author}{C.~Maurini},
\newblock \bibinfo{title}{Nucleation under multi-axial loading in variational
  phase-field models of brittle fracture},
\newblock \bibinfo{journal}{International Journal of Fracture}
  \bibinfo{volume}{237} (\bibinfo{year}{2021}).
\bibitem[{Braides(1998)}]{Braides:1998}
\bibinfo{author}{A.~Braides},
\newblock \bibinfo{title}{Approximation of free-discontinuity problems}
  (\bibinfo{year}{1998}).
\bibitem[{Pham et~al.(2010)Pham, Amor, Marigo, and Maurini}]{Pham:2010}
\bibinfo{author}{K.~Pham}, \bibinfo{author}{H.~Amor}, \bibinfo{author}{J.-J.
  Marigo}, \bibinfo{author}{C.~Maurini},
\newblock \bibinfo{title}{Gradient damage models and their use to approximate
  brittle fracture},
\newblock \bibinfo{journal}{International Journal of Damage Mechanics}
  \bibinfo{volume}{20} (\bibinfo{year}{2010}) \bibinfo{pages}{618--652}.
\bibitem[{Pham et~al.(2011)Pham, Marigo, and Maurini}]{Pham:2011}
\bibinfo{author}{K.~Pham}, \bibinfo{author}{J.-J. Marigo},
  \bibinfo{author}{C.~Maurini},
\newblock \bibinfo{title}{The issues of the uniqueness and the stability of the
  homogeneous response in uniaxial tests with gradient damage models},
\newblock \bibinfo{journal}{Journal of the Mechanics and Physics of Solids}
  \bibinfo{volume}{59} (\bibinfo{year}{2011}) \bibinfo{pages}{1163--1190}.
\bibitem[{Kuhn et~al.(2015)Kuhn, Schlüter, and Müller}]{Kuhn:2015}
\bibinfo{author}{C.~Kuhn}, \bibinfo{author}{A.~Schlüter},
  \bibinfo{author}{R.~Müller},
\newblock \bibinfo{title}{On degradation functions in phase field fracture
  models},
\newblock \bibinfo{journal}{Computational Materials Science}
  \bibinfo{volume}{108} (\bibinfo{year}{2015}) \bibinfo{pages}{374--384}.
  \bibinfo{note}{Selected Articles from Phase-field Method 2014 International
  Seminar}.
\bibitem[{Lorentz et~al.(2012)Lorentz, Cuvilliez, and
  Kazymyrenko}]{Lorentz:2012}
\bibinfo{author}{E.~Lorentz}, \bibinfo{author}{S.~Cuvilliez},
  \bibinfo{author}{K.~Kazymyrenko},
\newblock \bibinfo{title}{Modelling large crack propagation: From gradient
  damage to cohesive zone models},
\newblock \bibinfo{journal}{International Journal of Fracture}
  \bibinfo{volume}{178} (\bibinfo{year}{2012}).
\bibitem[{Tanne(2017)}]{Tanne:thesis}
\bibinfo{author}{E.~Tanne}, \bibinfo{title}{Variational phase-field models from
  brittle to ductile fracture : nucleation and propagation}, Ph.D. thesis,
  \bibinfo{year}{2017}.
\bibitem[{Miehe et~al.(2010)Miehe, Hofacker, and Welschinger}]{Miehe:20102}
\bibinfo{author}{C.~Miehe}, \bibinfo{author}{M.~Hofacker},
  \bibinfo{author}{F.~Welschinger},
\newblock \bibinfo{title}{A phase field model for rate-independent crack
  propagation: Robust algorithmic implementation based on operator splits},
\newblock \bibinfo{journal}{Computer Methods in Applied Mechanics and
  Engineering} \bibinfo{volume}{199} (\bibinfo{year}{2010})
  \bibinfo{pages}{2765--2778}.
\bibitem[{Gerasimov and De~Lorenzis(2015)}]{Gerasimov:2015}
\bibinfo{author}{T.~Gerasimov}, \bibinfo{author}{L.~De~Lorenzis},
\newblock \bibinfo{title}{A line search assisted monolithic approach for
  phase-field computing of brittle fracture},
\newblock \bibinfo{journal}{Computer Methods in Applied Mechanics and
  Engineering} \bibinfo{volume}{312} (\bibinfo{year}{2015}).
\bibitem[{Farrell and Maurini(2015)}]{Farrell:2015}
\bibinfo{author}{P.~Farrell}, \bibinfo{author}{C.~Maurini},
\newblock \bibinfo{title}{Linear and nonlinear solvers for variational
  phase-field models of brittle fracture},
\newblock \bibinfo{journal}{International Journal for Numerical Methods in
  Engineering} \bibinfo{volume}{109} (\bibinfo{year}{2015}).
\bibitem[{Lengsfeld et~al.(1998)Lengsfeld, Schmitt, Alter, Kaminsky, and
  Leppek}]{Lengsfeld:1998}
\bibinfo{author}{M.~Lengsfeld}, \bibinfo{author}{J.~Schmitt},
  \bibinfo{author}{P.~Alter}, \bibinfo{author}{J.~Kaminsky},
  \bibinfo{author}{R.~Leppek},
\newblock \bibinfo{title}{Comparison of geometry-based and ct voxel-based
  finite element modelling and experimental validation},
\newblock \bibinfo{journal}{Medical Engineering {\&} Physics}
  \bibinfo{volume}{20} (\bibinfo{year}{1998}) \bibinfo{pages}{515--522}.
\bibitem[{Arndt et~al.(2021)Arndt, Bangerth, Blais, Fehling, Gassm{\"o}ller,
  Heister, Heltai, K{\"o}cher, Kronbichler, Maier, Munch, Pelteret, Proell,
  Simon, Turcksin, Wells, and Zhang}]{Arndt:2021}
\bibinfo{author}{D.~Arndt}, \bibinfo{author}{W.~Bangerth},
  \bibinfo{author}{B.~Blais}, \bibinfo{author}{M.~Fehling},
  \bibinfo{author}{R.~Gassm{\"o}ller}, \bibinfo{author}{T.~Heister},
  \bibinfo{author}{L.~Heltai}, \bibinfo{author}{U.~K{\"o}cher},
  \bibinfo{author}{M.~Kronbichler}, \bibinfo{author}{M.~Maier},
  \bibinfo{author}{P.~Munch}, \bibinfo{author}{J.-P. Pelteret},
  \bibinfo{author}{S.~Proell}, \bibinfo{author}{K.~Simon},
  \bibinfo{author}{B.~Turcksin}, \bibinfo{author}{D.~Wells},
  \bibinfo{author}{J.~Zhang},
\newblock \bibinfo{title}{The \texttt{deal.II} library, version 9.3},
\newblock \bibinfo{journal}{Journal of Numerical Mathematics}
  \bibinfo{volume}{29} (\bibinfo{year}{2021}) \bibinfo{pages}{171--186}.
\bibitem[{Balay et~al.(1997)Balay, Gropp, McInnes, and Smith}]{PETSC}
\bibinfo{author}{S.~Balay}, \bibinfo{author}{W.~D. Gropp},
  \bibinfo{author}{L.~C. McInnes}, \bibinfo{author}{B.~F. Smith},
\newblock \bibinfo{title}{Efficient management of parallelism in object
  oriented numerical software libraries},
\newblock in: \bibinfo{editor}{E.~Arge}, \bibinfo{editor}{A.~M. Bruaset},
  \bibinfo{editor}{H.~P. Langtangen} (Eds.), \bibinfo{booktitle}{Modern
  Software Tools in Scientific Computing}, \bibinfo{publisher}{Birkh{\"{a}}user
  Press}, \bibinfo{year}{1997}, pp. \bibinfo{pages}{163--202}.
\bibitem[{Corrêa et~al.(2022)Corrêa, Hao, Neerup, Almeida, Shi, Thomsen, and
  Fosbøl}]{Correa:2022}
\bibinfo{author}{L.~F. Corrêa}, \bibinfo{author}{J.~Hao},
  \bibinfo{author}{R.~Neerup}, \bibinfo{author}{S.~Almeida},
  \bibinfo{author}{M.~Shi}, \bibinfo{author}{K.~Thomsen},
  \bibinfo{author}{P.~L. Fosbøl},
\newblock \bibinfo{title}{Review of barium sulphate solubility measurements},
\newblock \bibinfo{journal}{Geothermics} \bibinfo{volume}{104}
  (\bibinfo{year}{2022}) \bibinfo{pages}{102465}.
\bibitem[{Shakoorioskooie et~al.(2022)Shakoorioskooie, Griffa, Leemann, Zboray,
  and Lura}]{Shakoorioskooie:2022}
\bibinfo{author}{M.~Shakoorioskooie}, \bibinfo{author}{M.~Griffa},
  \bibinfo{author}{A.~Leemann}, \bibinfo{author}{R.~Zboray},
  \bibinfo{author}{P.~Lura},
\newblock \bibinfo{title}{Quantitative analysis of the evolution of {ASR}
  products and crack networks in the context of the concrete mesostructure},
\newblock \bibinfo{journal}{Cement and Concrete Research} \bibinfo{volume}{162}
  (\bibinfo{year}{2022}) \bibinfo{pages}{106992}.
\bibitem[{Monteiro et~al.(1985)Monteiro, Maso, and Ollivier}]{Monteiro1985}
\bibinfo{author}{P.~J.~M. Monteiro}, \bibinfo{author}{J.~C. Maso},
  \bibinfo{author}{J.~P. Ollivier},
\newblock \bibinfo{title}{{The aggregate-mortar interface}},
\newblock \bibinfo{journal}{Cement and Concrete Research} \bibinfo{volume}{15}
  (\bibinfo{year}{1985}) \bibinfo{pages}{953--958}.
\bibitem[{Maso(1992)}]{Maso1992}
\bibinfo{author}{J.~Maso},
\newblock \bibinfo{title}{{Interfaces in cementitious composites}},
\newblock in: \bibinfo{editor}{J.~Maso} (Ed.), \bibinfo{booktitle}{Rilem
  proceedings 18}, \bibinfo{publisher}{E \& FN Spon}, \bibinfo{year}{1992}.
\bibitem[{Carrara and {De Lorenzis}(2017)}]{Carrara2017b}
\bibinfo{author}{P.~Carrara}, \bibinfo{author}{L.~{De Lorenzis}},
\newblock \bibinfo{title}{{Chloride diffusivity of the interfacial transition
  zone and bulk paste in concrete from microscale analysis}},
\newblock \bibinfo{journal}{Modelling and Simulation in Materials Science and
  Engineering} \bibinfo{volume}{25} (\bibinfo{year}{2017})
  \bibinfo{pages}{045011}.
\bibitem[{Husem(2003)}]{Husem:2003}
\bibinfo{author}{M.~Husem},
\newblock \bibinfo{title}{The effects of bond strengths between lightweight and
  ordinary aggregate-mortar, aggregate-cement paste on the mechanical
  properties of concrete},
\newblock \bibinfo{journal}{Materials Science and Engineering: A}
  \bibinfo{volume}{363} (\bibinfo{year}{2003}) \bibinfo{pages}{152--158}.
\bibitem[{Yang(1998)}]{Yang:1998}
\bibinfo{author}{C.~Yang},
\newblock \bibinfo{title}{Effect of the transition zone on the elastic moduli
  of mortar},
\newblock \bibinfo{journal}{Cement and Concrete Research} \bibinfo{volume}{28}
  (\bibinfo{year}{1998}) \bibinfo{pages}{727--736}.
\bibitem[{Jebli et~al.(2018)Jebli, Jamin, Malachanne, Garcia-Diaz, and {El
  Youssoufi}}]{Jebli:2018}
\bibinfo{author}{M.~Jebli}, \bibinfo{author}{F.~Jamin},
  \bibinfo{author}{E.~Malachanne}, \bibinfo{author}{E.~Garcia-Diaz},
  \bibinfo{author}{M.~{El Youssoufi}},
\newblock \bibinfo{title}{Experimental characterization of mechanical
  properties of the cement-aggregate interface in concrete},
\newblock \bibinfo{journal}{Construction and Building Materials}
  \bibinfo{volume}{161} (\bibinfo{year}{2018}) \bibinfo{pages}{16--25}.
\bibitem[{Scrivener et~al.(2004)Scrivener, Crumbie, and
  Laugesen}]{Scrivener2004}
\bibinfo{author}{K.~L. Scrivener}, \bibinfo{author}{A.~K. Crumbie},
  \bibinfo{author}{P.~Laugesen},
\newblock \bibinfo{title}{{The interfacial transition zone (ITZ) between cement
  paste and aggregate in concrete}},
\newblock \bibinfo{journal}{Interface Science} \bibinfo{volume}{12}
  (\bibinfo{year}{2004}) \bibinfo{pages}{411--421}.
\bibitem[{Carrara and {De Lorenzis}(2017)}]{Carrara2017}
\bibinfo{author}{P.~Carrara}, \bibinfo{author}{L.~{De Lorenzis}},
\newblock \bibinfo{title}{{Consistent identification of the interfacial
  transition zone in simulated cement microstructures}},
\newblock \bibinfo{journal}{Cement and Concrete Composites}
  \bibinfo{volume}{80} (\bibinfo{year}{2017}) \bibinfo{pages}{224--234}.
\bibitem[{262(2022)}]{SIA_262_2022}
\bibinfo{author}{S.~262}, \bibinfo{title}{Prüfung von {Festbeton} - {Teil} 13:
  {Bestimmung} des {Elastizitätsmoduls} unter {Druckbelastung}
  ({Sekantenmodul})}, \bibinfo{type}{Standard} \bibinfo{number}{SIA 262.263 -
  SN EN 12390-13:2021}, Schweizer Norm, \bibinfo{address}{Zürich},
  \bibinfo{year}{2022}.
\bibitem[{Institution(2019)}]{EN12390}
\bibinfo{author}{E.~S. Institution}, \bibinfo{title}{EN 12390-13. Testing
  Hardened Concrete: Part 13. Determination of secant modulus of elasticity in
  compression}, \bibinfo{number}{Part 13}, \bibinfo{publisher}{European
  Standards Institution}, \bibinfo{year}{2019}.
\bibitem[{Wang et~al.(2021)Wang, Xing, Jivkov, Li, and
  Engelberg}]{WangXing:2021}
\bibinfo{author}{J.~Wang}, \bibinfo{author}{l.~Xing},
  \bibinfo{author}{A.~Jivkov}, \bibinfo{author}{Q.~Li},
  \bibinfo{author}{D.~Engelberg},
\newblock \bibinfo{title}{Interfacial transition zones in concrete meso-scale
  models - balancing physical realism and computational efficiency},
\newblock \bibinfo{journal}{Construction and Building Materials}
  \bibinfo{volume}{293} (\bibinfo{year}{2021}) \bibinfo{pages}{123332}.
\bibitem[{Chen et~al.(2024)Chen, Zhang, Wang, Zhao, and Wang}]{Chen:2024}
\bibinfo{author}{Q.~Chen}, \bibinfo{author}{J.~Zhang},
  \bibinfo{author}{Z.~Wang}, \bibinfo{author}{T.~Zhao},
  \bibinfo{author}{Z.~Wang},
\newblock \bibinfo{title}{A review of the interfacial transition zones in
  concrete: Identification, physical characteristics, and mechanical
  properties},
\newblock \bibinfo{journal}{Engineering Fracture Mechanics}
  \bibinfo{volume}{300} (\bibinfo{year}{2024}) \bibinfo{pages}{109979}.
\bibitem[{Xiao et~al.(2013)Xiao, Li, Sun, Lange, and Shah}]{Xiao:2013}
\bibinfo{author}{J.~Xiao}, \bibinfo{author}{W.~Li}, \bibinfo{author}{Z.~Sun},
  \bibinfo{author}{D.~A. Lange}, \bibinfo{author}{S.~P. Shah},
\newblock \bibinfo{title}{Properties of interfacial transition zones in
  recycled aggregate concrete tested by nanoindentation},
\newblock \bibinfo{journal}{Cement and Concrete Composites}
  \bibinfo{volume}{37} (\bibinfo{year}{2013}) \bibinfo{pages}{276--292}.
\bibitem[{Gupta and Seshagiri~Rao(1998)}]{Gupta:1998}
\bibinfo{author}{A.~Gupta}, \bibinfo{author}{K.~Seshagiri~Rao},
\newblock \bibinfo{title}{Index properties of weathered rocks:
  inter-relationships and applicability},
\newblock \bibinfo{journal}{Bulletin of Engineering Geology and the
  Environment} \bibinfo{volume}{57} (\bibinfo{year}{1998})
  \bibinfo{pages}{161--172}.
\bibitem[{Gupta and Rao(2000)}]{Gupta:2000}
\bibinfo{author}{A.~Gupta}, \bibinfo{author}{K.~S. Rao},
\newblock \bibinfo{title}{Weathering effects on the strength and deformational
  behaviour of crystalline rocks under uniaxial compression state},
\newblock \bibinfo{journal}{Engineering geology} \bibinfo{volume}{56}
  (\bibinfo{year}{2000}) \bibinfo{pages}{257--274}.
\bibitem[{Thomas and Slate(1963)}]{Thomas:1963}
\bibinfo{author}{T.~Thomas}, \bibinfo{author}{F.~O. Slate},
\newblock \bibinfo{title}{Tensile bond strength between aggregate and cement
  paste or mortar},
\newblock in: \bibinfo{booktitle}{Journal Proceedings},
  volume~\bibinfo{volume}{60}, \bibinfo{year}{1963}, pp.
  \bibinfo{pages}{465--486}.
\bibitem[{Tschegg et~al.(1995)Tschegg, Rotter, Roelfstra, Bourgund, and
  Jussel}]{Tschegg:1995}
\bibinfo{author}{E.~K. Tschegg}, \bibinfo{author}{H.~M. Rotter},
  \bibinfo{author}{P.~E. Roelfstra}, \bibinfo{author}{U.~Bourgund},
  \bibinfo{author}{P.~Jussel},
\newblock \bibinfo{title}{Fracture mechanical behavior of aggregate–cement
  matrix interfaces},
\newblock \bibinfo{journal}{Journal of Materials in Civil Engineering}
  \bibinfo{volume}{7} (\bibinfo{year}{1995}) \bibinfo{pages}{199--203}.
\bibitem[{Rao and Prasad(2002)}]{Rao:2002}
\bibinfo{author}{G.~Rao}, \bibinfo{author}{B.~Prasad},
\newblock \bibinfo{title}{Influence of the roughness of aggregate surface on
  the interface bond strength},
\newblock \bibinfo{journal}{Cement and Concrete Research} \bibinfo{volume}{32}
  (\bibinfo{year}{2002}) \bibinfo{pages}{253--257}.
\bibitem[{Irie et~al.(2022)Irie, Spin-Neto, Borges, Wenzel, and
  Soares}]{Milena:2022}
\bibinfo{author}{M.~Irie}, \bibinfo{author}{R.~Spin-Neto},
  \bibinfo{author}{J.~Borges}, \bibinfo{author}{A.~Wenzel},
  \bibinfo{author}{P.~Soares},
\newblock \bibinfo{title}{Effect of data binning and frame averaging for
  micro-ct image acquisition on the morphometric outcome of bone repair
  assessment},
\newblock \bibinfo{journal}{Scientific Reports} \bibinfo{volume}{12}
  (\bibinfo{year}{2022}).
\bibitem[{{RX solutions}(2023)}]{X-Act}
\bibinfo{author}{{RX solutions}}, \bibinfo{title}{X-act}, \bibinfo{year}{2023}.
  \URLprefix \url{www.rxsolutions.fr}, \bibinfo{note}{\textit{vers.} 23.04.1,
  2023-09-20}.
\bibitem[{Schulze et~al.(2011)Schulze, Heil, Groß, Bruellmann, Dranischnikow,
  Schwanecke, and Schoemer}]{Schulze}
\bibinfo{author}{R.~Schulze}, \bibinfo{author}{U.~Heil},
  \bibinfo{author}{D.~Groß}, \bibinfo{author}{D.~Bruellmann},
  \bibinfo{author}{E.~Dranischnikow}, \bibinfo{author}{U.~Schwanecke},
  \bibinfo{author}{E.~Schoemer},
\newblock \bibinfo{title}{Artefacts in cbct: a review},
\newblock \bibinfo{journal}{Dento maxillo facial radiology}
  \bibinfo{volume}{40} (\bibinfo{year}{2011}) \bibinfo{pages}{265--73}.
\bibitem[{Barrett and Keat(2004)}]{Barrett:2004}
\bibinfo{author}{J.~Barrett}, \bibinfo{author}{N.~Keat},
\newblock \bibinfo{title}{Artifacts in ct: Recognition and avoidance},
\newblock \bibinfo{journal}{Radiographics : a review publication of the
  Radiological Society of North America, Inc} \bibinfo{volume}{24}
  (\bibinfo{year}{2004}) \bibinfo{pages}{1679--91}.
\bibitem[{Paganin et~al.(2002)Paganin, Mayo, Gureyev, Miller, and
  Wilkins}]{Paganin_2002}
\bibinfo{author}{D.~Paganin}, \bibinfo{author}{S.~C. Mayo},
  \bibinfo{author}{T.~E. Gureyev}, \bibinfo{author}{P.~R. Miller},
  \bibinfo{author}{S.~W. Wilkins},
\newblock \bibinfo{title}{Simultaneous phase and amplitude extraction from a
  single defocused image of a homogeneous object},
\newblock \bibinfo{journal}{Journal of Microscopy} \bibinfo{volume}{206}
  (\bibinfo{year}{2002}) \bibinfo{pages}{33--40}.
\bibitem[{{Thermofischer Scientific}(2023)}]{Avizo}
\bibinfo{author}{{Thermofischer Scientific}}, \bibinfo{title}{Avizo 3d},
  \bibinfo{year}{2023}. \URLprefix \url{www.thermofisher.com},
  \bibinfo{note}{\textit{vers.} 2023.2, 2023-11-20}.
\bibitem[{Buades et~al.(2011)Buades, Coll, and Morel}]{Buades:2011}
\bibinfo{author}{A.~Buades}, \bibinfo{author}{B.~Coll}, \bibinfo{author}{J.-M.
  Morel},
\newblock \bibinfo{title}{{Non-Local Means Denoising}},
\newblock \bibinfo{journal}{{Image Processing On Line}} \bibinfo{volume}{1}
  (\bibinfo{year}{2011}) \bibinfo{pages}{208--212}.
\bibitem[{Beucher and Meyer(1993)}]{Beucher:1993}
\bibinfo{author}{S.~Beucher}, \bibinfo{author}{F.~Meyer}, \bibinfo{title}{The
  Morphological Approach to Segmentation: The Watershed Transformation}, volume
  \bibinfo{volume}{Vol. 34}, \bibinfo{year}{1993}, p.
  \bibinfo{pages}{433–481}.
\bibitem[{Meyer(1994)}]{Meyer1994}
\bibinfo{author}{F.~Meyer},
\newblock \bibinfo{title}{{Topographic distance and watershed lines}},
\newblock \bibinfo{journal}{Signal Processing} \bibinfo{volume}{38}
  (\bibinfo{year}{1994}) \bibinfo{pages}{113--125}.
\bibitem[{Cavalli et~al.(2016)Cavalli, Griffa, Bressi, Partl, Tebaldi, and
  Poulikakos}]{Michele:2016}
\bibinfo{author}{M.~C. Cavalli}, \bibinfo{author}{M.~Griffa},
  \bibinfo{author}{S.~Bressi}, \bibinfo{author}{M.~Partl},
  \bibinfo{author}{G.~Tebaldi}, \bibinfo{author}{L.~Poulikakos},
\newblock \bibinfo{title}{Multiscale imaging and characterization of the effect
  of mixing temperature on asphalt concrete containing recycled components},
\newblock \bibinfo{journal}{Journal of Microscopy} \bibinfo{volume}{264}
  (\bibinfo{year}{2016}) \bibinfo{pages}{22--33}.
\bibitem[{Erdoğan et~al.(2007)Erdoğan, Garboczi, and Fowler}]{Erdogan:2007}
\bibinfo{author}{S.~Erdoğan}, \bibinfo{author}{E.~Garboczi},
  \bibinfo{author}{D.~Fowler},
\newblock \bibinfo{title}{Shape and size of microfine aggregates: {X}-ray
  microcomputed tomography vs. laser diffraction},
\newblock \bibinfo{journal}{Powder Technology} \bibinfo{volume}{177}
  (\bibinfo{year}{2007}) \bibinfo{pages}{53--63}.
\bibitem[{Ketcham and Mote(2019)}]{Ketcham_2019}
\bibinfo{author}{R.~A. Ketcham}, \bibinfo{author}{A.~S. Mote},
\newblock \bibinfo{title}{Accurate {Measurement} of {Small} {Features} in
  {X}‐{Ray} {CT} {Data} {Volumes}, {Demonstrated} {Using} {Gold} {Grains}},
\newblock \bibinfo{journal}{Journal of Geophysical Research: Solid Earth}
  \bibinfo{volume}{124} (\bibinfo{year}{2019}) \bibinfo{pages}{3508--3529}.
\bibitem[{Leclerc et~al.(2012)Leclerc, Périé, Hild, and Roux}]{Leclerc:2012}
\bibinfo{author}{H.~Leclerc}, \bibinfo{author}{J.-N. Périé},
  \bibinfo{author}{F.~Hild}, \bibinfo{author}{S.~Roux},
\newblock \bibinfo{title}{Digital volume correlation: What are the limits to
  the spatial resolution?},
\newblock \bibinfo{journal}{Mecanique and Industries} \bibinfo{volume}{13}
  (\bibinfo{year}{2012}) \bibinfo{pages}{361--371}.
\bibitem[{Bay(2008)}]{Bay2008}
\bibinfo{author}{B.~K. Bay},
\newblock \bibinfo{title}{Methods and applications of digital volume
  correlation},
\newblock \bibinfo{journal}{Journal of Strain Analysis for Engineering Design}
  \bibinfo{volume}{43} (\bibinfo{year}{2008}) \bibinfo{pages}{745--760}.
\bibitem[{Rührnschopf and Klingenbeck(2011)}]{Ruhrnschopf:2011}
\bibinfo{author}{E.-P. Rührnschopf}, \bibinfo{author}{K.~Klingenbeck},
\newblock \bibinfo{title}{A general framework and review of scatter correction
  methods in x-ray cone-beam computerized tomography. part 1: Scatter
  compensation approaches},
\newblock \bibinfo{journal}{Medical Physics} \bibinfo{volume}{38}
  (\bibinfo{year}{2011}) \bibinfo{pages}{4296--4311}.
\bibitem[{R{\"{u}}hrnschopf and Klingenbeck(2011)}]{Ruhrnschopf2011b}
\bibinfo{author}{E.~P. R{\"{u}}hrnschopf}, \bibinfo{author}{K.~Klingenbeck},
\newblock \bibinfo{title}{{A general framework and review of scatter correction
  methods in cone beam CT. Part 2: Scatter estimation approaches}},
\newblock \bibinfo{journal}{Medical Physics} \bibinfo{volume}{38}
  (\bibinfo{year}{2011}) \bibinfo{pages}{5186--5199}.
\bibitem[{Lifton et~al.(2015)Lifton, Malcolm, and McBride}]{Lifton2015}
\bibinfo{author}{J.~J. Lifton}, \bibinfo{author}{A.~A. Malcolm},
  \bibinfo{author}{J.~W. McBride},
\newblock \bibinfo{title}{{An experimental study on the influence of scatter
  and beam hardening in x-ray CT for dimensional metrology}},
\newblock \bibinfo{journal}{Measurement Science and Technology}
  \bibinfo{volume}{27} (\bibinfo{year}{2015}).
\bibitem[{Lorensen and Cline(1998)}]{Lorensen1998}
\bibinfo{author}{W.~E. Lorensen}, \bibinfo{author}{H.~E. Cline},
\newblock \bibinfo{title}{{Marching cubes: a high resolution 3D surface
  construction algorithm}},
\newblock in: \bibinfo{booktitle}{Seminal graphics},
  volume~\bibinfo{volume}{21}, \bibinfo{publisher}{ACM}, \bibinfo{address}{New
  York, NY, USA}, \bibinfo{year}{1998}, pp. \bibinfo{pages}{347--353}.
\bibitem[{Mathur et~al.(2023)Mathur, Brozovich, and Rausch}]{mathur_2023_brief}
\bibinfo{author}{M.~Mathur}, \bibinfo{author}{J.~M. Brozovich},
  \bibinfo{author}{M.~K. Rausch},
\newblock \bibinfo{title}{A brief note on building augmented reality models for
  scientific visualization},
\newblock \bibinfo{journal}{Finite Elements in Analysis and Design}
  \bibinfo{volume}{213} (\bibinfo{year}{2023}).

\end{thebibliography}

\end{document}